\PassOptionsToPackage{dvipsnames}{xcolor}
\documentclass[article,nojss,nofooter]{jss}\usepackage[]{graphicx}\usepackage[]{xcolor}
\makeatletter
\def\maxwidth{ %
  \ifdim\Gin@nat@width>\linewidth
    \linewidth
  \else
    \Gin@nat@width
  \fi
}
\makeatother

\definecolor{fgcolor}{rgb}{0.345, 0.345, 0.345}

\usepackage{framed}
\makeatletter
 {\par\unskip\endMakeFramed%
 \at@end@of@kframe}
\makeatother

\definecolor{shadecolor}{rgb}{.97, .97, .97}
\definecolor{messagecolor}{rgb}{0, 0, 0}
\definecolor{warningcolor}{rgb}{1, 0, 1}
\definecolor{errorcolor}{rgb}{1, 0, 0}
\newenvironment{knitrout}{}{} % an empty environment to be redefined in TeX

\usepackage{alltt}

\usepackage{orcidlink,thumbpdf,lmodern}
\usepackage{booktabs}
\usepackage{longtable}
\usepackage{colortbl}
\usepackage{float}
\usepackage{macros}
\AtBeginDocument{\setlength{\LTcapwidth}{\linewidth}}
\definecolor{fgcolor}{rgb}{0.345, 0.345, 0.345}

\author{Leticia W\"uthrich {\normalfont and} Torsten Hothorn\\
        Universit\"at Z\"urich}
\Plainauthor{Wüthrich \& Hothorn}

\title{Joint Estimation of Marginal and Heterogeneous Treatment Effects}
\Plaintitle{Joint Estimation of Marginal and Heterogeneous Treatment Effects}
\Shorttitle{Joint Estimation of Marginal and Heterogeneous Treatment Effects}

\Abstract{
  Randomized clinical trials typically aim to estimate a marginal treatment effect. While covariate adjustment can improve precision, it may change the estimand in nonlinear models due to noncollapsibility, leading to conditional rather than marginal treatment effects. At the same time, identifying prognostic and predictive covariates is important for understanding treatment effect heterogeneity and informing clinical decision-making. Keeping marginal interpretability while allowing efficiency gains and assessment of heterogeneity remains a methodological challenge.

  In this work, we extend nonparanormal adjusted marginal inference to allow for heterogeneous treatment effects. The proposed framework embeds the marginal treatment effect directly in a joint model for the outcome and baseline covariates. This construction preserves marginal interpretability while adjusting for potentially prognostic and/or predictive covariates. The method applies to continuous, binary, ordinal, and time-to-event outcomes and allows explicit estimation and ranking of prognostic and predictive covariates on a common scale.

  For continuous outcomes, we show that the asymptotic variance of the marginal treatment effect measured as Cohen's $d$ is never worse and often better under covariate adjustment than without adjustment. Efficiency gains are primarily driven by prognostic effects, with realistic predictive effects contributing little additional improvement. Simulation studies confirm these findings across outcome types and demonstrate unbiased and more efficient estimation of marginal effects for Cohen's $d$, log-odds ratios, and log-hazard ratios. Application to an acupuncture trial demonstrates that the method reproduces the original trial findings while improving efficiency and allowing ranking of prognostic and predictive covariates.
}

\Plainkeywords{noncollapsibility, covariate adjustment, prognostic covariate, predictive covariate, transformation model, randomized trial}

\Address{
  Leticia W\"uthrich, Torsten Hothorn\\
  Department of Biostatistics\\
  University of Zurich\\
  Hirschengraben 84, 8001 Zurich, Switzerland
}
\IfFileExists{upquote.sty}{\usepackage{upquote}}{}
\begin{document}

\graphicspath{{./figure/}}
\DeclareGraphicsExtensions{.pdf,.png}

%%%%%%%%%%%%%%%%%%%%%%%%%%%%%%%%%%%%%%%%%%%%%%%%%%%%%%%%%%%%%%%%%%%%%%

% LaTeX file for Chapter 02

\section{Introduction}

Randomized clinical trials are typically designed to answer a simple question: what is the average effect of assigning the new treatment rather than control in the trial population \citep{EMA2020:estimand}? This is the \emph{marginal treatment effect} and is defined with respect to the population represented by the trial's inclusion and exclusion criteria \citep{Lancker2024:covadj}. Under simple randomization and in the absence of loss to follow-up, this effect can be estimated without adjusting for baseline covariates, because treatment assignment is independent of baseline characteristics, and in expectation, randomization ensures covariate balance between treatment arms. However, in finite samples, especially in smaller trials, substantial imbalance may occur by chance, leading to increased variability of the treatment effect estimate and, in some cases, large deviations from the true effect \citep{Nguyen2017:bias}. For this reason, and more generally to increase efficiency, covariate adjustment is often recommended. 
% Adjusting for baseline covariates can improve precision by explaining additional variation in the outcome, thereby reducing the variance of the treatment effect estimator \citep{Lancker2024:covadj}. 
Recent regulatory guidance has renewed attention to when and how baseline covariates should be incorporated in primary analyses \citep{FDA2023:covadj}.

Although covariate adjustment is often motivated by gains in precision, baseline characteristics may also provide clinically meaningful information. Identifying patient characteristics that are associated with the outcome, or that modify the treatment effect, can inform treatment decisions and guide personalized medicine strategies \citep{Hauck1998:covadj, Kraemer2006:estimands, Sechidis2018:progpred}. In this setting, \emph{prognostic} covariates are those associated with the outcome irrespective of treatment, and may provide useful information about expected outcomes without affecting treatment choice. \emph{Predictive} covariates, in contrast, modify the treatment effect, meaning they interact with treatment and may guide individualized treatment decisions \citep{Ballman2015:progpred, Sechidis2018:progpred, Ondra2016:subgroup}.

An important methodological issue arises when including baseline covariates in the analysis: it can change the very nature of the estimand being estimated. The treatment effect becomes \emph{conditional}: it compares treatment groups at fixed covariate values and depends on which covariates are included in the model. For linear models, the marginal and conditional treatment effects coincide when there are no treatment--covariate interactions (ANCOVA); this is also the case in the presence of interactions (ANHECOVA) if covariates are centered \citep{Lin2013:hte, Ye2023:hte, Lancker2024:covadj}. In nonlinear models, this is generally not true due to \emph{noncollapsibility}: the conditional treatment effect differs from the marginal effect even in the absence of confounding \citep{Hauck1998:covadj, Greenland1999:noncollapsible, Rhian2021:noncollapsible}. This has been well documented for widely used models such as logistic and Cox regression, in which the log-odds ratio and log-hazard ratio are noncollapsible. However, noncollapsibility is not restricted to these models; it arises more generally whenever the expectation of a nonlinear transformation differs from the transformation of the expectation \citep{Greenland1999:noncollapsible, Rhian2021:noncollapsible}. For example, Cohen’s \(d\) can be interpreted as a shift on a probit scale within a transformation model framework \citep{Hothorn2017:transformation} and is therefore also noncollapsible \citep{Rhian2021:noncollapsible, Dandl2026:NAMI}. Consequently, in many nonlinear settings, adjusting for covariates changes the estimand from a marginal to a conditional effect unless additional steps are taken.

Approaches to handling baseline covariates in randomized trials generally follow one of two conceptual viewpoints. The first views covariate information primarily as a means of increasing efficiency but aims to preserve the marginal interpretation of the treatment effect. A popular family of methods within this perspective is based on semiparametric augmentation, in which the unadjusted treatment contrast is modified by adding a mean-zero augmentation term that increases efficiency without changing the marginal estimand \citep{Tsiatis2008:covadj, Zhang2008:covadj, Lu2008:covadj, Ye2024:covadj}. Standardization (or G-computation), highlighted in the FDA guidance, is an implementation that recovers the marginal contrast by fitting a conditional model, predicting outcomes under each treatment, and averaging these predictions over the empirical covariate distribution \citep{Zhang2008:margodds, Rhian2021:noncollapsible, Ye2023:gcomp, FDA2023:covadj, Lancker2024:covadj}. Inverse probability of treatment weighting (IPTW), augmented IPTW (AIPTW), and targeted maximum likelihood estimation (TMLE) provide further examples of such approaches \citep{vanDerLaan2006:tmle, Tackney2023:covadj, Bannick2025:aipw}. These methods focus primarily on efficient estimation of the marginal effect, and while some allow for treatment--covariate interactions \citep{Lancker2024:covadj, Tackney2023:covadj}, prognostic and predictive effects are not typically parameterized explicitly. Consequently, these methods do not directly provide interpretable measures of prognostic or predictive strength for baseline covariates.
% maybe say many of these methods are implemented in RobinCar package

The second conceptual viewpoint treats baseline covariates as primary objects of interest, focusing on the characterization of individual or subgroup-specific treatment effects, often using machine learning methods such as forest algorithms. One early approach, Virtual Twins, fits treatment-specific outcome models and derives conditional average treatment effects (CATEs) as differences in predicted outcomes. These estimates are then used for covariate ranking, without explicitly separating prognostic and predictive components \citep{Foster2011:forest}. More recent approaches, such as causal forests, directly estimate CATEs using orthogonalization techniques that treat prognostic effects as nuisance components \citep{Athey2019:forest}. Model-based forests also directly estimate CATEs, but use model-based recursive partitioning and simultaneously model prognostic and predictive effects \citep{Seibold2018:forest, Dandl2024:forest}. Other approaches do not estimate CATEs but instead focus more explicitly on covariate ranking \citep{Lipkovich2011:sides, Sechidis2018:progpred}. However, empirical studies suggest that many of these methods have difficulty distinguishing predictive from prognostic effects, particularly in smaller samples, when interaction signals are weak, or when prognostic effects are strong \citep{Sechidis2018:progpred, Hermansson2021:forest, Lipkovich2024:progpred}. Moreover, variable importance measures often lack interpretability in terms of clinically relevant effect sizes, and these methods are generally more suitable for exploratory rather than confirmatory settings, as they typically do not provide formal type I error control \citep{Ondra2016:subgroup}.

This work aims to bridge these two viewpoints by proposing a unified framework that (i) estimates a marginal treatment effect, (ii) adjusts for baseline covariates while separating and quantifying prognostic and predictive effects in an interpretable way, and (iii) accommodates a broad range of outcome types within a single modeling framework. 

We build on the nonparanormal adjusted marginal inference (NAMI) approach proposed by \citet{Dandl2026:NAMI}, which uses the nonparanormal likelihood framework \citep{Hothorn2025:nonparanormal} to adjust for prognostic covariates, and extend it to also accommodate predictive effects. This approach jointly models the outcome and baseline covariate distributions and, by construction, ensures that the treatment effect is collapsible, thereby preserving marginal interpretability. Therefore, rather than deriving a marginal effect from a conditional model, the marginal estimand is embedded directly in the model specification. Because the marginal distribution of the outcome is constructed using transformation models as described in \citet{Hothorn2017:transformation}, the method can accommodate different types of outcomes (continuous, binary, ordinal, time-to-event), different censoring mechanisms, and stratification within a unified framework. Prognostic and predictive effects are encoded in the correlation structure of the joint model and are quantified on a standardized latent scale, allowing direct comparison and ranking of covariates. Hypothesis testing for prognostic and predictive effects is possible within a frequentist framework, and adjustment for multiplicity can be applied when several covariates are assessed simultaneously.

The remainder of this work is organized as follows. We begin by reviewing univariate transformation models and the nonparanormal framework. We then introduce the extended model that allows for heterogeneous treatment effects and demonstrate its applicability across different outcome types: continuous outcomes with Cohen’s $d$ as the marginal effect, binary outcomes with the log-odds ratio, and survival outcomes with the log-hazard ratio. For Cohen’s $d$ in the presence of a single covariate, we derive its theoretical asymptotic variance and study its properties under varying strengths of prognostic and predictive effects. We subsequently conduct a simulation study to assess estimation accuracy, precision, and power of the marginal treatment effect across outcome types and levels of predictive power. Finally, we apply the proposed method to data from an acupuncture trial for chronic headaches, assessing its ability to reproduce the main trial results, accommodate different outcome types, quantify and rank prognostic and predictive effects, and improve efficiency of the marginal treatment effect.

%%%%%%%%%%%%%%%%%%%%%%%%%%%%%%%%%%%%%%%%%%%%%%%%%%%%%%%%%%%%%%%%%%%%%%

% LaTeX file for Chapter 03

\section{Motivating example}

As a motivating example, we consider the \texttt{anorexia} dataset from the \texttt{MASS} package \citep{Hand1993SmallDataSets, MASS}. This dataset contains information on 72 young female patients with anorexia nervosa, who were randomly assigned to one of three treatment groups: control (Cont, \(n = 26\)), cognitive behavioral therapy (CBT, \(n = 29\)), and family therapy (FT, \(n = 17\)). The primary outcome is the post-treatment weight, and the baseline weight before treatment is available as a covariate.

Suppose that the estimand of interest is the population-level marginal treatment effect, measured for instance by Cohen’s \(d\), comparing each active treatment group to the control group. The observed data points in Figure~\ref{fig:fig-mot_condensity} show the post-treatment weight against the pre-treatment weight by treatment group. While little to no correlation is observed in the control group, a positive correlation appears in both treatment groups. This could suggest that patients entering the study with higher pre-treatment weight may respond more favorably to active treatment.

Because treatment assignment is randomized, adjustment for pre-treatment weight is not required to estimate the marginal treatment effect. Nevertheless, incorporating baseline weight into the analysis may be clinically relevant, as it allows the investigation of treatment effect heterogeneity and may improve efficiency. Additionally, since the trial has a relatively small sample size, random baseline imbalances between treatment groups may occur by chance, increasing variability and making covariate adjustment helpful for improving precision.

%<<fig_anorexia, echo=FALSE, fig.pos='H', fig.cap="Post-treatment weight plotted against baseline weight in the anorexia %dataset, shown by treatment group: control (Cont), cognitive behavioral therapy (CBT), and family therapy (FT). Points %represent individual patients; solid lines show group-specific LOESS trends (span = 0.9, degree = 1).", fig.height=3.8, %fig.width=6, out.width="76%", fig.align='center'>>=
%fig_anorexia
%@

One strategy would be to include pre-treatment weight and its interaction with treatment in a linear regression model. Cohen’s \(d\) could then be calculated as the estimated mean difference divided by the residual standard deviation. However, this estimand would no longer be comparable to the marginal treatment effect, since Cohen’s \(d\) is a noncollapsible effect measure. Although the mean difference of post-treatment weight stays unchanged under appropriate covariate centering, including the pre-treatment weight in the model may explain more of the variability of the outcome, leading to a smaller residual standard deviation and thus a larger Cohen’s \(d\). %Nonetheless, an advantage of this method is that it clearly separates prognostic from predictive effects.

An alternative approach would be to use a causal forest, which can detect heterogeneity in treatment effects without having to explicitly specify an interaction term. However, this method does not estimate a marginal treatment effect by construction. Instead, it estimates the conditional average treatment effect (CATE) given the pre-treatment weight. Additionally, while this in principle identifies subgroups of patients that respond differently to treatment, the sample size in this dataset may be too small to reliably distinguish prognostic from predictive effects.

We instead propose a regression-based methodology that jointly models pre-treatment weight \(Y_0\) and post-treatment weight \(Y_1\) conditional on treatment group \(W\). This allows estimation of the marginal Cohen’s \(d\) while adjusting for and quantifying prognostic and predictive effects of pre-treatment weight. Compared to an unadjusted analysis, this approach yields a more precise and efficient marginal treatment effect by accounting for potential baseline imbalance and by explaining additional variation in the outcome.

\begin{knitrout}
\definecolor{shadecolor}{rgb}{0.969, 0.969, 0.969}\color{fgcolor}\begin{figure}[H]

{\centering \includegraphics[width=0.84\linewidth]{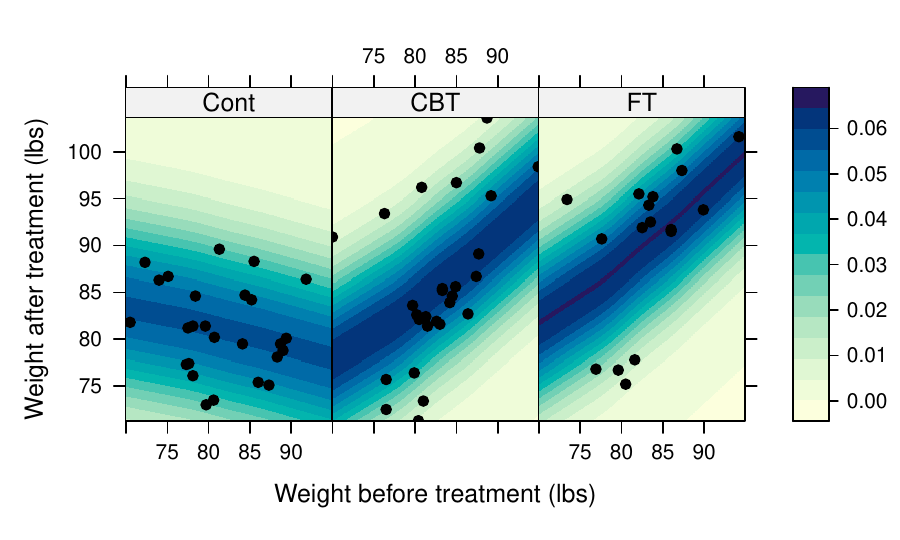} 

}

\caption[Estimated conditional density \(f_{Y_1 \mid Y_0, W}(y_1 \mid y_0, w)\) of post-treatment weight \(Y_1\) given pre-treatment weight \(Y_0\) and treatment group \(W\), derived from the joint model of \(Y_0, Y_1 \mid W\)]{Estimated conditional density \(f_{Y_1 \mid Y_0, W}(y_1 \mid y_0, w)\) of post-treatment weight \(Y_1\) given pre-treatment weight \(Y_0\) and treatment group \(W\), derived from the joint model of \(Y_0, Y_1 \mid W\). Treatment groups are control (Cont), cognitive behavioral therapy (CBT), and family therapy (FT). Color shading represents the model-based density, with darker regions indicating higher density. Black dots show the observed data.}\label{fig:fig-mot_condensity}
\end{figure}

\end{knitrout}

Applying this approach to the \texttt{anorexia} dataset, we find evidence of a marginal treatment effect for both active treatment groups compared to control. The estimated Cohen’s \(d\) is \(0.64\) for CBT and \(1.24\) for FT, indicating higher post-treatment weight under both interventions relative to control. Consistent with the weak association between pre- and post-treatment weight in the control group, we do not find evidence of a prognostic effect of pre-treatment weight. In contrast, there is evidence of a predictive effect in both active treatment groups, with a slightly stronger effect for FT. The direction of these effects suggests that patients with higher pre-treatment weight tend to experience a larger treatment benefit, a pattern also visible in the derived conditional densities shown in Figure~\ref{fig:fig-mot_condensity}.

%%%%%%%%%%%%%%%%%%%%%%%%%%%%%%%%%%%%%%%%%%%%%%%%%%%%%%%%%%%%%%%%%%%%%%

% LaTeX file for Chapter 04

\section{Methods}
\label{sec:methods}

This section introduces a novel modeling framework for estimating marginal treatment effects in randomized controlled trials (RCTs) while adjusting for covariates and covariate--treatment interactions. The proposed approach is based on the general nonparanormal (Gaussian copula) modeling framework of \citet{Liu2009:nonparanormal}, and extends the covariate-adjustment methodology in \citet{Dandl2026:NAMI} to additionally allow for covariate--treatment interactions.

We first introduce univariate transformation models as a flexible framework for estimating marginal treatment effects across different outcome types. Then, we extend this framework to a multivariate nonparanormal model that incorporates covariates and allows for heterogeneous treatment effects while preserving the marginal interpretation of the treatment effect.

%%%%%%%%%%%%%%%%%%%%%%%%%%%%%%%%%
\subsection{Univariate transformation models}
%%%%%%%%%%%%%%%%%%%%%%%%%%%%%%%%%

Transformation models as described in \citet{BoxCox1964} and \citet{Hothorn2017:transformation} offer a unified framework that encompasses many familiar regression models—such as linear, logistic, and Cox models. They bridge fully parametric models, which achieve interpretability often at the cost of restrictive distributional assumptions, and nonparametric models, which are more flexible but harder to interpret. By transforming the distribution of the outcome into a desired scale and defining regression parameters as shift effects on this transformed scale, transformation models decouple the interpretation of effects from the outcome distribution. Therefore, they retain the flexibility of nonparametric models while still providing interpretable effect estimates as in parametric models.

A key characteristic of transformation models is that they model the full outcome distribution rather than a single summary statistic such as the expectation. This enables the incorporation of many different outcome types, even in the presence of censored or truncated data, into one common framework. Thus, many classical, as well as novel models, can be understood as transformation models.

Let \(\rY \in \samY\) denote an at least ordered outcome of interest, \(W \in \{0,1\}\) a binary treatment indicator denoting two groups (control \(W=0\) and treatment \(W=1\)), and \(\rX = (X_1,\ldots,X_{J-1})\) a vector of \(J-1\) baseline covariates. In a randomized controlled trial (RCT), the propensity score $\pi = \Prob(W = 1)$ is constant and does not rely on covariates, thus \(W \indep \rX\). The main estimand is the marginal treatment effect \(\tau\), which describes how the distribution of \(\rY\) differs between treatment groups.

We denote the conditional cumulative distribution function (CDF) of \(\rY\) in arm \(w\) by \(F_w(y) := \Prob(\rY \le y \mid W = w)\). The conditional CDF given $W = 0$ (``control'') is \(F_0(y) := \Prob(\rY \le y \mid W = 0)\), and the conditional CDF given $W = 1$ (``treated'') is \(F_1(y) := \Prob(\rY \le y \mid W = 1)\). The treatment effect \(\tau := \tau(F_0,F_1)\) quantifies the discrepancy between these two distributions. Because \(F_0(y)\) and \(F_1(y)\) do not rely on the covariates \(\rX\), \(\tau\) reflects the \emph{marginal} treatment effect. With randomization of the treatment assignment, \(\tau\) can be estimated from \(\rY\) and \(W\) alone, ignoring covariates.

The following CDF specifies an unadjusted transformation model for the outcome \(\rY\) given treatment group \(W = 0\) or \(W = 1\):

\begin{equation}
F_w(y) = \Prob(\rY \le y \mid W = w)
= G\!\big(h(y) - \tau w\big)
= G\!\big(h(y \mid w)\big).
\label{eq:univ-tm}
\end{equation}

Here, \(G: \R \to [0,1]\) is a fixed CDF with a parameter-free, log-concave, absolutely continuous density; \(h: \samY \to \R\) is a monotone nondecreasing transformation function that maps the outcome distribution of the control group onto the quantile scale of \(G\) (that is, \(h(y) = G^{-1}(F_0(y))\)); and \(\tau\) is a scalar shift parameter representing the treatment effect on the transformed latent scale. Under control, this yields \(F_0(y) = G\!\big(h(y)\big)\), while under treatment, \(F_1(y) = G\!\big(h(y) - \tau\big)\).

Because a monotone transformation \(h(\cdot)\) always exists that maps any distribution \(F_0(y)\) to another, this formulation separates the shape of the outcome distribution, captured by \(h\), from the shift induced by treatment, captured by \(\tau\). The interpretation of \(\tau\) depends on the choice of \(G\): for example, \(\tau\) corresponds to a log-odds ratio when \(G\) is the logistic CDF, and to a log-hazard ratio when \(G\) is the Gumbel CDF. Moreover, specific combinations of \(G\) and \(h\) can impose distributional assumptions and recover well-known classical models. For instance, if \(G\) is the standard normal CDF and \(h(y)\) is linear, we obtain a normal linear model. Similarly, if \(G\) is the Gumbel CDF and \(h(y)\) is log-linear, we obtain the Weibull model. Section~\ref{sec:applications} illustrates some of these model applications; a more comprehensive list of examples is given in Table~1 of \citet{Hothorn2017:transformation}.

The treatment effect \(\tau\) in model~\eqref{eq:univ-tm} represents the \emph{marginal} effect of \(W\) on \(\rY\). To improve precision and capture treatment effect heterogeneity, covariates \(\rX\) and their interactions with treatment can be incorporated into the model as linear predictors. This gives rise to the adjusted linear transformation model:

\begin{equation}
F(y \mid W = w, \rX = \rx)
= G\!\big(h_{\rx}(y) - \tau_{\rx} w - \tilde{\rx}^\top \betavec - w\, \tilde{\rx}^\top \gammavec \big),
\label{eq:cond-tm}
\end{equation}

where \(\tilde{\rx}\) denotes an appropriate coding of the covariates, \(\betavec\) captures prognostic effects, \(\gammavec\) captures predictive effects, and \(\tau_{\rx}\) represents the \emph{conditional} treatment effect.

However, this model is generally \emph{noncollapsible}: integrating over \(\rX\) does not recover the marginal model~\eqref{eq:univ-tm}, that is,

\begin{equation*}
\int F(y \mid W = w, \rX = \rx)\, f_{\rX}(\rx)\, d\rx
\;\neq\;
G\!\big(h(y) - \tau w\big).
\end{equation*}

The conditional treatment parameter \(\tau_{\rx}\) therefore differs from the marginal effect \(\tau\). To retain the marginal interpretation of \(\tau\) while incorporating prognostic and predictive effects, we now extend the framework from univariate to multivariate transformation models based on a Gaussian-copula construction, where heterogeneous treatment effects are represented through the covariance structure rather than explicit regression parameters.

%%%%%%%%%%%%%%%%%%%%%%%%%%%%%%%%%
\subsection{Nonparanormal models}
\label{sec:mmlt}
%%%%%%%%%%%%%%%%%%%%%%%%%%%%%%%%%

The nonparanormal model of \citet{Dandl2026:NAMI}, building on the multivariate transformation model framework of \citet{Klein2022:multivariate}, extends the univariate transformation model to the joint distribution of the outcome \(Y\) and its covariates \(\rX = (X_1, \dots, X_{J-1})\). The idea is to model all components jointly on a latent normal scale while preserving the marginal outcome distribution \(F_w(y)\) as in~\eqref{eq:univ-tm}. This allows estimation of the marginal treatment effect \(\tau\) in the presence of covariates and covariate–treatment interactions. The resulting novel conditional distribution \(F(y \mid W = w, \rX = \rx)\) features the treatment effect \(\tau\) in a collapsible form and provides a clear separation between prognostic and predictive effects.

In this framework, both the outcome \(Y\) and covariates \(\rX\) are mapped to a latent standard normal scale through a monotone, nondecreasing transformation function \(h_j: \samY_j \to \R\). First, the marginal covariate distributions are parameterized as unconditional transformation models, \(\Prob(X_j \le x_j) \;=\; \Phi\!\big(h_j(x_j)\big), j = 1, ..., J-1\), where \(\Phi\) denotes the CDF of the standard normal distribution. The outcome \(Y\) is parametrized conditionally on treatment \(W = w\) via:

\begin{equation*}
h_J(y \mid w)
\;=\;
\Phi^{-1}\!\Big(G\!\big(h(y) - \tau w\big)\Big),
\end{equation*}

where \(h_J(y \mid w)\) maps \(Y\) to the standard normal scale while embedding a ``second'' transformation via \(G\) and \(h(y)\). This formulation enables modeling of the marginal treatment effect \(\tau\) not only on the probit scale (as in the normal linear model) but also on other scales, such as the log-odds scale (with \(G\) being the logistic CDF) or the log-hazard scale (with \(G\) being the Gumbel CDF). In other words, although \(h_J(y \mid w)\) is modeled on a latent normal scale, which is necessary for the joint modeling, the distribution of \(h(y)\) can follow any distribution specified by \(G\), offering flexibility in the interpretation of \(\tau\).

This choice of \(\Phi\) is particularly appealing because it provides a direct link to Gaussian copulas \citep{Song2009:gaussiancopula} and the nonparanormal model \citep{Liu2009:nonparanormal}. Under this model, the joint distribution of the transformed covariates and outcome is multivariate normal. The multivariate transformation function \(\hvec : \mathcal{X}_1 \times \dots \times \mathcal{X}_{J-1} \times \mathcal{Y} \to \R^J\), defined as \(\mZ(w) \;=\; \hvec(\rX, Y \mid W=w) \;=\; \big(h_1(X_1),\, \dots,\, h_{J-1}(X_{J-1}),\, h_J(Y \mid w)\big)^\top\) formulates the joint conditional CDF of the covariates and outcome given treatment as

\begin{equation}
\Prob(\rX \le \rx,\, Y \le y \mid W = w)
= \Phi_{\mSigma(w)}\!\big(\hvec(\rx, y \mid W=w)\big),
\label{eq:joint-cdf}
\end{equation}

where \(\Phi_{\mSigma(w)}\) denotes the \(J\)-dimensional normal CDF with a \(J \times J\) treatment-dependent correlation matrix \(\mSigma(w)\). 
%When all variables are absolutely continuous, this can be written as: \(\mZ(w) \;\sim\; \mathcal{N}_J\!\big(0,\, \mSigma(w)\big)\).

Prognostic and predictive effects are captured through the joint dependence structure of the latent variables via \(\mSigma(w)\). Only the last row and column of \(\mSigma(w)\) depend on the treatment assignment \(w\), since the \(1\!:\!(J-1) \times 1\!:\!(J-1)\) block corresponds to the covariates, which are independent of \(w\) by randomization. Consequently, prognostic effects are contained in the last row and column of \(\mSigma(0)\), whereas predictive effects are reflected in the discrepancy between the corresponding elements of \(\mSigma(1)\) and \(\mSigma(0)\). If the \(\mSigma(w)\) does not depend on \(w\), i.e., \(\mSigma(w) = \mSigma\) for all \(w\), the model reduces to the nonparanormal adjusted inference model (NAMI) as described in \citet{Dandl2026:NAMI}, which allows adjustment for prognostic covariate effects but does not accommodate predictive (treatment--covariate interaction) effects.

While the marginal distributions of \(Y\) and \(\rX\) may be non-normal, or even discrete, their joint dependence is Gaussian after transformation. This combination of flexible marginal distributions with a parametric Gaussian copula defines the ``nonparanormal'' model. In other words, the marginals are considered nonparametric because the transformation functions \(h_j\) avoid imposing distributional assumptions on the original scales of \(Y\) and \(\rX\), whereas the joint dependence is considered parametric because the transformed variables are assumed to be jointly Gaussian.

%% --------------------------------------------------------------
\subsubsection[Parameterization of mSigma(w)]{Parameterization of \texorpdfstring{$\mSigma(w)$}{mSigma(w)}} %and \texorpdfstring{$\mOmega(w)$}{mOmega(w)}}
\label{subsec:mmlt-omega}
%% --------------------------------------------------------------

To ensure identifiability and that the treatment effect \(\tau\) retains its \emph{marginal} interpretation, i.e., that the univariate model~\eqref{eq:univ-tm} can be recovered from the multivariate formulation~\eqref{eq:joint-cdf}, we impose \(\operatorname{diag}\bigl(\mSigma(w)\bigr) \equiv 1\). This constraint is most conveniently imposed by parameterizing \(\mSigma(w)\) through its inverse Cholesky factor \(\mOmega(w)\), using the factorization \(\mSigma(w) = \mOmega(w)^{-1}\, \mOmega(w)^{-\top}\), which guarantees the positive definiteness of \(\mSigma(w)\). The lower-triangular \(J \times J\) matrix \(\mOmega(w) = \big(\omega_{j j'}^{(w)}\big)\) has positive diagonal entries \(\omega_{jj}^{(w)} > 0\) for \(j = 1, \dots, J\), and lower-triangular elements \(\omega_{j j'}^{(w)}\) for \(1 \le j' < j \le J\).
%% this last sentence is 'too similar' to NAMI paper

For computational convenience, we further reparameterize \(\mOmega(w)\) in terms of unconstrained parameters \(\mLambda(w) = \mLambda(\lambdavec(w))\), a unit lower-triangular matrix with ones on the diagonal and free parameters in the strict lower triangle, \(\lambdavec(w) = \big(\lambda_{21}, \dots, \lambda_{J,J-1}^{(w)}\big)^\top \in \RR^{J(J - 1)/2}\). The resulting matrix takes the form

\begin{equation}
\begin{aligned}
\mLambda(w)
  &= \mLambda(\lambdavec(w)) 
   = \big(\lambda_{jj'}^{(w)}\big)_{1 \le j' < j \le J}
  &= \begin{pmatrix}
     1 \\
     \lambda_{21}       & 1        &        & 0 \\
     \lambda_{31}       & \lambda_{32} & 1 \\
     \vdots             & \vdots   &        & \ddots \\
     \lambda_{J1}^{(w)} & \lambda_{J2}^{(w)} & \cdots & \lambda_{J,J-1}^{(w)} & 1
      \end{pmatrix}.
\end{aligned}
\label{eq:mLambda}
\end{equation}

Following Section~2, Option~2 of \citet{Hothorn2025:nonparanormal}, which enforces the constraint \(\operatorname{diag}\bigl(\mSigma(w)\bigr) \equiv 1\), the matrix \(\mOmega(w)\) is obtained from \(\mLambda(w)\), and can be partitioned into four blocks where only the two lower blocks depend on \(w\), as:

\begin{align*}
\mOmega(w)
  \;=\; \mLambda(w)\,
     \Big(\operatorname{diag}\big(\mLambda(w)^{-1}\mLambda(w)^{-\top}\big)\Big)^{1/2}
 \;=\; 
\left(
\begin{array}{c|c}
  \mOmega_{1:(J-1),\,1:(J-1)} & 0_{1J} \\
  \hline
  \omegavec(w)_{J,\,1:(J-1)} & \omega^{(w)}_{JJ}
\end{array}
\right)
\end{align*}

We denote the diagonal elements of \(\mOmega(w)\) by \(\sigmavec(w)^{-1} = \operatorname{diag}\big(\mOmega(w)\big)\), explicitly \( \sigmavec(w)^{-1} = \bigl(1,\, \omega_{22},\, \dots,\, \omega_{J-1,J-1},\, \omega_{JJ}^{(w)}\bigr)^\top \), where only the last element depends on \(w\). As shown in Section~\ref{subsec:mmlt-cond}, \(\sigma_j^{(w)}\) represents the conditional standard deviation of the \(j\)th latent variable given the first \(j-1\) variables, for \(j > 1\). Consequently, \(\sigma_1 = 1\) because the marginal variances are normalized to one. Since the covariates \(\rX\) do not depend on the treatment assignment \(W = w\), only \(\sigma_J^{(w)}\) varies with \(w\), capturing treatment--dependent heterogeneity in the residual variability of the outcome \(Y\).

To clearly separate prognostic from predictive effects, \(\mLambda(w)\) can be further decomposed into a baseline (prognostic) component \(\mLambda(\lambdavec)\) and a treatment-dependent (predictive) component \(\mGamma(\gammavec)\), where only the last row \(J\) depends on \(w\) with \(\lambda_{Jj}^{(w)} = \lambda_{Jj} + w \gamma_{j}\):

\begin{equation}
\begin{aligned}
\mLambda(w)
  &= \mLambda(\lambdavec) + w\, \mGamma(\gammavec)
  &= \begin{pmatrix}
       1                                                                 \\
       \lambda_{21}       & 1              &  &  0                       \\
       \lambda_{31}       & \lambda_{32}   &  1                          \\
       \vdots             & \vdots         &  & \ddots                   \\
       \lambda_{J1}       & \lambda_{J2}   & \cdots & \lambda_{J,J-1}    & 1
     \end{pmatrix} + w\,
     \begin{pmatrix}
       0                                                             \\
       0                  & 0              &        &   0            \\
       0                  & 0              &        0                \\
       \vdots             & \vdots         &        & \ddots         \\
       \gamma_{1}         & \gamma_{2}     & \cdots & \gamma_{J-1}   & 0
     \end{pmatrix}.
\end{aligned}
\label{eq:mLambda-gamma}
\end{equation}

Here \(\lambda_{J1}, \dots, \lambda_{J,J-1}\) represent the \emph{prognostic} effects and \(\gamma_{1}, \dots, \gamma_{J-1}\) represent the \emph{predictive} effects. For the control arm (\(W = 0\)), the last row of \(\mLambda(0)\) contains only the prognostic effects. For the treatment arm (\(W = 1\)), the last row of \(\mLambda(1)\) includes the sum of both prognostic and predictive components.

%% --------------------------------------------------------------
\subsubsection{Derivation of the conditional model}
\label{subsec:mmlt-cond}
%% --------------------------------------------------------------

As discussed in \citet{Klein2022:multivariate}, a multivariate distribution can always be expressed as a sequence of conditional distributions. Therefore, starting from the joint distribution of \(Y\) and \(\rX\) given treatment \(W\) in model~\eqref{eq:joint-cdf}, we can derive the conditional distribution of \(Y\) given \(\rX\) and \(W\); analogous derivations can be obtained for any variable \(j\) conditional on its predecessors \(1, \dots, j-1\) for \(j > 1\). This can be done by first whitening the joint model, that is, remove its dependence structure through multiplication of the inverse Cholesky factor \(\mOmega(w)\) with the latent vector \(\mZ(w) \;=\; \hvec(\rX, Y \mid W = w) \;=\; \big(h_1(X_1),\, \dots,\, h_{J-1}(X_{J-1}),\, h_J(Y \mid w)\big)^\top\). When all variables are absolutely continuous, this yields \(\boldsymbol{\varepsilon}(w) \;=\; \mOmega(w)\,\mZ(w) \;\sim\; \mathcal{N}_J(0, \mE_J)\), a system of \(J\) independent equations, where \(\mE_J\) denotes the \(J \times J\) identity matrix. The last equation of this system reads:

\begin{equation}
\varepsilon_J^{(w)}
\;=\;
\sum_{j=1}^{J-1} \omega_{Jj}^{(w)}\,h_j(X_j)
\;+\;
\omega_{JJ}^{(w)}\,h_J(Y \mid w).
\label{eq:epsilon_J}
\end{equation}

Because \(\varepsilon_J^{(w)} \sim \mathcal{N}(0,1)\) and the transformation \(h_J(\cdot)\) is monotone nondecreasing, the rank order of \(Y\) is preserved in~\eqref{eq:epsilon_J} for fixed \(\rX = \rx\). That is, inequalities on the outcome \(Y\) correspond one-to-one with inequalities on the transformed variable. Hence, for absolutely continuous \(\rX\), the conditional distribution of \(Y\) given \(\rX = \rx\) and \(W = w\) follows as

\begin{equation}
\Prob\!\big(Y \le y \mid W = w, \rX = \rx\big)
\;=\;
\Phi\!\left(
\sum_{j=1}^{J-1} \omega_{Jj}^{(w)}\,h_j(x_j)
\;+\;
\omega_{JJ}^{(w)}\,h_J(y \mid w)
\right),
\label{eq:cond-cdf}
\end{equation}

Using the decomposition \(\mLambda(w) = \mLambda(\lambdavec) + w\,\mGamma(\gammavec)\) from~\eqref{eq:mLambda-gamma}, the conditional CDF can be expressed explicitly in terms of prognostic and predictive components:

\begin{equation}
\Prob\!\big(Y \le y \mid W = w, \rX = \rx\big)
\;=\;
\Phi\!\left(
\sum_{j=1}^{J-1} \sigma_j^{-1}\big(\lambda_{Jj} + w\,\gamma_j\big)\,h_j(x_j)
\;+\;
\big(\sigma_J^{(w)}\big)^{-1}\,h_J(y \mid w)
\right),
\label{eq:cond-cdf-add}
\end{equation}

where \(\sigma_j^{-1} = \omega_{jj}\) for \(j < J\) and \(1/\sigma_J^{(w)} = \omega_{JJ}^{(w)}\). Here, \(\sigma_j^{-1}\lambda_{Jj}\) represents the \emph{prognostic} effect of covariate \(X_j\) on the latent scale, while \(\sigma_j^{-1}\gamma_j\) captures its \emph{predictive} effect.

By rearranging equation~\eqref{eq:epsilon_J} to isolate \(h_J(Y \mid w)\), we see that the last row of \(\mOmega(w)\) encodes the regression coefficients of the transformed covariates \(h_1(X_1), \dots, h_{J-1}(X_{J-1})\) on the transformed outcome \(h_J(Y \mid w)\):

\begin{equation*}
h_J(Y \mid w)
\;=\;
-\frac{\sum_{j=1}^{J-1} \omega_{Jj}^{(w)}\,h_j(X_j)}{\omega_{JJ}^{(w)}}
\;+\;
\frac{1}{\omega_{JJ}^{(w)}}\,\varepsilon_J^{(w)}.
\end{equation*}

%where \(\hvec_{-J}(\rX) = \big(h_1(X_1), \dots, h_{J-1}(X_{J-1})\big)^\top\) and \(\mOmega_{J,-J}(w) = \big(\omega_{J1}^{(w)}, \dots, \omega_{J,J-1}^{(w)}\big)\) denotes the last row of \(\mOmega(w)\) excluding its diagonal element. 
Because \(\varepsilon_J^{(w)} \sim \mathcal{N}(0,1)\), the conditional expectation and variance of the transformed outcome given the covariates are

\begin{equation*}
\mathbb{E}\!\left[h_J(Y \mid w) \mid \rX = \rx\right]
\;=\;
-\frac{\sum_{j=1}^{J-1} \omega_{Jj}^{(w)}\,h_j(X_j)}{\omega_{JJ}^{(w)}},
\qquad
\Var\!\left[h_J(Y \mid w) \mid \rX = \rx\right]
\;=\;
\frac{1}{\big(\omega_{JJ}^{(w)}\big)^{2}}.
\end{equation*}

And the conditional distribution of \(h_J(Y \mid w)\) given \(\rX = \rx\) is

\begin{equation}
h_J(Y \mid w)\,\big|\,\rX = \rx
\;\sim\;
\mathcal{N}\!\left(
-\frac{\sum_{j=1}^{J-1} \omega_{Jj}^{(w)}\,h_j(X_j)}{\omega_{JJ}^{(w)}},
\;\frac{1}{\big(\omega_{JJ}^{(w)}\big)^{2}}
\right).
\label{eq:cond-hJ-dist}
\end{equation}

The conditional regression coefficient of covariate \(X_j\) on the latent outcome scale can be written as \(\beta_j^{(w)} = -\omega_{Jj}^{(w)} / \omega_{JJ}^{(w)} = -\omega_{Jj}^{(w)}\,\sigma_J^{(w)} = -(\sigma_J^{(w)} / \sigma_j)(\lambda_{Jj} + w\,\gamma_j)\). The prognostic effect of \(X_j\) is given by \(\beta_j^{(0)}\), while the predictive effect is \(\beta_j^{(1)} - \beta_j^{(0)}\). By construction, \(h_j(X_j)\) is mapped to a standard normal scale and is monotone; therefore, quantiles are preserved, and a unit increase in \(h_j(X_j)\) corresponds to a quantile shift on the original scale of \(X_j\). For example, this includes moving from the median of \(X_j\) (\(\Phi^{-1}(0.5) = 0\)) to its 84\textsuperscript{th} percentile (\(\Phi^{-1}(0.84) \approx 1\)). Thus, \(\beta_j^{(w)}\) measures the change in the latent outcome associated with such a quantile shift in \(X_j\), holding all other covariates fixed. Because \(\Var\!\big(h_j(X_j)\big) = 1\) for \(j = 1, \dots, J-1\), the coefficients are on a common scale, which allows covariates to be ranked and compared directly according to their prognostic strength \(\lvert \omega_{Jj}^{(0)} \rvert\) and predictive strength \(\lvert \omega_{Jj}^{(1)} - \omega_{Jj}^{(0)} \rvert\).

The conditional standard deviation of the latent outcome given the covariates is \(\sigma_J^{(w)} = \big(\omega_{JJ}^{(w)}\big)^{-1}\), which may differ between treatment arms. The gain in explained variability obtained by adjusting for covariates and covariate--treatment interactions is quantified by the coefficient of determination \(R(w)^2 \;=\; 1 - \big(\omega_{JJ}^{(w)}\big)^{-2}\).

%% --------------------------------------------------------------
\subsubsection{Derivation of the marginal model}
\label{subsec:mmlt-marg}
%% --------------------------------------------------------------

The key property of this nonparanormal model is that it preserves the marginal distribution of the outcome \(Y\) unconditional on \(\rX\) as in~\eqref{eq:univ-tm}. Consequently, the treatment effect \(\tau\) retains its original marginal interpretation even after adjusting for covariates and covariate--treatment interactions. This property follows directly from the unit-variance constraint, \(\operatorname{diag}\bigl(\mSigma(w)\bigr) \equiv 1\). To recover the marginal model, we integrate the conditional distribution in~\eqref{eq:cond-cdf} over the distribution of \(\rX\) given \(W\):

\begin{align*}
\Prob\!\big(Y \le y \mid W = w\big)
&= 
\int \Prob\!\big(Y \le y \mid W = w, \rX = \rx\big)\,
    f_{\rX}(\rx)\, d\rx \\[0.8em]
&= 
\E_{\rX}\!\Big[
    \Phi\!\Big(
      \underbrace{\sum_{j=1}^{J-1} \omega_{Jj}^{(w)}\,h_j(x_j)}_{=: S}
      \;+\;
      \omega^{(w)}_{JJ}\,h_J(y\mid w)
    \Big)
  \Big] \\[0.6em]
&=\; \Phi\!\left(\frac{\omega^{(w)}_{JJ}\,h_J(y\mid w)}{\sqrt{1+\Var(S)}}\right).
\end{align*}

%%% MAYBE REFERENCE SOMETHING FOR THESE EQUATIONS

%The last equality uses the elementary Normal–probit averaging identity: if \(S\sim\mathcal N(0,s^2)\), then \(\E\{\Phi(a+S)\}=\Phi\!\big(a/\sqrt{1+s^2}\big)\). (Equivalently, \(\E\{\Phi(a+S)\}=\Pr(U\le a+S)=\Pr(U-S\le a)\) with \(U\sim\mathcal N(0,1)\) independent of \(S\), hence \(U-S\sim\mathcal N(0,1+s^2)\).)

\(\Var(S)\) can be found by rearranging~\eqref{eq:epsilon_J} as \( \omega_{JJ}^{(w)}\,h_J(Y \mid w) \;=\; \varepsilon_J^{(w)} \;-\; S \) and taking the variance. Because \(\Var\big(\varepsilon_J^{(w)}\big)=1\), \(\Var\big(h_J(Y\mid w)\big)=1\) by the unit-variance constraint, and \(\varepsilon_J^{(w)}\) is independent of \(S\), we obtain

\begin{align*}
\Var\!\left(\omega_{JJ}^{(w)}\,h_J(Y \mid w)\right)
\;=\;  \Var(\varepsilon_J^{(w)}) + \Var(S) 
\implies
\big(\omega_{JJ}^{(w)}\big)^2
\;=\;  1 + \Var(S)%, \\
\iff
\Var(S)
\;=\; \big(\omega_{JJ}^{(w)}\big)^2 - 1.
\end{align*}

This implies \(\sqrt{1+\Var(S)}=\omega_{JJ}^{(w)}\), and averaging the conditional model~\eqref{eq:cond-cdf} over \(\rX\) recovers the same marginal model as in~\eqref{eq:univ-tm}:

\begin{equation*}
\Prob\!\big(Y \le y \mid W = w\big)
  \;=\;
  \Phi\!\big(h_J(y \mid w)\big)
  \;=\;
  \Phi\!\Bigl(
      \Phi^{-1}\!\big(
          G\!\big(h(y) - \tau w\big)
      \big)
  \Bigr)
  \;=\;
  G\!\big(h(y) - \tau w\big).
\end{equation*}

Therefore, this novel conditional model is \emph{collapsible} and the treatment effect \(\tau\) remains \emph{marginally} interpretable.

%% --------------------------------------------------------------
\subsubsection{Summary of parameter interpretation}
\label{subsec:mmlt-interpretation}
%% --------------------------------------------------------------

This novel nonparanormal framework models the joint distribution of the outcome and the covariates, from which both the marginal distribution of the outcome and the conditional distribution of the outcome given \(\rX\) can be derived. The parameters of both these models can be meaningfully interpreted, as summarized below.

\textbf{Marginal treatment effect:} represented by \(\tau\); it retains its marginal interpretation from the univariate marginal model~\eqref{eq:univ-tm}. Its interpretation depends on the choice of \(G\): for example, \(\tau\) can represent Cohen's \(d\), a log-odds ratio, or a log-hazard ratio \citep[see Table~1]{Hothorn2017:transformation}.

\textbf{Prognostic and predictive strength:} contained in the last row of \(\mOmega(w)\), that is, \(\omega_{Jj}^{(w)}\) for \(j = 1, \dots, J-1\). \(\lvert \omega_{Jj}^{(0)} \rvert\) is the prognostic strength of covariate \(X_j\), while \(\lvert \omega_{Jj}^{(1)} - \omega_{Jj}^{(0)} \rvert\) is its predictive strength. These quantities are directly comparable across covariates. Hypothesis tests for prognostic and predictive effects are performed using the unconstrained parameters in \(\mLambda(w)\): for example, \(H_0: \lambda_{Jj} = 0\) tests whether covariate \(X_j\) is prognostic, while \(H_0: \gamma_j = 0\) tests whether \(X_j\) is predictive. Note that the signs of these coefficients are reversed relative to the corresponding correlations (a positive coefficient corresponds to a negative correlation and vice versa).

\textbf{Conditional regression parameters:} extracted from the conditional distribution of \(h_J(Y \mid w)\) given \(\rX = \rx\) in~\eqref{eq:cond-hJ-dist} as \(\beta_j^{(w)} = -\omega_{Jj}^{(w)} / \omega_{JJ}^{(w)}\), where \(\beta_j^{(0)}\) is the prognostic effect of \(X_j\) and \(\beta_j^{(1)} - \beta_j^{(0)}\) is its predictive effect.

\textbf{Conditional standard deviation and coefficient of determination:} again extracted from~\eqref{eq:cond-hJ-dist}, the conditional standard deviation is \(\sigma_J^{(w)} = 1 / \omega_{JJ}^{(w)}\), and the corresponding coefficient of determination is \(R(w)^2 = 1 - \big(\sigma_J^{(w)}\big)^{2} \;=\; 1 - \big(\omega_{JJ}^{(w)}\big)^{-2}\). Both of these vary by treatment assignment.

%% --------------------------------------------------------------
\subsubsection{Likelihood construction}
\label{subsec:mmlt-likelihood}
%% --------------------------------------------------------------

Parameter estimation is performed by maximum likelihood as described in \citet{Hothorn2025:nonparanormal}, optimizing the exact joint likelihood. An example of the likelihood construction for a normally distributed outcome and a single normally distributed covariate, with the marginal treatment effect expressed as Cohen’s $d$, is provided in Appendix~\ref{sec:A_seproofs}.

%%%%%%%%%%%%%%%%%%%%%%%%%%%%%%%%%%%%%
\subsection{Specific model applications}
\label{sec:applications}
%%%%%%%%%%%%%%%%%%%%%%%%%%%%%%%%%%%%%

The choice of \(G\) and \(h\) in the unadjusted transformation model~\eqref{eq:univ-tm} depends on the outcome type and on the desired interpretation of the marginal treatment effect \(\tau\). In this section, we consider the same model applications as in \citet{Dandl2026:NAMI} and we show how (i) continuous, (ii) ordinal and binary, and (iii) time-to-event outcomes can all be embedded within this unified framework through appropriate choices of \(G\) and \(h\). These transformation models recover widely used regression models such as normal linear regression, (proportional odds) logistic regression, and a fully parametrized version of the Cox proportional hazards model. For the special case of a normally distributed outcome with a single normally distributed covariate, we additionally derive closed-form expressions for the standard error of Cohen’s \(d\) under different configurations of prognostic and predictive effects.

%% --------------------------------------------------------------
\subsubsection{Continuous outcomes}
\label{subsec:application-linear}
%% --------------------------------------------------------------

For a continuous, normally distributed outcome \(Y \in \mathbb{R}\), choosing \(G = \Phi\) together with a linear transformation function \(h(y) = \vartheta_1 + \vartheta_2\, y\) recovers the classical normal linear regression model. This is because imposing a linear transformation function on the standard normal scale is equivalent to assuming that the outcome is normally distributed. The distribution of \(Y\) is \(Y \mid W=0 \;\sim\; \mathcal{N}\!\bigl(-\vartheta_1/\vartheta_2,\; \vartheta_2^{-2}\bigr)\) in the control arm and \(Y \mid W=1 \;\sim\; \mathcal{N}\!\bigl(-(\vartheta_1-\tau)/\vartheta_2,\; \vartheta_2^{-2}\bigr)\) in the treatment arm. The unadjusted transformation model becomes:

\begin{equation}
F_w(y) \;=\; \Phi\!\bigl(\vartheta_1 + \vartheta_2\, y - \tau w\bigr).
\label{eq:normal-tm}
\end{equation}

The marginal treatment effect \(\tau\) represents the standardized difference in means, that is, Cohen’s \(d\), which is noncollapsible because it corresponds to a shift on the probit scale, a nonlinear link function \citep{Rhian2021:noncollapsible}. In contrast, the unstandardized mean difference, \(\mathbb{E}(Y \mid W=1) - \mathbb{E}(Y \mid W=0) = \tau / \vartheta_2\), is collapsible, as it represents a shift on the original outcome scale under the identity (linear) link. More intuitively, Cohen’s \(d\) is the mean difference divided by the standard deviation; when covariates explain part of the outcome variability, the residual standard deviation in the conditional model is smaller, which leads to a larger standardized mean difference.

The normality assumption can be relaxed by allowing a more flexible transformation function \(h\), for example using Bernstein polynomials of order \(M\) \citep{Hothorn2017:transformation}, rather than restricting it to be linear. However, removing this assumption comes at a cost: the treatment effect \(\tau\) no longer corresponds to Cohen’s \(d\) on the original outcome scale, but instead, it represents a mean difference on the latent standard normal scale. To obtain a more intuitive effect measure, \(\tau\) can be mapped to the probability that an individual under control has a smaller outcome than an independently selected individual under treatment:
\(
\text{AUC} = \Prob(Y_0 < Y_1) = \Phi\!\left(\tau \,/\, \sqrt{2}\right),
\)
where \(Y_0\) and \(Y_1\) denote independent potential outcomes under control and treatment, respectively \citep{Sewak2023:AUC}. 

To obtain the corresponding multivariate transformation model adjusting for covariates \(\rX\), a second transformation of the outcome (as described in Section~\ref{sec:mmlt}) is not required because the marginal model for \(Y\) already uses \(G = \Phi\), meaning that we can directly write \(h_J(y \mid w) = h(y) - \tau w\). The corresponding conditional distribution of \(Y\) given \(W\) and continuous covariates \(\rX\) is

\begin{equation}
\Prob\!\big(Y \le y \mid W = w, \rX = \rx\big)
\;=\;
\Phi\!\left(
\sum_{j=1}^{J-1} \omega_{Jj}^{(w)}\,h_j(x_j)
\;+\;
\omega_{JJ}^{(w)}\,\big(h(y) - \tau w)\big)
\right).
\label{eq:cond-normal}
\end{equation}

By construction, the treatment effect \(\tau\) in~\eqref{eq:cond-normal} is collapsible: integrating over the covariates recovers a marginal treatment effect as in~\eqref{eq:normal-tm}.

% =========================================
\paragraph{Standard error for Cohen’s \(d\)}
\label{subsubsec:se-cohensd}

For the case of a normally distributed outcome and one normally distributed covariate, closed-form expressions for the standard error of Cohen’s \(d\) can be derived under different scenarios of prognostic and predictive effects. This special case is equivalent to a normal linear model with covariate main effects and covariate–treatment interactions, where covariates are appropriately centered and the residual standard deviation is allowed to differ between treatment arms. % when gamma = -lambda/2, the variance is the same in both groups
Proofs of the results stated in the following Lemmas are provided in Section~\ref{sec:A_seproofs} of the Appendix, and simulations validating agreement between these standard errors and the software implementation are shown in Section~\ref{sec:A_sim_consistency}. 

Since both the outcome and the covariate are normally distributed, the transformation functions \(h_1\) and \(h_J\) are linear; we denote \(h_1(x) = \vartheta_{21} + \vartheta_{22}\, x\) for the covariate and \(h_J(y \mid w) = \vartheta_{11} + \vartheta_{12}\, y - \tau w\) for the outcome.

% ==========
\paragraph{Lemma 1 \citep{Dandl2026:NAMI}.}\textit{Standard error of Cohen’s \(d\) without prognostic or predictive effects.}

\textit{Consider the model \(Y \mid W = w \sim \mathcal{N}\!\bigl(-(\vartheta_{11} - \tau w)/\vartheta_{12},\; \vartheta_{12}^{-2}\bigr)\), where the marginal treatment effect \(\tau\) corresponds to Cohen’s \(d\). In a balanced trial with sample size \(N\) per arm, the unadjusted standard error of \(\tau\) is
\(
\mathrm{SE}(\tau)
\;=\;
\sqrt{\frac{1}{N}\left(\frac{\tau^2}{4} + 2\right)}. 
\)
}

% ==========
\paragraph{Lemma 2 \citep{Dandl2026:NAMI}.}\textit{Standard error of Cohen’s \(d\) with prognostic effect only.}

\textit{Let \(Y \mid W = w \sim \mathcal{N}\!\bigl(-(\vartheta_{11} - \tau w)/\vartheta_{12},\; \vartheta_{12}^{-2}\bigr)\) and \(X_1 \sim \mathcal{N}\!\bigl(-\vartheta_{21}/\vartheta_{22},\; \vartheta_{22}^{-2}\bigr)\), with a joint distribution specified by a Gaussian copula with correlation \(\rho = -\lambda/\sqrt{1+\lambda^2}\), where \(\lambda := \lambda_{21}\) is the single unconstrained copula parameter in~\eqref{eq:mLambda-gamma}. Assume identical correlation in both treatment arms (no predictive effect), that is, \(\gamma_1 = 0\). Then, in a balanced trial with sample size \(N\) per arm, the standard error of \(\tau\) after covariate adjustment is
\(
\mathrm{SE}(\tau, \lambda)
\;=\;
\sqrt{\frac{1}{N}\left(\frac{\tau^2}{4} + \frac{2}{\lambda^2 + 1}\right)}.
\)
}

% ==========
\paragraph{Lemma 3.}\textit{Standard error of Cohen’s \(d\) with predictive effect only.}

\textit{Let \(Y \mid W = w \sim \mathcal{N}\!\bigl(-(\vartheta_{11} - \tau w)/\vartheta_{12},\; \vartheta_{12}^{-2}\bigr)\) and \(X_1 \sim \mathcal{N}\!\bigl(-\vartheta_{21}/\vartheta_{22},\; \vartheta_{22}^{-2}\bigr)\), with a joint distribution specified by a treatment-dependent Gaussian copula such that \(\rho_0=0\) in control and \(\rho_1 = -\gamma/\sqrt{1+\gamma^2}\) in treatment, where \(\gamma := \gamma_{1}\) is the single unconstrained copula parameter in~\eqref{eq:mLambda-gamma}. Assume no prognostic effect, that is, \(\lambda_{21} = 0\). Then, in a balanced trial with sample size \(N\) per arm, the standard error of \(\tau\) after adjusting for the covariate--treatment interaction is
\(
\mathrm{SE}(\tau, \gamma)
\;=\;
\sqrt{\frac{1}{N}\left(\frac{\tau^2}{4} + \frac{\gamma^2 + 4}{\gamma^2 + 2}\right)}.
\)
}

% ==========
\paragraph{Lemma 4.}\textit{Standard error of Cohen’s \(d\) with prognostic and predictive effects.}

\textit{Let \(Y \mid W = w \sim \mathcal{N}\!\bigl(-(\vartheta_{11} - \tau w)/\vartheta_{12},\; \vartheta_{12}^{-2}\bigr)\) and
\(X_1 \sim \mathcal{N}\!\bigl(-\vartheta_{21}/\vartheta_{22},\; \vartheta_{22}^{-2}\bigr)\), with a joint distribution specified by a treatment-dependent Gaussian copula such that \(\rho_0 = - \lambda / \sqrt{1+\lambda^2}\) in control and \(\rho_1 = - (\lambda+\gamma) / \sqrt{1+(\lambda+\gamma)^2}\) in treatment, where \(\lambda := \lambda_{21}\) and \(\gamma := \gamma_{1}\) are the prognostic and predictive copula parameters in~\eqref{eq:mLambda-gamma}. Then, in a balanced trial with sample size \(N\) per arm, the standard error of \(\tau\) after adjusting for the covariate and covariate--treatment interaction is
\(
\mathrm{SE}(\tau, \lambda, \gamma)
\;=\;
\sqrt{\frac{1}{N}\left(\frac{\tau^2}{4}
+ \frac{\gamma^2 + 4}{2\lambda^2 + 2\lambda\gamma + \gamma^2 + 2}\right)}.
\)
}

The squared standard error ratio between the adjusted and unadjusted analyses, \(\big(\mathrm{SE}_{\text{adj}} / \mathrm{SE}_{\text{unadj}}\big)^2\), represents the relative sample size required under adjustment to achieve the same power as the unadjusted analysis. As shown in Section~\ref{sec:A_seprops} of the Appendix, when only prognostic effects of \(X_1\) are present, \(\mathrm{SE}(\tau, \lambda)\) decreases monotonically as \(|\lambda|\) increases (Figure~\ref{fig:fig-SEprog}). Likewise, when only predictive effects are present, \(\mathrm{SE}(\tau, \gamma)\) decreases monotonically with increasing \(|\gamma|\) (Figure~\ref{fig:fig-SEpred}). The efficiency gain is larger for prognostic than for predictive effects: for example, to obtain a 25\% reduction in required sample size at fixed power with true treatment effect \(\tau = 0.5\), the prognostic-only case would require an outcome–covariate correlation of approximately \(|\rho| = 0.51\), whereas the predictive-only case would require a correlation in the treatment arm of approximately \(|\rho_1| = 0.83\).

When both prognostic and predictive effects are present, the standard error \(\mathrm{SE}(\tau, \lambda, \gamma)\) is always less than or equal to the unadjusted standard error \(\mathrm{SE}(\tau)\) (Figure~\ref{fig:fig-SEprog_pred}). The worst-case scenario occurs when \(\lambda = -\gamma/2\), corresponding to correlations of equal magnitude but opposite sign in the two treatment arms; in this situation, no efficiency gain is achieved and \(\mathrm{SE}(\tau, \lambda, \gamma) = \mathrm{SE}(\tau)\). In all other configurations, covariate adjustment leads to a reduction in standard error, with prognostic effects yielding larger and more rapid efficiency gains than predictive effects.

%% --------------------------------------------------------------
\subsubsection{Ordinal and binary outcomes}
\label{subsec:application-binary}
%% --------------------------------------------------------------

In the case of discrete, ordered outcomes \(Y \in \{y_1 < \dots < y_K\}\), the log-odds ratio interpretation of \(\tau\) can be obtained by choosing \(G\) as the cumulative distribution function of the standard logistic distribution (the inverse logit function, \(\operatorname{logit}^{-1}\)). The transformation function \(h(y_k) = \vartheta_k\) is a step function representing the log odds of outcome \(y_k\) in the control group, with steps at \(y_k\) for \(k = 1, \dots, K-1\). When \(K > 2\), the unadjusted model corresponds to an ordinal logistic regression model with

\begin{equation}
\Prob(Y \le y_k \mid W = w)
\;=\;
\operatorname{logit}^{-1}(\vartheta_k - \tau w),
\label{eq:binary-tm}
\end{equation}

which assumes proportional odds. For \(K = 2\), this reduces to a binary logistic regression model with \(\Prob(Y \le y_1 \mid W = w) = \Prob(Y = y_1 \mid W = w) = \operatorname{logit}^{-1}(\vartheta_1 - \tau w)\).

% In the case of an ordinal outcome, the treatment effect tau is more intuitively interpreted as the log odds of probability of a higher outcome category in treatment vs. control. In the binary outcome, it can be interpreted as the log odds of y_2 in treatment vs. control. 

Because this model does not use \(G = \Phi\), a second transformation is necessary to obtain the multivariate transformation model; the distribution is mapped onto the latent standard normal via
\(
h_J(y_k \mid w)
\;=\;
\Phi^{-1}\!\bigl(\operatorname{logit}^{-1}(\vartheta_k - \tau w)\bigr).
\)
The resulting conditional distribution of a binary or ordinal outcome given treatment \(W\) and continuous covariates \(\rX\) is therefore

\begin{equation}
\Prob(Y \le y_k \mid W = w, \rX = \rx)
\;=\;
\Phi\!\left(
\sum_{j=1}^{J-1} \omega_{Jj}^{(w)}\,h_j(x_j)
\;+\;
\omega_{JJ}^{(w)}\,\Phi^{-1}\!\bigl(\operatorname{logit}^{-1}(\vartheta_k - \tau w)\bigr)
\right),
\label{eq:cond-binary}
\end{equation}

where the treatment effect \(\tau\) is a log-odds ratio that is, by construction, collapsible. Integrating over the covariates in~\eqref{eq:cond-binary} recovers a marginal treatment effect \(\tau\) as in~\eqref{eq:binary-tm}.

%% --------------------------------------------------------------
\subsubsection{Survival outcomes}
\label{subsec:application-survival}
%% --------------------------------------------------------------

Finally, we consider a time-to-event outcome \(Y \in \mathbb{R}^+\). The log-hazard ratio interpretation of \(\tau\) can be obtained by setting \(G\) as the cumulative distribution function of the Gumbel distribution (the inverse complementary log-log function, \(\operatorname{cloglog}^{-1}\)). This corresponds to the following unadjusted model

\begin{equation}
\Prob(Y \le y \mid W = w)
\;=\;
\operatorname{cloglog}^{-1}\!\bigl(h(y) - \tau w\bigr),
\label{eq:survival-tm}
\end{equation}

which assumes proportional hazards. Different parameterizations of the transformation function give rise to different survival models. For example, the log-linear specification \(h(y) = \vartheta_1 + \vartheta_2 \log(y)\) imposes a distributional assumption on the survival times and recovers the Weibull model. If no specific distributional assumption is desired, a flexible transformation function \(h(y)\) can be used instead, for instance based on Bernstein polynomials, yielding a fully parametric Cox model.

As for ordinal outcomes, a second transformation to the latent standard normal scale is required in the multivariate transformation model. Defining
\(
h_J(y \mid w)
\;=\;
\Phi^{-1}\!\bigl(\operatorname{cloglog}^{-1}(h(y) - \tau w)\bigr)
\)
leads to the corresponding conditional distribution of the survival times given treatment \(W\) and continuous covariates \(\rX\) as:

\begin{equation}
\Prob(Y \le y \mid W = w, \rX = \rx)
\;=\;
\Phi\!\left(
\sum_{j=1}^{J-1} \omega_{Jj}^{(w)}\,h_j(x_j)
\;+\;
\omega_{JJ}^{(w)}\,\Phi^{-1}\!\bigl(\operatorname{cloglog}^{-1}(h(y) - \tau w)\bigr)
\right).
\label{eq:cond-survival}
\end{equation}

where the treatment effect \(\tau\) is a log-hazard ratio that is, by construction, collapsible. Integrating over the covariates in~\eqref{eq:cond-survival} recovers a marginal treatment effect \(\tau\) as in~\eqref{eq:survival-tm}.

%% maybe eventually show how stratification can be incorporated
%% maybe mention mutual exclusivity between conditional univariate model

%%%%%%%%%%%%%%%%%%%%%%%%%%%%%%%%%%%%%%%%%%%%%%%%%%%%%%%%%%%%%%%%%%%%%%

% LaTeX file for Chapter 05

\section{Simulation studies}
\label{sec:results}

%%%%%%%%%%%%%%%%%%%%%%%%%%%%%%%%%
\subsection{Data-generating process and models}
\label{sec:results-sim-setup}
%%%%%%%%%%%%%%%%%%%%%%%%%%%%%%%%%

This simulation study evaluates the performance of nonparanormal adjusted marginal inference with heterogeneous treatment effects (NAMI-HTE) in estimating marginal treatment effects and detecting predictive covariates. The design extends the covariate-adjustment (prognostic-only) framework of \citet{Dandl2026:NAMI} to allow for treatment effect heterogeneity through predictive parameters. Each scenario was replicated 10,000 times.

In all experiments, the treatment indicator \(W\) followed a Bernoulli distribution \(W \sim B(1,0.5)\), corresponding to a balanced randomized trial. We considered three outcome types and four covariates. Outcomes were generated from the following conditional distribution functions \(\Prob(Y \le y \mid W = w, \rX = \rx)\):

\begin{eqnarray*}
\begin{cases}
\text{normal:} &
\Phi\!\left[
   \sum_{j=1}^{4} \omega_{5j}^{(w)} h_j(x_j)
   + \omega_{55}^{(w)}\bigl(\eparm_1 + \eparm_2 y - \tau w \bigr)
\right],
\quad y \in \mathbb{R},
\\[0.4em]
\text{binary:} &
\Phi\!\left[
   \sum_{j=1}^{4} \omega_{5j}^{(w)} h_j(x_j)
   + \omega_{55}^{(w)}\,\Phi^{-1}\!\{\expit(\eparm_1 - \tau w)\}
\right],
\quad y \in \{0,1\},
\\[0.4em]
\text{survival:} &
\Phi\!\left[
   \sum_{j=1}^{4} \omega_{5j}^{(w)} h_j(x_j)
   + \omega_{55}^{(w)}\,\Phi^{-1}\!\{\cloglog^{-1}(\eparm_1 + \eparm_2 \log(y) - \tau w)\}
\right],
\quad y \in \mathbb{R}^+ .
\end{cases}
\end{eqnarray*}

Following \citet{Dandl2026:NAMI}, we fixed \(\eparm_1 = 0\) and \(\eparm_2 = 1\). Under this parametrization, the first model generates a normally distributed outcome \(Y \sim \mathcal{N}(w\tau,1)\), so that \(\tau\) corresponds to Cohen’s \(d\). The second yields a Bernoulli outcome where \(\tau\) is a log-odds ratio, and the third produces a Weibull outcome where \(\tau\) is a log-hazard ratio. For the survival setting, we imposed independent right-censoring with a censoring probability of 70\%. 

The true marginal treatment effect was set to either \(\tau = 0\) or \(\tau = 0.5\). The sample size was chosen to achieve approximately \(60\%\) power for testing \(H_0:\tau=0\) in an unadjusted analysis under a true effect of \(\tau=0.5\). This yielded per-arm sample sizes of \(N=41\) for the normally distributed outcome, \(N=161\) for the binary outcome, and \(N=131\) for the survival outcome \citep[see Appendix~B for derivations]{Dandl2026:NAMI}.

\begin{table}[ht]
\centering
\caption{Latent correlations between covariates and the outcome in simulation studies, shown by treatment arm and predictive strength $\gamma$ of $X_1$.}
\label{tab:latent-cor}
\begin{tabular}{lcccc}
\toprule
 & \multicolumn{1}{c}{$Y\mid W=0$} & \multicolumn{3}{c}{$Y\mid W=1$} \\
\cmidrule(lr){2-2}\cmidrule(lr){3-5}
 & \text{for any $\gamma$} & $\gamma=0$ & $\gamma=0.25$ & $\gamma=0.5$ \\
\midrule
$X_1$ & -0.24 & -0.24 & -0.45 & -0.60 \\
$X_2$ & -0.18 & -0.18 & -0.12 & -0.06 \\
$X_3$ & -0.13 & -0.13 & -0.09 & -0.04 \\
$X_4$ & -0.10 & -0.10 & -0.07 & -0.03 \\
\bottomrule
\end{tabular}
\end{table}

We generated four continuous covariates \(X_1,\dots,X_4\). Covariate \(X_1\) followed a \(\chi^2\) distribution with five degrees of freedom, while \(X_2, X_3,\) and \(X_4\) were \(t\)-distributed with two, three, and four degrees of freedom, respectively. The covariates were correlated with \(\lambda_{jj'} = 0.25\) for all \(j = 2,\dots,4\) and \(j' = 1,\dots,j-1\). All four covariates had a prognostic effect with \(\lambda_{5j}=0.25\) for \(j=1,\dots,4\), whereas only \(X_1\) was allowed to be predictive, with \(\gamma_1\in\{0,0.25,0.5\}\). The remaining predictive parameters were set to zero. This parameterization induces the latent correlations between covariates and the outcome shown in Table~\ref{tab:latent-cor}.

For each simulated dataset, the marginal treatment effect \(\tau\) and its standard error were estimated using (i) an unadjusted marginal inference transformation model (MI) as in~\eqref{eq:univ-tm}, and (ii) a nonparanormal marginal inference model with heterogeneous treatment effects (NAMI-HTE) as in~\eqref{eq:joint-cdf}. Empirical power under a true effect \(\tau = 0.5\) for testing \(H_0:\tau = 0\), as well as empirical size under \(\tau = 0\), was evaluated using a two-sided Wald test at significance level \(\alpha = 0.05\).

Predictive effects were estimated within the NAMI-HTE model. To assess whether any covariates were predictive, we tested the joint null hypothesis \(\gamma_1=\dots=\gamma_4=0\) using multiplicity-adjusted \(p\)-values. Empirical power under \(\gamma_1 = 0.25\) or \(\gamma_1 = 0.5\), and size under \(\gamma_1=0\), were evaluated at \(\alpha=0.05\). Further implementation details are provided in Appendix~\ref{subsec:A_sim_setup}.

%%%%%%%%%%%%%%%%%%%%%%%%%%%%%%%%%
\subsection{Simulation results}
%%%%%%%%%%%%%%%%%%%%%%%%%%%%%%%%%

\begin{knitrout}
\definecolor{shadecolor}{rgb}{0.969, 0.969, 0.969}\color{fgcolor}\begin{figure}[H]

{\centering \includegraphics[width=0.9\linewidth]{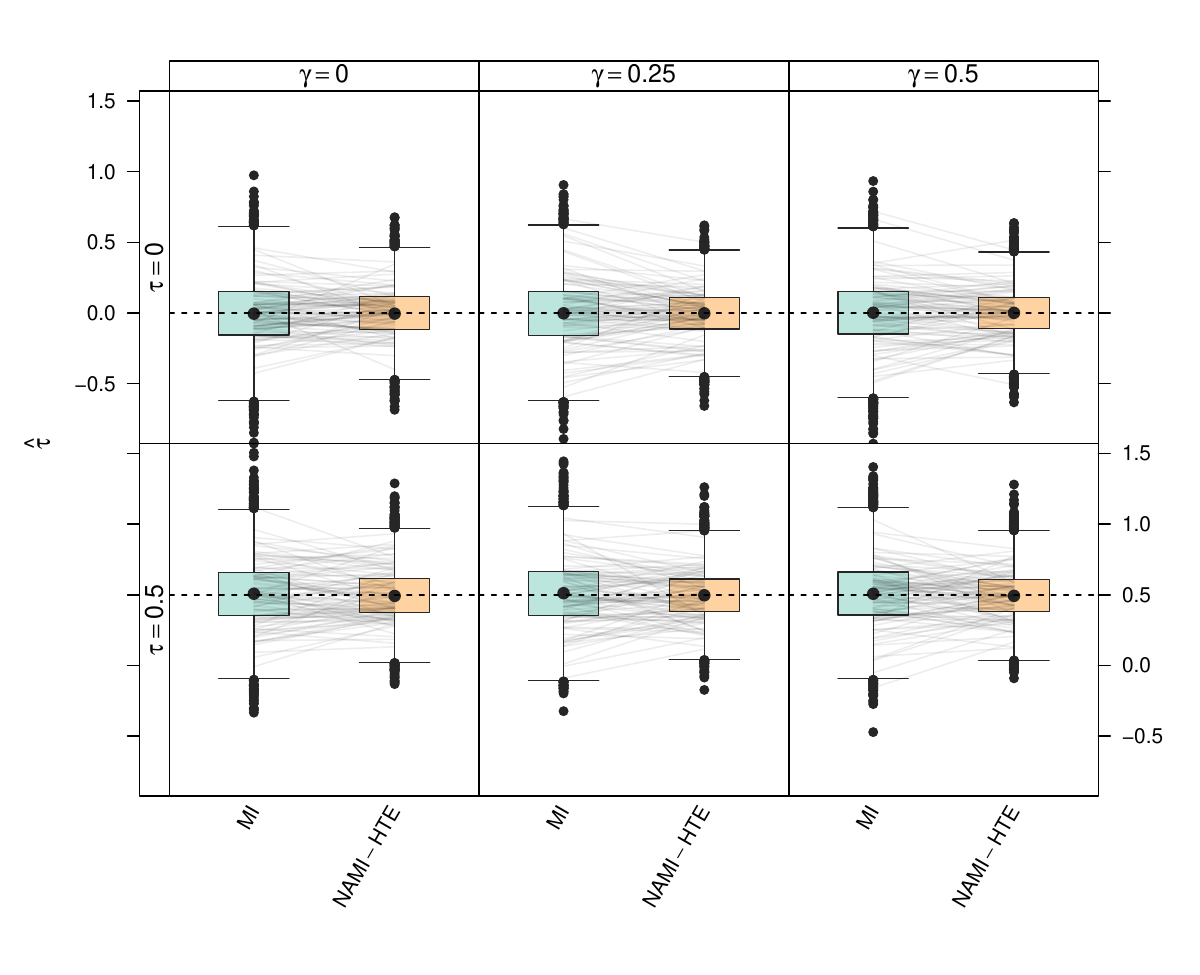} 

}

\caption[Estimated marginal treatment effect \(\hat{\tau}\) (Cohen's \(d\)) in simulations with a continuous normally distributed outcome]{Estimated marginal treatment effect \(\hat{\tau}\) (Cohen's \(d\)) in simulations with a continuous normally distributed outcome. Rows correspond to the true treatment effect \(\tau\) and columns to the true predictive effect \(\gamma\) of \(X_1\). Results are shown for unadjusted marginal inference (MI) and the nonparanormal adjusted marginal inference model with heterogeneous treatment effects (NAMI-HTE). The dashed horizontal line indicates the true value of \(\tau\).}\label{fig:fig-tauhat_continuous}
\end{figure}

\end{knitrout}

Figure~\ref{fig:fig-tauhat_continuous} shows the empirical distribution of the estimated marginal treatment effect for the continuous outcome under the unadjusted (MI) and the adjusted (NAMI-HTE) models. The corresponding estimated standard errors are shown in Figure~\ref{fig:fig-SEtau_continuous}. Empirical power for testing \(H_0:\tau=0\) is reported in Table~\ref{tbl-tau_power}, and empirical power for testing \(H_0:\gamma=0\) are reported in Table~\ref{tbl-gamma_power}. Complete results for all outcome types, including estimates of both \(\tau\) and \(\gamma\), their empirical size, and standard errors of \(\tau\), are provided in Appendix~\ref{subsec:A_sim_results}.

Across all simulation settings, the estimated marginal treatment effect is centered at the true value under both MI and NAMI-HTE, as seen in Figure~\ref{fig:fig-tauhat_continuous}. This indicates that adjustment does not introduce bias in the estimation of \(\tau\), even in the presence of predictive effects. The main difference between the two approaches is precision: the distribution of \(\hat{\tau}\) is visibly narrower under NAMI-HTE than under MI. The same pattern is observed for binary and survival outcomes (Figures~\ref{fig:fig-tauhat_binary} and~\ref{fig:fig-tauhat_survival}).

\begin{knitrout}
\definecolor{shadecolor}{rgb}{0.969, 0.969, 0.969}\color{fgcolor}\begin{figure}[H]

{\centering \includegraphics[width=0.9\linewidth]{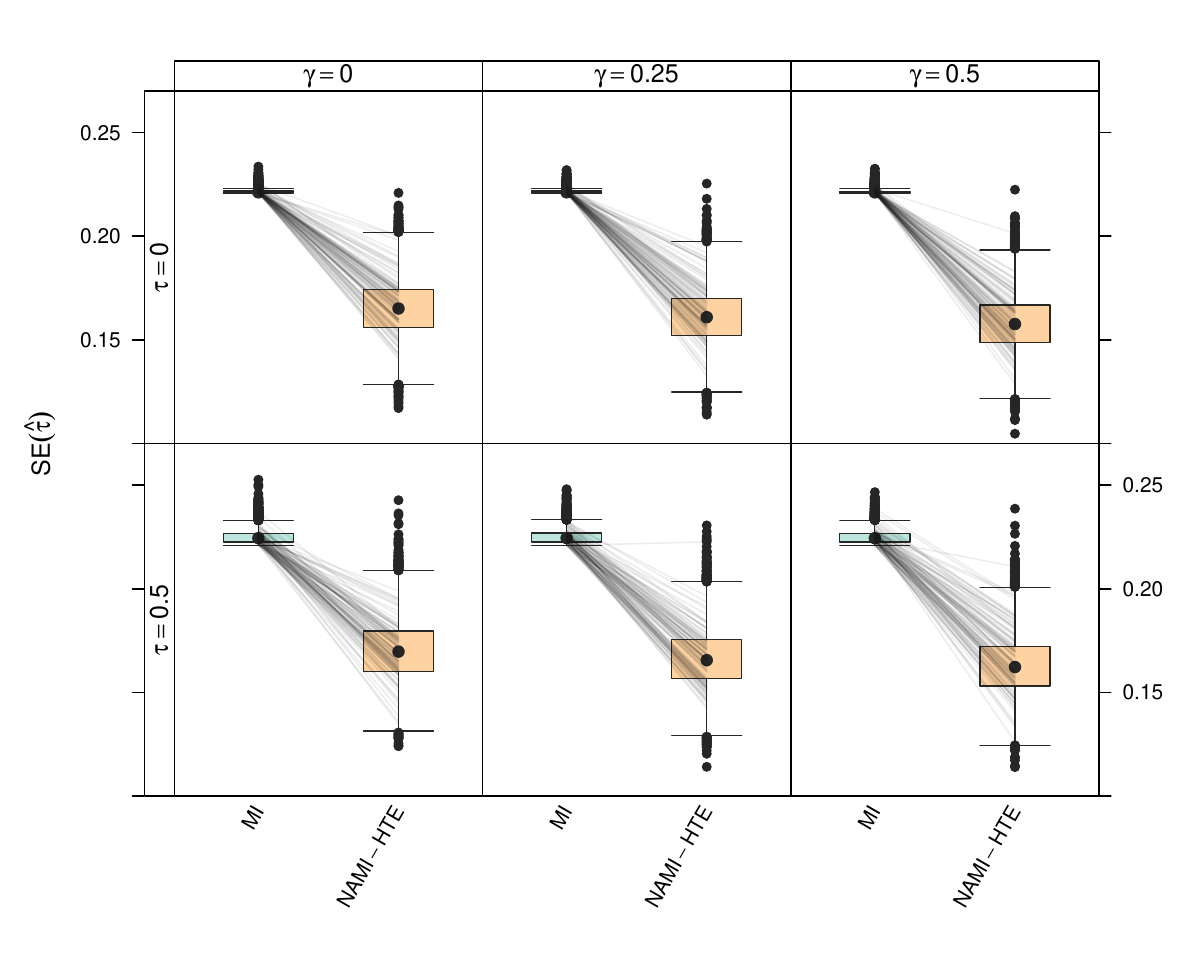} 

}

\caption[Estimated standard error \(\mathrm{SE}(\hat{\tau})\) of the marginal treatment effect \(\tau\) (Cohen's \(d\)) in simulations with a continuous normally distributed outcome]{Estimated standard error \(\mathrm{SE}(\hat{\tau})\) of the marginal treatment effect \(\tau\) (Cohen's \(d\)) in simulations with a continuous normally distributed outcome. Rows correspond to the true treatment effect \(\tau\) and columns to the true predictive effect \(\gamma\) of \(X_1\). Results are shown for unadjusted marginal inference (MI) and the nonparanormal adjusted marginal inference model with heterogeneous treatment effects (NAMI-HTE).}\label{fig:fig-SEtau_continuous}
\end{figure}

\end{knitrout}

The improvement in precision is reflected in the estimated standard errors. For all outcome types, the standard error of \(\hat{\tau}\) is smaller under NAMI-HTE than under MI. This gain is already present when \(\gamma=0\), indicating that it is driven by prognostic information in the covariates. Increasing the predictive strength of \(X_1\) leads to only a modest additional reduction in the standard error under NAMI-HTE, while the standard error under MI remains unchanged. This pattern is consistent across continuous, binary, and survival outcomes (Figures~\ref{fig:fig-SEtau_continuous}, \ref{fig:fig-SEtau_binary}, and~\ref{fig:fig-SEtau_survival}).

Consistent with the smaller standard errors, the adjusted model achieves higher power (Table~\ref{tbl-tau_power}). Under MI, the power to detect a treatment effect of \(\tau=0.5\) is approximately 60\% across all scenarios, as intended by the design. Under NAMI-HTE, power increases substantially due to adjustment. For the continuous outcome without predictive effects, power increases to approximately 83\%. When the predictive effect increases, power rises slightly further, but the additional gain is small. Similar trends are observed for binary and survival outcomes.

Empirical sizes for testing \(H_0:\tau=0\) under a true null effect are close to the nominal 0.05 level for MI across all outcomes (Table~\ref{tbl-tau_size}). For NAMI-HTE, the nominal level is maintained for binary outcomes. In the continuous and survival outcome settings, empirical sizes are slightly above 0.05, which could be explained by a lack of normal approximation of the Wald statistic. This is shown for continuous outcomes in the QQ-plot in Figure~\ref{fig:fig-tstatQQ_continuous2}, where deviations from normality at the tails can be seen when \(N=41\). When the per-arm sample size is increased to \(N=500\), the approximation improves and type I error control is restored (Table~\ref{tbl-tau_sizeN}).

\begin{table}[H]
\centering
\caption{Empirical power for testing $H_0: \tau = 0$ using a Wald test ($\alpha=0.05$) under true effect $\tau=0.5$. Results are shown across configurations of outcome type, model (MI: unadjusted marginal inference, NAMI-HTE: nonparanormal adjusted marginal inference with heterogeneous treatment effects) and true predictive effect $\gamma$ of $X_1$.}
\label{tbl-tau_power}
\small
\renewcommand{\arraystretch}{1.15}
\setlength{\tabcolsep}{4.5pt}
\begin{tabular}{llccc}
\toprule
Outcome & Model & \multicolumn{3}{c}{Power} \\
\cmidrule(lr){3-5}
 &  & $\gamma=0$ & $\gamma=0.25$ & $\gamma=0.50$ \\
\midrule
continuous & MI & 0.620 & 0.624 & 0.619 \\
 & NAMI-HTE & 0.834 & 0.854 & 0.866 \\
\arrayrulecolor{gray!35}\midrule\arrayrulecolor{black}
binary & MI & 0.607 & 0.610 & 0.595 \\
 & NAMI-HTE & 0.783 & 0.792 & 0.805 \\
\arrayrulecolor{gray!35}\midrule\arrayrulecolor{black}
survival & MI & 0.600 & 0.609 & 0.599 \\
 & NAMI-HTE & 0.792 & 0.810 & 0.820 \\
\bottomrule
\end{tabular}
\end{table}

For predictive effects, the estimator of \(\gamma_1\) is centered at the true value across all outcome types and treatment effect settings (Figures~\ref{fig:fig-gammahat_continuous}, \ref{fig:fig-gammahat_binary}, and~\ref{fig:fig-gammahat_survival}). However, empirical power to detect predictive effects is limited. For \(\gamma_1=0.25\), power remains below 10\% across outcomes, and for \(\gamma_1=0.5\), power remains below 40\% (Table~\ref{tbl-gamma_power}). 
%Power is somewhat higher for binary outcomes than for continuous or survival outcomes, but remains moderate overall. 
The presence or absence of a marginal treatment effect does not impact the ability to detect predictive effects. Empirical sizes for testing predictive effects are below the nominal 0.05 level in all scenarios; this conservativeness is expected, as inference is based on a joint test of all predictive parameters with multiplicity adjustment.

\begin{table}[H]
\centering
\caption{Empirical power for testing $H_0: \gamma = 0$ using a Wald test ($\alpha=0.05$) under true effect $\gamma=0.25$ or $\gamma=0.5$. Results are shown across configurations of outcome type, true marginal treatment effect $\tau$, and true predictive effect $\gamma$ of $X_1$ models, only for the NAMI-HTE (nonparanormal adjusted marginal inference with heterogeneous treatment effects) model.}
\label{tbl-gamma_power}
\small
\renewcommand{\arraystretch}{1.15}
\setlength{\tabcolsep}{4.5pt}
\begin{tabular}{lllcc}
\toprule
Outcome & Model & $\tau$ & \multicolumn{2}{c}{Power} \\
\cmidrule(lr){4-5}
 &  &  & $\gamma=0.25$ & $\gamma=0.50$ \\
\midrule
continuous & NAMI-HTE & 0 & 0.068 & 0.227 \\
 & NAMI-HTE & 0.5 & 0.072 & 0.218 \\
\arrayrulecolor{gray!35}\midrule\arrayrulecolor{black}
binary & NAMI-HTE & 0 & 0.088 & 0.367 \\
 & NAMI-HTE & 0.5 & 0.080 & 0.348 \\
\arrayrulecolor{gray!35}\midrule\arrayrulecolor{black}
survival & NAMI-HTE & 0 & 0.086 & 0.373 \\
 & NAMI-HTE & 0.5 & 0.083 & 0.359 \\
\bottomrule
\end{tabular}
\end{table}

%%%%%%%%%%%%%%%%%%%%%%%%%%%%%%%%%%%%%%%%%%%%%%%%%%%%%%%%%%%%%%%%%%%%%%

% LaTeX file for Chapter 06

\section{Application to the acupuncture trial}
\label{sec:application}

In this section, we apply the proposed NAMI-HTE framework to a publicly available randomized clinical trial on acupuncture for treating chronic headache \citep{Vickers2004:acupuncture, Vickers2006:data}, which has also been analyzed in previous work investigating treatment effect heterogeneity \citep{Su2018:random, Svensson2026:shapley}. The aim of this application is twofold. First, we replicate the main findings from the original trial analysis and demonstrate that this method can reproduce results obtained from classical models. Second, we illustrate how this joint modeling of the outcome and baseline covariates can be used to learn about prognostic and predictive effects, handle missing follow-up data, and adapt to different outcome scales, all while keeping \(\tau\) marginally interpretable. The R code to reproduce this analysis is available at \url{https://github.com/lwuthrich/NAMI-HTE-acupuncture}.

The trial evaluated whether acupuncture reduces chronic headache burden compared to usual care. A total of \(n = 401\) patients were randomized in a 1:1 ratio to receive either acupuncture over a 3-month period (\(n=205\)) or usual care (\(n=196\)). Randomization was minimized on age, sex, headache diagnosis (migraine versus tension-type), and headache chronicity. The primary outcome was a patient-reported headache score, measured four times per day over four weeks and averaged by week; this was assessed at 12 months after study entry and at baseline. Although the original score was based on a 6-point Likert scale, the primary analysis treated it as continuous because it was repeatedly measured and averaged over many observations, yielding a more granular and approximately continuous variable.

The main analysis of the original trial \citep{Vickers2004:acupuncture} used an ANCOVA model adjusting for the baseline headache score and minimization covariates. This corresponds to a normal linear model with adjustment for covariate main effects (prognostic effects) but without treatment--covariate interactions (predictive effects). The analysis was based on complete cases, excluding \(100\) participants without follow-up data and leaving \(161\) treated and \(140\) control patients.

\begin{table}[h]
\centering
\caption{Overview of the models fitted in the acupuncture trial application.}
\label{tab:model_features}
\small  
\renewcommand{\arraystretch}{1.15}
\begin{tabular}{lccccccc}
\toprule
Model 
& \shortstack{Prognostic\\ adjustment} 
& \shortstack{Predictive\\ adjustment} 
& \shortstack{All\\ datapoints} 
& \shortstack{Assumes\\ normality} 
& Continuous 
& Ordinal \\
\midrule
\arrayrulecolor{gray!35}
\texttt{m1} &  &  &  & \checkmark & \checkmark &  \\ \hline
\texttt{m2} & \checkmark &  &  & \checkmark & \checkmark &  \\ \hline
\texttt{m3} & \checkmark & \checkmark &  & \checkmark & \checkmark &  \\ \hline
\texttt{m4} & \checkmark & \checkmark & \checkmark & \checkmark & \checkmark &  \\ \hline
\texttt{m5} & \checkmark & \checkmark & \checkmark &  & \checkmark &  \\ \hline
\texttt{m6} & \checkmark & \checkmark & \checkmark &  &  & \checkmark \\
\arrayrulecolor{black}
\bottomrule
\end{tabular}
\end{table}

We fit a sequence of models (m1--m6), summarized in Table~\ref{tab:model_features}. Model \texttt{m1} is an unadjusted normal linear model based on complete cases. Model \texttt{m2} adjusts for prognostic effects of the baseline covariates and therefore corresponds to the original ANCOVA analysis. Model \texttt{m3} additionally allows for predictive effects, which enables us to study treatment effect heterogeneity. Model \texttt{m4} uses all randomized patients by including baseline information from participants with missing follow-up outcomes. Model \texttt{m5} relaxes the normality assumption of the outcome, and model \texttt{m6} treats the endpoint as ordinal.

All models use a probit link so that the marginal treatment effect \(\tau\) has a Cohen’s \(d\) interpretation. In models \texttt{m1}--\texttt{m4}, which assume normality of the outcome, this interpretation holds on the original outcome scale. % and Cohen’s d can be transformed into a mean difference on the original scale.
In models \texttt{m5} and \texttt{m6}, which do not rely on this assumption, Cohen’s \(d\) applies only on the latent standard normal scale and \(\tau\) can therefore be interpreted as a mean change in probit score. Estimation is performed by maximum likelihood as in \citet{Hothorn2025:nonparanormal}; further details are provided in Section~\ref{subsec:A_app_models} of the Supplement.

Table~\ref{tab:tau_results} reports the estimated marginal treatment effect \(\hat{\tau}\), its standard error, confidence interval, and \(p\)-value across models. The original analysis of the trial found a benefit of acupuncture compared to usual care, reporting an adjusted mean difference of \(-4.6\) (95\% CI: \(-7.1\) to \(-2.2\), \(p=0.0002\)) in the acupuncture group compared to control; the unadjusted mean difference was \(-6.1\). Our findings are consistent with these results: model \texttt{m1} estimates Cohen’s \(d = -0.4\), with a corresponding unadjusted mean difference of \(-6.1\). In model \texttt{m2}, which mirrors the original ANCOVA analysis, we estimate Cohen’s \(d = -0.3\), with a corresponding adjusted mean difference of \(-4.65\).

\begin{table}[h]
\centering
\caption{Estimated marginal treatment effect $\hat{\tau}$ across models, with corresponding standard errors, 95\% Wald confidence intervals, and $p$-values.}
\label{tab:tau_results}
\small
\renewcommand{\arraystretch}{1.15}
\begin{tabular}{llcccr}
\toprule
Model 
& \shortstack{Interpretation of $\tau$}
& $\hat{\tau}$
& $\mathrm{SE}(\hat{\tau})$
& \shortstack{95\% CI}
& $p$-value \\
\midrule
\arrayrulecolor{gray!35}
\texttt{m1} & \shortstack{Cohen’s $d$}
& -0.40 & 0.12 & From -0.63 to -0.17 & 0.0006 \\ \hline
\texttt{m2} & \shortstack{Cohen’s $d$}
& -0.30 & 0.09 & From -0.48 to -0.13 & 0.0005 \\ \hline
\texttt{m3} & \shortstack{Cohen’s $d$}
& -0.30 & 0.09 & From -0.47 to -0.13 & 0.0005 \\ \hline
\texttt{m4} & \shortstack{Cohen’s $d$}
& -0.31 & 0.09 & From -0.48 to -0.14 & 0.0004 \\ \hline
\texttt{m5} & \shortstack{Latent Cohen’s $d$}
& -0.38 & 0.09 & From -0.57 to -0.20 & < 0.0001 \\ \hline
\texttt{m6} & \shortstack{Latent Cohen’s $d$}
& -0.43 & 0.10 & From -0.63 to -0.24 & < 0.0001 \\
\arrayrulecolor{black}
\bottomrule
\end{tabular}
\end{table}

Across all adjusted models (\texttt{m2}--\texttt{m6}), the estimated marginal treatment effect remains in the same direction and of similar magnitude, with smaller standard errors than in the unadjusted model (\texttt{m1}). Allowing predictive effects in \texttt{m3} or including all randomized participants in \texttt{m4} has little impact on \(\hat{\tau}\). Slightly larger changes are observed when the outcome is modeled differently. Because headache scores are bounded and strictly positive and may therefore be skewed, the normality assumption may be unrealistic. Model \texttt{m5} relaxes this assumption by using a flexible transformation of the outcome distribution and estimates a mean difference in probit score of \(-0.38\). Model \texttt{m6} instead treats the endpoint as ordinal and estimates a mean difference in probit score for the probability of a worse outcome of \(-0.43\). In this specification, the standard error increases slightly, as expected under an ordinal model.

\begin{table}[H]
\centering
\caption{Ranking of covariates by prognostic and predictive importance across models. Rankings are based on the last row of the inverse Cholesky factor $\mOmega(w)$ of the treatment--dependent correlation matrix, where $w=0$ denotes the control group and $w=1$ denotes the acupuncture group. Prognostic importance is ranked using $\mOmega(0)$, and predictive importance using the difference between $\mOmega(1)$ and $\mOmega(0)$. Rank 1 indicates the strongest effect within each model. Significance stars denote multiplicity-adjusted $p$-values from the corresponding hypothesis tests for prognostic and predictive effects: $^{*}p<0.05$, $^{**}p<0.01$, and $^{***}p<0.001$. Covariates are abbreviated as follows: pk1 = baseline headache score, chr = headache chronicity, and mig = headache diagnosis (migraine or tension-type).}
\label{tab:rankings}
\small
\renewcommand{\arraystretch}{1.15}
\setlength{\tabcolsep}{3.5pt}
\begin{tabular}{llllll|llll}
\toprule
& \multicolumn{5}{c}{Prognostic ranking} & \multicolumn{4}{c}{Predictive ranking} \\
\cmidrule(l){2-6}\cmidrule(r){7-10}
Rank & \texttt{m2} & \texttt{m3} & \texttt{m4} & \texttt{m5} & \texttt{m6}  & \texttt{m3} & \texttt{m4} & \texttt{m5} & \texttt{m6} \\
\midrule
\arrayrulecolor{gray!35}
1 & pk1*** & pk1*** & pk1*** & pk1*** & pk1*** & pk1 & pk1 & pk1* & pk1 \\ \hline
2 & mig & sex & sex & chr & sex & sex & mig & sex & sex \\ \hline
3 & chr & age & age & mig & chr & mig & sex & mig & mig \\ \hline
4 & age & chr & chr & sex & mig & chr & chr & chr & chr \\ \hline
5 & sex & mig & mig & age & age & age & age & age & age \\
\arrayrulecolor{black}
\bottomrule
\end{tabular}
\end{table}

Table~\ref{tab:rankings} shows the covariate rankings with respect to prognostic and predictive importance, while Table~\ref{tab:lambda_gamma_results} reports the corresponding effect estimates used for hypothesis testing, with \(\lambda_j\) representing the prognostic effect of covariate \(X_j\) and \(\gamma_j\) representing its predictive effect. Across all adjusted models, baseline headache score is consistently the top-ranked prognostic covariate. In \texttt{m3}, for example, the prognostic effect \(\lambda_{\texttt{pk1}}=-1.1\) corresponds to a latent correlation of \(0.73\). % in the control arm. 
The direction of this effect indicates that patients with higher baseline headache burden tend to report higher headache scores at follow-up, irrespective of treatment assignment. The statistical evidence for this prognostic association remains strong across models, even after multiplicity adjustment.

Predictive effects, introduced in \texttt{m3}--\texttt{m6} through a treatment--dependent copula structure, are also ranked highest for the baseline headache score. Although the direction and magnitude of the estimated predictive effect are consistent across models, most models fail to reject the null of no predictive effect after multiplicity adjustment. The only model showing some evidence for a predictive effect is \texttt{m5}, which relaxes the normality assumption. This could be explained by the greater flexibility of the outcome specification. In this model, the estimated predictive effect for the baseline headache score is \(\gamma_{\texttt{pk1}}=0.39\), with latent correlations of \(0.7\) in the control arm and \(0.5\) in the treatment arm. The direction of this effect, together with the conditional densities of follow-up headache score given baseline headache score shown in Figure~\ref{fig:fig-app_condensity}, suggests that patients with more severe baseline headache may derive greater benefit from acupuncture. This would align with the findings of \citet{Svensson2026:shapley} on the same dataset, and with the meta-analysis of \citet{Witt2019:acupuncture}, which reported larger treatment effects of acupuncture among patients with more severe pain at baseline compared to controls.

\begin{knitrout}
\definecolor{shadecolor}{rgb}{0.969, 0.969, 0.969}\color{fgcolor}\begin{figure}[H]

{\centering \includegraphics[width=0.9\linewidth]{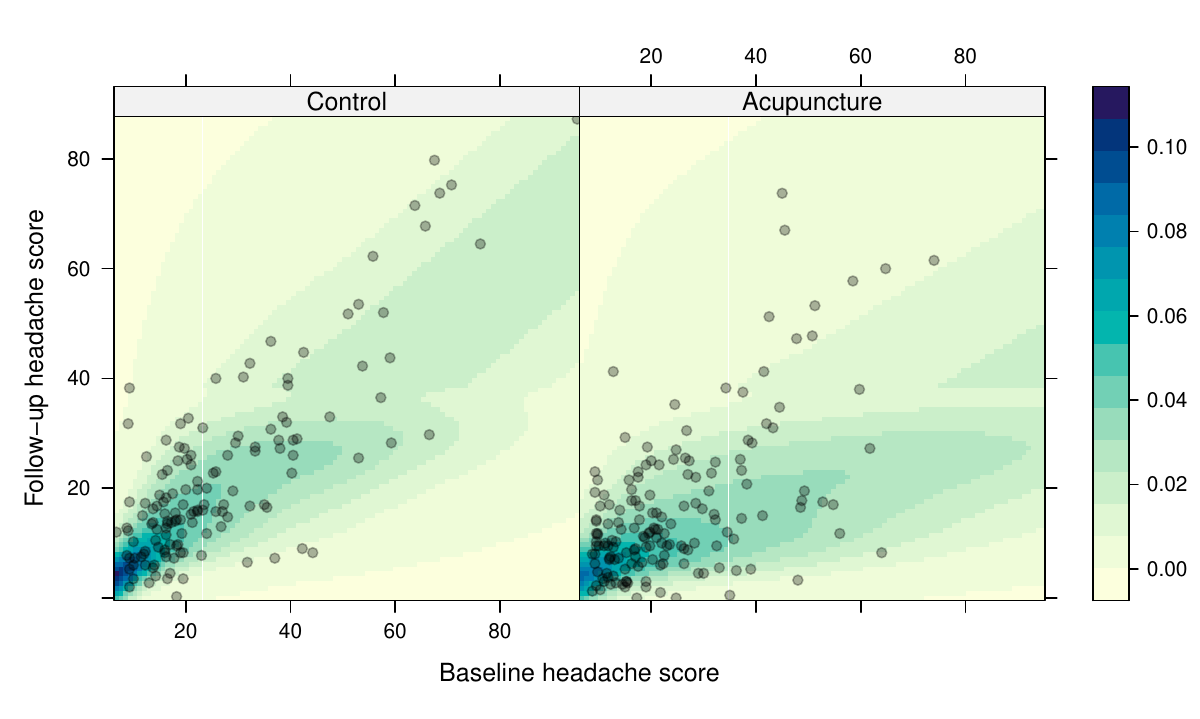} 

}

\caption{Estimated conditional density \(f_{Y_{12} \mid Y_1, W}(y_{12} \mid y_1, w)\) of follow-up headache score \(Y_{12}\) given baseline headache score \(Y_1\) and treatment group \(W\). The density is derived from the joint model \texttt{m5} by marginalization over the other baseline covariates and then conditioning on \(Y_1\). Color shading represents the model-based density, with darker regions indicating higher density. Dots show the observed data.}\label{fig:fig-app_condensity}
\end{figure}

\end{knitrout}

Overall, this application shows that the proposed method estimates a marginal treatment effect \(\tau\) consistent with a benefit of acupuncture compared to control, and that this conclusion is robust across alternative outcome assumptions and scales. Covariate adjustment improves precision, primarily through prognostic adjustment, while predictive adjustment adds little further reduction in standard error. At the same time, this method allows direct interpretation of covariate effects: baseline headache score is the dominant prognostic factor across all models and the most plausible candidate effect modifier. However, the evidence for treatment effect heterogeneity is weak and remains sensitive to the assumed outcome distribution.

%%%%%%%%%%%%%%%%%%%%%%%%%%%%%%%%%%%%%%%%%%%%%%%%%%%%%%%%%%%%%%%%%%%%%%

% LaTeX file for Chapter 07

\section{Discussion}

We presented and evaluated nonparanormal adjusted marginal inference with heterogeneous treatment effects (NAMI-HTE) as a framework for jointly estimating both marginal treatment effects and treatment effect heterogeneity. Unlike traditional approaches, the proposed framework avoids the usual compartmentalization between these two objectives, therefore preserving marginal interpretability while simultaneously adjusting for and quantifying prognostic and predictive effects of covariates.

For continuous outcomes, we showed theoretically that the asymptotic variance of Cohen’s $d$ is never worse, and often smaller, when adjusting for covariates compared to an unadjusted analysis. Efficiency gains are mainly driven by prognostic effects, while realistic predictive effects add only modest additional improvement. Simulation studies under correct model specification confirmed these theoretical results and showed the same pattern for continuous, binary, and survival outcomes. The marginal treatment effect, expressed as Cohen’s $d$, a log-odds ratio, or a log-hazard ratio, remained collapsible under covariate adjustment and was estimated without bias and with improved precision. 

The prognostic and predictive effects are explicitely estimated on a standardized scale, allowing direct comparison and ranking between covariates. Formal hypothesis testing with multiplicity adjustment is possible, allowing confirmatory assessment of treatment--covariate interactions with controlled type~I error. This contrasts with more exploratory forest-based methods for detecting heterogeneous treatment effects \citep{Athey2019:forest, Seibold2018:forest, Dandl2024:forest}, which are usually not designed for confirmatory inference \citep{Ondra2016:subgroup}.

A considerable advantage of this framework compared with existing approaches is its generality and flexibility. It can be applied to continuous, binary, ordinal, and time-to-event outcomes, accommodates different assumptions on the outcome distribution, and allows incorporation of various censoring mechanisms (right-, left-, and interval-censoring), missing outcomes, and stratified randomization. Although general approaches for covariate-adjusted marginal treatment effect estimation exist \citep{Zhang2008:covadj}, implementations are typically tailored to specific settings. To the best of our knowledge, no existing method unites this range of outcome types, censoring mechanisms, and design features within a single framework. 

A limitation is the weak power to detect predictive effects, which are often of primary clinical interest. Trials are typically powered to detect the marginal treatment effect, not treatment--covariate interactions. As a result, predictive effects may be difficult to detect with realistic sample sizes. This limitation is not specific to this framework but reflects a general property of interaction testing \citep{Ondra2016:subgroup}; similar challenges have been observed for forest-based methods, which may struggle to separate prognostic and predictive effects in finite samples \citep{Sechidis2018:progpred, Hermansson2021:forest, Lipkovich2024:progpred}. Another limitation is the reliance on a latent Gaussian correlation structure. This implies that associations between outcome and covariates are assumed to be monotone on the original scales, and linear on the latent scale. If the true relationship is non-monotone, for example increasing and then decreasing, this assumption is violated. %Such patterns could be addressed by categorizing the covariate, but this may lead to loss of information and reduced efficiency. 

Future work should investigate the robustness of the method under model misspecification. Two types are particularly relevant: (i) misspecification of the marginal outcome model, for example through incorrect distributional assumptions, and (ii) misspecification of the correlation structure, for example in the case of non-monotonicity. Regulatory guidance emphasizes that covariate adjustment methods should provide valid inference under approximately the same minimal assumptions required for unadjusted estimation in randomized trials \citep{FDA2023:covadj}. It would therefore be important to assess robustness under misspecification and to compare performance with augmentation-based approaches for the estimation of the marginal treatment effect. 

%%%%%%%%%%%%%%%%%%%%%%%%%%%%%%%%%%%%%%%%%%%%%%%%%%%%%%%%%%%%%%%%%%%%%%

\bibliography{biblio}

\newpage

\begin{appendix}

% LaTeX file for Chapter 01

%%%%%%%%%%%%%%%%%%%%%%%%%%%%%%%%%%%%%%%%%%%%%%%%
\section{Proof of Lemmas}
\label{sec:A_seproofs}
%%%%%%%%%%%%%%%%%%%%%%%%%%%%%%%%%%%%%%%%%%%%%%%%

Consider a binary treatment indicator \(W\in\{0,1\}\), a normally distributed outcome

\[
Y\mid W=w \sim \mathcal{N}\!\left(
-\frac{\vartheta_{11}+(w-\tfrac12)\tau}{\vartheta_{12}},\;
\vartheta_{12}^{-2}
\right),
\]
and a normally distributed covariate
\(X_1\sim \mathcal{N}\!\bigl(-\vartheta_{21}/\vartheta_{22},\;\vartheta_{22}^{-2}\bigr)\).

Since \(X\) and \(Y\) are normal, we parametrize them using linear transformation functions as in Section~\ref{subsec:application-linear}. To simplify the Fisher information matrix, we encode the treatment effect as a symmetric shift \(-\tfrac12\tau\) in arm \(W=0\) and \(+\tfrac12\tau\) in arm \(W=1\); this makes \(\tau\) orthogonal to the outcome intercept \(\vartheta_{11}\). We also index the variables so that \(j=1\) corresponds to \(Y\) and \(j=2\) to \(X\). This gives

\[
h_1(y\mid w)=\vartheta_{11}+\vartheta_{12}\,y+(w-\tfrac12)\tau,
\qquad
h_2(x)=\vartheta_{21}+\vartheta_{22}\,x,
\]

where \(\tau\) corresponds to Cohen’s \(d\).

Treatment-specific dependence is parameterized through a lower-triangular unit-diagonal matrix \(\mLambda(w)\) as in Section~\ref{subsec:mmlt-omega}. With one covariate, the relevant dependence parameter in arm \(w\) reduces to a scalar \(\lambda^{(w)} \;=\; \lambda + w \, \gamma\). As done for \(\tau\), we reparametrize this dependence structure as \(\lambda^{(0)}=\lambda_0-\tfrac12\gamma\) in arm \(W=0\) and \(\lambda^{(1)}=\lambda_0+\tfrac12\gamma\) in arm \(W=1\) to simplify the Fisher information matrix; this ensures that \(\lambda_0\) and \(\gamma\) are orthogonal. This gives

\[
\lambda^{(w)}=\lambda_0+(w-\tfrac12)\gamma .
\]

The bivariate model is
\[
\hvec(Y,X\mid W=w)
=\big(h_1(Y\mid W=w),h_2(X)\big)^\top
\sim \mathcal{N}_2\!\bigl(0,\;\mLambda(w)^{-1}\mLambda(w)^{-\top}\bigr),
\]

with covariance

\[
\mSigma(w)=
\begin{pmatrix}
1 & -\lambda^{(w)}\\
-\lambda^{(w)} & (\lambda^{(w)})^2+1
\end{pmatrix}.
\]

This parametrization differs slightly from that in Section~\ref{subsec:mmlt-omega} but leads to the same likelihood and the same inference for \(\vartheta_{11},\vartheta_{12},\tau\); it is used here because it makes the Fisher calculations simpler.

Since both variables are continuous, the joint density can be defined as \(f_{Y,X\mid W}(y,x\mid w) = f_{Y\mid W}(y\mid w)\,f_{X\mid Y,W}(x\mid y,w)\). The marginal distribution of \(Y\mid W=w\) is

\begin{align*}
F_{Y\mid W}(y\mid w)
&=\Phi\bigl(\vartheta_{11}+\vartheta_{12}y+(w-\tfrac12)\tau\bigr),\\
f_{Y\mid W}(y\mid w)
&=\phi\bigl(\vartheta_{11}+\vartheta_{12}y+(w-\tfrac12)\tau\bigr)\,\vartheta_{12},\\
\log f_{Y\mid W}(y\mid w)
&\propto
-\tfrac12\bigl(\vartheta_{11}+\vartheta_{12}y+(w-\tfrac12)\tau\bigr)^2
+\log(\vartheta_{12}).
\end{align*}

Using the same logic as in Section~\ref{subsec:mmlt-cond}, the conditional distribution of \(X\mid Y=y,W=w\) is

\begin{align*}
F_{X\mid Y,W}(x\mid y,w)
&=\Phi\Big(
\vartheta_{21}+\vartheta_{22}x
+\lambda^{(w)}\bigl(\vartheta_{11}+\vartheta_{12}y+(w-\tfrac12)\tau\bigr)
\Big),\\
f_{X\mid Y,W}(x\mid y,w)
&=\phi\Big(
\vartheta_{21}+\vartheta_{22}x
+\lambda^{(w)}\bigl(\vartheta_{11}+\vartheta_{12}y+(w-\tfrac12)\tau\bigr)
\Big)\,\vartheta_{22},\\
\log f_{X\mid Y,W}(x\mid y,w)
&\propto
-\tfrac12\Big(
\vartheta_{21}+\vartheta_{22}x
+\lambda^{(w)}\bigl(\vartheta_{11}+\vartheta_{12}y+(w-\tfrac12)\tau\bigr)
\Big)^2
+\log(\vartheta_{22}).
\end{align*}

For a single observation, the log-likelihood for \(\mTheta=(\vartheta_{11},\vartheta_{12},\tau, \vartheta_{21},\vartheta_{22},\lambda_0,\gamma)^\top\) is

\[
\ell(\mTheta)
=
\log f_{Y\mid W}(y\mid w)
+
\log f_{X\mid Y,W}(x\mid y,w).
\]

Let \(-\mathcal H\) denote the observed Fisher information,

\[
-\mathcal H
= -\frac{\partial^2}{\partial\mTheta\,\partial\mTheta^\top}\,\ell(\mTheta).
\]

The expected Fisher information in arm \(w\) is \(\mI = \Ex_{Y,X\mid W=w}[-\mathcal H]\); this can be evaluated analytically using only the low-order moments \(\mathbb E(Y)\), \(\mathbb E(Y^2)\), \(\mathbb E(X)\), \(\mathbb E(X^2)\) and \(\mathbb E(XY)\) in each arm. These moments follow directly from the linear transformations defining \(Y\) and \(X\), together with the identities \(E(U^2)=\Var(U)+E(U)^2$ and $E(UV)=\Cov(U,V)+E(U)E(V)\). Additionally, since the transformations are linear, $\mSigma(w)$ represents the correlation matrix of $(Y,X)$; the variances and covariances are obtained by multiplying its elements with $1/\vartheta_{12}^2$ and $1/\vartheta_{22}^2$.

With \(\mathbb E(Y) = -(\vartheta_{11}+(w-\tfrac12)\tau)/\vartheta_{12}\) and \(\Var(Y) = \vartheta_{12}^{-2}\), we obtain 

\[
\mathbb E(Y^2) = \frac{1}{\vartheta_{12}^2} + 
\left(\frac{\vartheta_{11}+(w-\tfrac12)\tau}{\vartheta_{12}}\right)^2 .
\]

Similarly, with \(\mathbb E(X) = -\vartheta_{21}/\vartheta_{22}\) and \(\Var(X) = \left((\lambda^{(w)})^2+1\right)/\vartheta_{22}^2\), we obtain 

\[
\mathbb E(X^2) = \frac{(\lambda^{(w)})^2+1}{\vartheta_{22}^2} + 
\left(\frac{\vartheta_{21}}{\vartheta_{22}}\right)^2.
\]

The covariance is \(\Cov(X,Y) = -\lambda^{(w)}/(\vartheta_{12}\vartheta_{22})\), hence

\[
\mathbb E(XY) = 
-\frac{\lambda^{(w)}}{\vartheta_{12}\vartheta_{22}}
+
\frac{\vartheta_{21}}{\vartheta_{22}}
\left(
\frac{\vartheta_{11}+(w-\tfrac12)\tau}{\vartheta_{12}}\right) .
\]

These five quantities are sufficient to compute \(\mI\) in each arm. With \(N\) observations per arm, the total expected information is

\[
\mI_N
=
N\Big(
\Ex_{Y,X\mid W=0}[-\mathcal H]
+
\Ex_{Y,X\mid W=1}[-\mathcal H]
\Big).
\]

To obtain the standard errors, we first reorder \(\mI_N\) so that \(\etavec=(\lambda_0,\gamma,\tau)^\top\) comes first and partition as

\[
\mI_N
=
N\begin{pmatrix}
D & B^\top\\
B & A
\end{pmatrix},
\]

where \(D\) is the \(3\times3\) block for \(\etavec\), \(A\) is the \(4\times4\) block for \(\xivec=(\vartheta_{11},\vartheta_{12},\vartheta_{21},\vartheta_{22})^\top\), and \(B\) contains the cross-information between \(\etavec\) and \(\xivec\). With this ordering, the blocks take the following form:

\[
D=
\begin{pmatrix}
2 & 0 & 0\\[0.4em]
0 & \tfrac12 & 0\\[0.4em]
0 & 0 & \frac{4\lambda_0^2+\gamma^2+4}{8}
\end{pmatrix},
\qquad
B=
\begin{pmatrix}
0 & 0 & \dfrac{\gamma}{2} & 2
\\[1em]
-\dfrac{2\lambda_0}{\vartheta_{22}} &
-\dfrac{\gamma}{2\vartheta_{22}} &
-\dfrac{\gamma\,\vartheta_{21}}{2\vartheta_{22}} &
-\dfrac{2\vartheta_{21}}{\vartheta_{22}}
\\[1em]
0 & 0 & \gamma\,\lambda_0 & 2\lambda_0
\\[0.6em]
\dfrac{2\lambda_0}{\vartheta_{12}} &
\dfrac{\gamma}{2\vartheta_{12}} & 
\frac{8\gamma\, \lambda_0\, \vartheta_{11} +\left(4\lambda_0^2+\gamma^2+4\right)\tau} {8\vartheta_{12}} &
-\frac{4\lambda_0\,\vartheta_{11}+\gamma\,\tau}{2\vartheta_{12}}
\end{pmatrix},
\]

\[
A=
\resizebox{\textwidth}{!}{$
\begin{pmatrix}
2 &
-\dfrac{2\vartheta_{21}}{\vartheta_{22}} &
2\lambda_{0} &
-\dfrac{4\lambda_{0}\,\vartheta_{11}+\gamma\,\tau}
{2\vartheta_{12}}
\\[1.2em]

-\dfrac{2\vartheta_{21}}{\vartheta_{22}} &
\dfrac{4\vartheta_{21}^{2}+4\lambda_{0}^{2}+\gamma^{2}+8}
{2\vartheta_{22}^{2}} &
-\dfrac{2\lambda_{0}\,\vartheta_{21}}{\vartheta_{22}} &
\dfrac{(4\lambda_{0}\,\vartheta_{11}+\gamma\,\tau)\vartheta_{21}
-4\lambda_{0}^{2}-\gamma^{2}}
{2\vartheta_{12}\,\vartheta_{22}}
\\[1.3em]

2\lambda_{0} &
-\dfrac{2\lambda_{0}\vartheta_{21}}{\vartheta_{22}} &
\dfrac{4\lambda_{0}^{2}+\gamma^{2}+4}{2} &
-\dfrac{(4\lambda_{0}^{2}+\gamma^{2}+4)\vartheta_{11}
+2\gamma\,\lambda_{0}\,\tau}
{2\vartheta_{12}}
\\[1.3em]

-\dfrac{4\lambda_{0}\,\vartheta_{11}+\gamma\,\tau}
{2\vartheta_{12}} &
\dfrac{(4\lambda_{0}\,\vartheta_{11}+\gamma\,\tau)\vartheta_{21}
-4\lambda_{0}^{2}-\gamma^{2}}
{2\vartheta_{12}\,\vartheta_{22}} &
-\left(
\dfrac{(4\lambda_{0}^{2}+\gamma^{2}+4)\vartheta_{11}
+2\gamma\,\lambda_{0}\,\tau}
{2\vartheta_{12}}
\right) &
\dfrac{
(16\lambda_{0}^{2}+4\gamma^{2}+16)\vartheta_{11}^{2}
+16\gamma\,\lambda_{0}\,\tau\,\vartheta_{11}
+(4\lambda_{0}^{2}+\gamma^{2}+4)\tau^{2}
+16\lambda_{0}^{2}+4\gamma^{2}+32
}
{8\vartheta_{12}^{2}}
\end{pmatrix}
$}.
\]

Let \(M := D - B^\top A^{-1}B\) denote the Schur complement of \(A\) in \(\mI_N\). By block inversion, the asymptotic covariance of \(\etavec\) is

\[
\Var(\etavec)
\;\approx\;
\frac{1}{N}\,M^{-1}
\;=\;
\frac{1}{N}\,
\begin{pmatrix}
\frac{\lambda_0^2+1}{2} &
\frac{\gamma\lambda_0}{2} &
-\frac{\lambda_0\tau}{4}\\[0.4em]
\frac{\gamma\lambda_0}{2} &
\frac{\gamma^2+4}{2} &
-\frac{\gamma\tau}{4}\\[0.4em]
-\frac{\lambda_0\tau}{4} &
-\frac{\gamma\tau}{4} &
\frac{\tau^2}{4}+
\frac{2(\gamma^2+4)}{4\lambda_0^2+\gamma^2+4}
\end{pmatrix}.
\]

Reparametrizing with \(\lambda_0=\lambda+\tfrac12\gamma\) gives

\begin{equation}
\Var(\lambda,\gamma,\tau)
\approx
\frac{1}{N}
\begin{pmatrix}
\frac{\lambda^2+2}{2} &
\frac{\gamma\lambda-2}{2} &
-\frac{\lambda\tau}{4}\\[0.4em]
\frac{\gamma\lambda-2}{2} &
\frac{\gamma^2+4}{2} &
-\frac{\gamma\tau}{4}\\[0.4em]
-\frac{\lambda\tau}{4} &
-\frac{\gamma\tau}{4} &
\frac{\tau^2}{4}+
\frac{\gamma^2+4}{2\lambda^2+2\lambda\gamma+\gamma^2+2}
\end{pmatrix}.
\label{eq:var_matrix}
\end{equation}

The asymptotic variance of \(\tau\) is the \((3,3)\) element of~\eqref{eq:var_matrix}, namely

\[
\Var(\tau)
\approx
\frac{1}{N}\,
\left(
\frac{\tau^2}{4}+
\frac{\gamma^2+4}{2\lambda^2+2\lambda\gamma+\gamma^2+2} \right),
\qquad
\mathrm{SE}(\tau)=\sqrt{\Var(\tau)}.
\]

This expression corresponds to Lemma~4 in Section~\ref{subsubsec:se-cohensd}. Setting \(\lambda=\gamma=0\) recovers the unadjusted case (Lemma~1), setting \(\gamma=0\) gives the prognostic-only case (Lemma~2), and setting \(\lambda=0\) gives the predictive-only case (Lemma~3). Derivations for Lemma~1 and Lemma~2 are also provided in the appendix of \citet{Dandl2026:NAMI}. Note that the asymptotic variances of \(\lambda\) and \(\gamma\) are only valid for this specific setup. Additionally, the model depends on the ordering of the variables; changing their order leads to different parameter values \citep{Barratt2023:optm}.

%%%%%%%%%%%%%%%%%%%%%%%%%%%%%%%%%%%%%%%%%%%%%%%%
\section{Theoretical properties of Lemmas}
\label{sec:A_seprops}
%%%%%%%%%%%%%%%%%%%%%%%%%%%%%%%%%%%%%%%%%%%%%%%%

As shown in Section~\ref{sec:A_seproofs}, closed-form expressions for the asymptotic standard error of the marginal treatment effect (Cohen’s \(d\)) are available in the case of a single continuous normal covariate and a continuous normal outcome (Lemmas~1–4). These results allow us to study how adjustment affects precision as a function of the strength of prognostic and predictive effects.

To quantify efficiency gains, we consider the ratio of squared standard errors between the adjusted and unadjusted analyses, \(\left(\mathrm{SE}_{\text{adj}}/\mathrm{SE}_{\text{unadj}}\right)^2\), 
which can be interpreted as the relative sample size required under adjustment to achieve the same power as the unadjusted analysis. Values below 1 indicate improved efficiency due to adjustment.

Figure~\ref{fig:fig-SEprog} considers the case of a purely prognostic covariate (\(\gamma=0\)), so that the correlation between \(X\) and \(Y\) is identical in both treatment arms. Figure~\ref{fig:fig-SEpred} considers a purely predictive covariate (\(\lambda=0\)), where the correlation between \(X\) and \(Y\) is present only in the treatment arm. Figure~\ref{fig:fig-SEprog_pred} shows the general case in which the covariate may be both prognostic and predictive, allowing the correlation between \(X\) and \(Y\) to differ between arms. In all figures, the effect size is fixed at \(\tau=0.5\).

For interpretability, the axes are expressed in terms of the arm-specific correlations between \(X\) and \(Y\) rather than the copula parameters \(\lambda\) and \(\gamma\). These correlations are directly interpretable on the original scale of the covariate and outcome.

In the purely prognostic case, adjustment always reduces the standard error unless the correlation is zero. The gain in efficiency increases with the absolute value of the correlation, meaning that stronger prognostic information improves precision of the marginal treatment effect. In the purely predictive case, adjustment also reduces the standard error whenever the treatment-specific correlation is nonzero. However, the gain in efficiency increases more slowly compared to the prognostic case, indicating that predictive information contributes less to precision than equally strong prognostic information. In the general case where the covariate may be both prognostic and predictive, adjustment reduces the standard error except when both effects are zero or when the correlations in the two arms have equal magnitude but opposite sign. The latter corresponds to \(\lambda = -\gamma/2\), where adjustment does not improve precision. Outside this boundary, incorporating covariate information leads to a smaller asymptotic variance of the marginal treatment effect.

\begin{knitrout}
\definecolor{shadecolor}{rgb}{0.969, 0.969, 0.969}\color{fgcolor}\begin{figure}[H]

{\centering \includegraphics[width=0.77\linewidth]{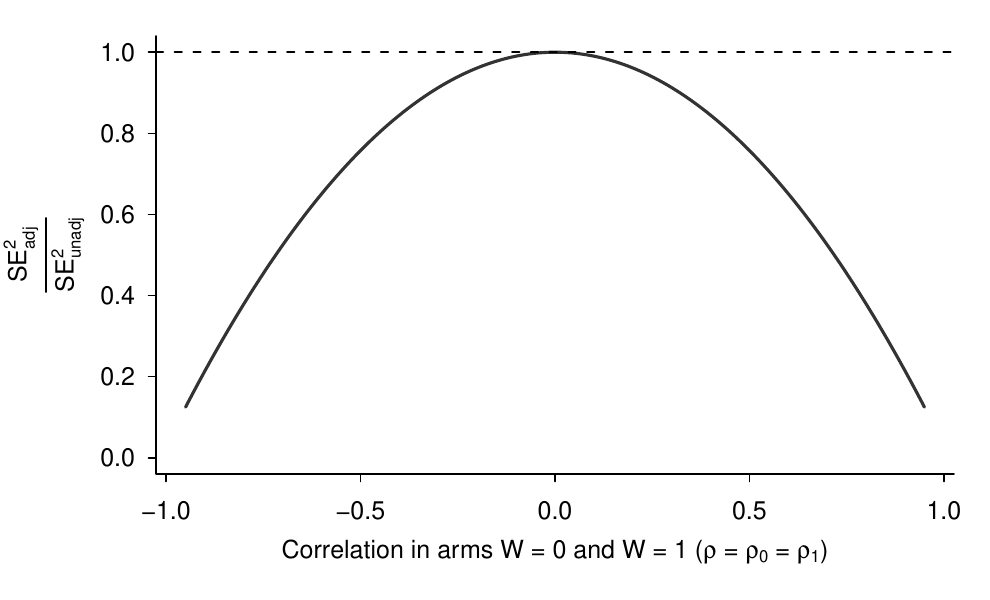} 

}

\caption{Theoretical ratio of squared standard errors \((\mathrm{SE}_{\text{adj}}/\mathrm{SE}_{\text{unadj}})^2\) for Cohen's \(d\) with a normal outcome and a purely prognostic normal covariate (\(\gamma = 0\)). The horizontal axis shows the correlation between \(X\) and \(Y\) in both treatment arms (\(\rho = \rho_0 = \rho_1\)), where \(\rho = -\lambda/\sqrt{1+\lambda^2}\). The dashed black line indicates equal efficiency (ratio of 1). The treatment effect is fixed at \(\tau = 0.5\).}\label{fig:fig-SEprog}
\end{figure}

\end{knitrout}

\begin{knitrout}
\definecolor{shadecolor}{rgb}{0.969, 0.969, 0.969}\color{fgcolor}\begin{figure}[H]

{\centering \includegraphics[width=0.77\linewidth]{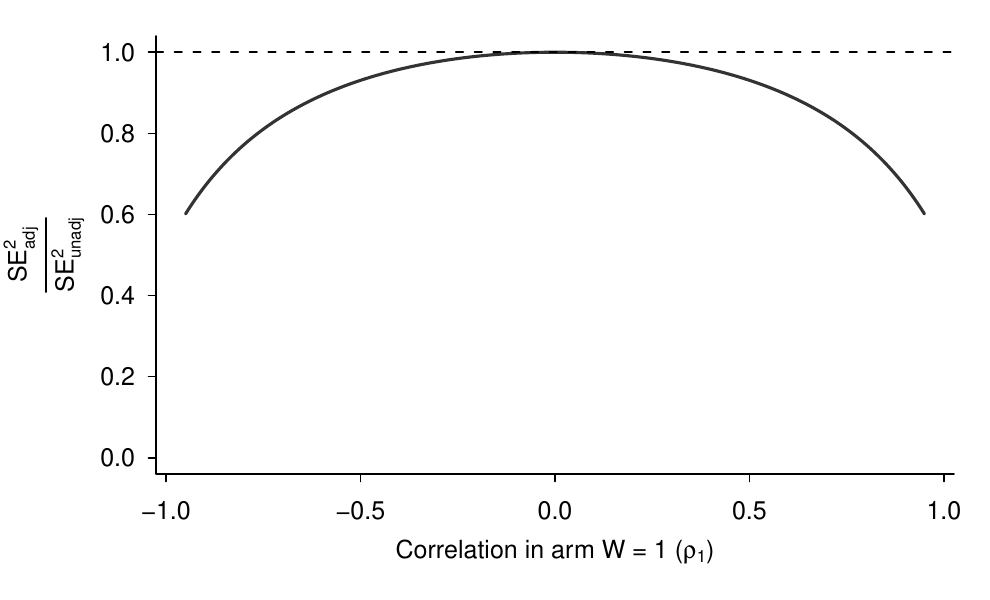} 

}

\caption{Theoretical ratio of squared standard errors \((\mathrm{SE}_{\text{adj}}/\mathrm{SE}_{\text{unadj}})^2\) for Cohen's \(d\) with a normal outcome and a purely predictive normal covariate (\(\lambda = 0\)). The horizontal axis shows the correlation between \(X\) and \(Y\) in the treatment arm (\(\rho_1\)), where \(\rho_1 = -\gamma/\sqrt{1+\gamma^2}\). The dashed black line indicates equal efficiency (ratio of 1). The treatment effect is fixed at \(\tau = 0.5\).}\label{fig:fig-SEpred}
\end{figure}

\end{knitrout}

\begin{knitrout}
\definecolor{shadecolor}{rgb}{0.969, 0.969, 0.969}\color{fgcolor}\begin{figure}[H]

{\centering \includegraphics[width=0.55\linewidth]{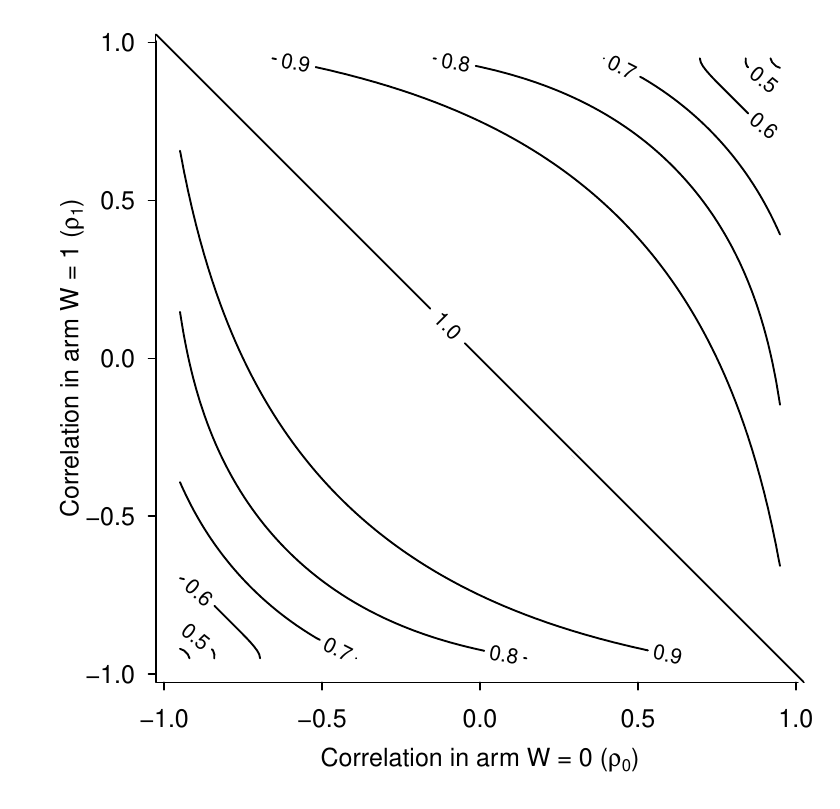} 

}

\caption{Theoretical ratio of squared standard errors \((\mathrm{SE}_{\text{adj}}/\mathrm{SE}_{\text{unadj}})^2\) for Cohen's \(d\) with a normal outcome and a normal with both prognostic and predictive effects. The horizontal axis shows the correlation between \(X\) and \(Y\) in the treatment arm (\(\rho_1\)), and the vertical axis shows the correlation in the control arm (\(\rho_0\)), where \(\rho_{w} = -(\lambda + w\,\gamma)/\sqrt{1+(\lambda + w\,\gamma)^2}\). Contours represent the relative efficiency of the adjusted estimator. The treatment effect is fixed at \(\tau = 0.5\).}\label{fig:fig-SEprog_pred}
\end{figure}

\end{knitrout}

%%%%%%%%%%%%%%%%%%%%%%%%%%%%%%%%%%%%%%%%%%%%%%%%
\section{Simulation studies}
%%%%%%%%%%%%%%%%%%%%%%%%%%%%%%%%%%%%%%%%%%%%%%%%

%% --------------------------------------------------------------
\subsection*{Implementation details}
\label{subsec:A_sim_setup}
%% --------------------------------------------------------------

All analyses were performed in \proglang{R} version~4.5.1 \citep{R}. All NAMI-HTE models in Section~\ref{sec:results} were fitted by maximum likelihood under the nonparanormal framework described in \citet{Hothorn2025:nonparanormal}, using the \texttt{Mmlt} function from the \texttt{tram} package \citep{pkg:tram}. Inference for the predictive copula parameters \(\gamma\) was based on Wald tests with multiplicity adjustment implemented via \texttt{glht} from the \texttt{multcomp} package \citep{pkg:multcomp}.

The marginal models for the outcome \(Y\mid W\) were used both for the unadjusted marginal inference (MI) analysis and as input to the adjusted NAMI-HTE model, since \texttt{tram::Mmlt} requires marginal transformation models for all variables (both outcome and covariates). The marginal specification of \(Y\mid W\) was therefore identical in the MI and NAMI-HTE analyses. For the continuous outcome, generated as standard normal, we used a linear transformation model fitted with \texttt{tram::Lm}, corresponding to~\eqref{eq:normal-tm}. For the binary outcome, we used \texttt{tram::Polr}, where the transformation function is a step function with one jump, corresponding to a binary logistic regression model as in~\eqref{eq:binary-tm}. For the survival outcome, we used \texttt{tram::Coxph} as in~\eqref{eq:survival-tm}; the transformation function was modeled using a Bernstein polynomial of order six, yielding a fully parameterized Cox-type model.

Marginal models for all covariates \(X_j,\; j = 1, ..., 4\) were also specified as transformation models and used as input to \texttt{tram::Mmlt}. Each covariate was modeled using \texttt{tram::BoxCox}, corresponding to \(F(x_j) \;=\; \Phi\!\bigl(h(x_j)\bigr)\), where \(h(x_j)\) is a flexible Bernstein polynomial of order six. This specification does not impose a distributional assumption on the covariates.

For the survival setting, outcomes were subject to independent right-censoring. We generated censoring times \(C\) from the following conditional distribution, as in \citet{Dandl2026:NAMI}:
\(
\Prob(C \le c \mid W = w, \rX = \rx) =
\Phi\!\left[
\sum_{j=1}^{4} \omega_{5j}^{(w)} h_j(x_j)
+ \omega_{55}^{(w)}
\Phi^{-1}\!\left\{
\cloglog^{-1}\bigl(\eparm_1 + \eparm_2 \log(c) - \gamma - \tau w\bigr)
\right\}
\right],
\)
with \(\eparm_1 = 0\) and \(\eparm_2 = 1\). An observation was censored if \(C < Y\). The parameter \(\gamma\) determines the noncensoring probability through \(\Prob(Y < C \mid W = w, \rX = \rx) = \mathrm{logit}^{-1}(\gamma)\). In our experiments, \(\gamma\) was chosen to yield a noncensoring probability of approximately 0.3, corresponding to heavy censoring.
%% WHEN RERUNNING SIMULATIONS ON SURVER: MAKE SURE TO CHANGE DIVISION IN SIM.S BY 1 INSTEAD OF 1.5, so that theta2 = 1.5

%% --------------------------------------------------------------
\subsection*{Further results}
\label{subsec:A_sim_results}
%% --------------------------------------------------------------

\begin{table}[H]
\centering
\caption{Empirical size for testing $H_0: \tau = 0$ using a Wald test ($\alpha=0.05$) under true effect $\tau=0$. Results are shown across configurations of outcome type, model (MI: unadjusted marginal inference, NAMI-HTE: nonparanormal adjusted marginal inference with heterogeneous treatment effects) and true predictive effect $\gamma$ of $X_1$.}
\label{tbl-tau_size}
\small
\renewcommand{\arraystretch}{1.15}
\setlength{\tabcolsep}{4.5pt}
\begin{tabular}{llccc}
\toprule
Outcome & Model & \multicolumn{3}{c}{Size} \\
\cmidrule(lr){3-5}
 &  & $\gamma=0$ & $\gamma=0.25$ & $\gamma=0.50$ \\
\midrule
continuous & MI & 0.052 & 0.053 & 0.052 \\
 & NAMI-HTE & 0.060 & 0.057 & 0.058 \\
\arrayrulecolor{gray!35}\midrule\arrayrulecolor{black}
binary & MI & 0.051 & 0.051 & 0.049 \\
 & NAMI-HTE & 0.048 & 0.053 & 0.053 \\
\arrayrulecolor{gray!35}\midrule\arrayrulecolor{black}
survival & MI & 0.048 & 0.052 & 0.052 \\
 & NAMI-HTE & 0.059 & 0.056 & 0.056 \\
\bottomrule
\end{tabular}
\end{table}\begin{table}[H]
\centering
\caption{Empirical size for testing $H_0: \gamma = 0$ using a Wald test ($\alpha=0.05$) under true effect $\gamma=0$. Results are shown across configurations of outcome type and true marginal treatment effect $\tau$, only for the NAMI-HTE (nonparanormal adjusted marginal inference with heterogeneous treatment effects) model.}
\label{tbl-gamma_size}
\small
\renewcommand{\arraystretch}{1.15}
\setlength{\tabcolsep}{4.5pt}
\begin{tabular}{lllc}
\toprule
Outcome & Model & $\tau$ & Size \\
\midrule
continuous & NAMI-HTE & 0 & 0.024 \\
 & NAMI-HTE & 0.5 & 0.024 \\
\arrayrulecolor{gray!35}\midrule\arrayrulecolor{black}
binary & NAMI-HTE & 0 & 0.011 \\
 & NAMI-HTE & 0.5 & 0.012 \\
\arrayrulecolor{gray!35}\midrule\arrayrulecolor{black}
survival & NAMI-HTE & 0 & 0.016 \\
 & NAMI-HTE & 0.5 & 0.015 \\
\bottomrule
\end{tabular}
\end{table}

\normalsize
{\let\small\normalsize
{\small
\renewcommand{\arraystretch}{1.15}
\setlength{\tabcolsep}{4.5pt}
\begin{longtable}{llllccc}
\caption{Empirical variability of $\hat{\tau}$ across simulations, summarized by the mean and median estimated standard error $SE(\hat{\tau})$ and the empirical standard deviation $SD(\hat{\tau})$. Results are shown across configurations of outcome type, model (MI: unadjusted marginal inference, NAMI-HTE: nonparanormal adjusted marginal inference with heterogeneous treatment effects), true marginal treatment effect $\tau$, and true predictive effect $\gamma$ of $X_1$.}\\
\label{tab:tau_variability}\\
\toprule
Outcome & Model & $\tau$ & $\gamma$ & \multicolumn{3}{c}{Variability measure of $\hat{\tau}$} \\
\cmidrule(lr){5-7}
 &  &  &  & Mean($SE(\hat{\tau})$) & Median($SE(\hat{\tau})$) & $SD(\hat{\tau})$ \\
\midrule
\endfirsthead
\multicolumn{7}{l}{\small\itshape Table~\thetable\ (continued)} \\
\toprule
Outcome & Model & $\tau$ & $\gamma$ & \multicolumn{3}{c}{Variability measure of $\hat{\tau}$} \\
\cmidrule(lr){5-7}
 &  &  &  & Mean($SE(\hat{\tau})$) & Median($SE(\hat{\tau})$) & $SD(\hat{\tau})$ \\
\midrule
\endhead
\midrule
\multicolumn{7}{r}{\small\itshape Continued on next page} \\
\endfoot
\bottomrule
\endlastfoot
continuous & MI & 0 & $\gamma=0$ & 0.222 & 0.221 & 0.227 \\
 & MI & 0 & $\gamma=0.25$ & 0.222 & 0.221 & 0.228 \\
 & MI & 0 & $\gamma=0.50$ & 0.222 & 0.221 & 0.224 \\
 & MI & 0.5 & $\gamma=0$ & 0.225 & 0.224 & 0.228 \\
 & MI & 0.5 & $\gamma=0.25$ & 0.225 & 0.224 & 0.229 \\
 & MI & 0.5 & $\gamma=0.50$ & 0.225 & 0.224 & 0.232 \\
 & NAMI-HTE & 0 & $\gamma=0$ & 0.165 & 0.165 & 0.174 \\
 & NAMI-HTE & 0 & $\gamma=0.25$ & 0.161 & 0.161 & 0.167 \\
 & NAMI-HTE & 0 & $\gamma=0.50$ & 0.158 & 0.158 & 0.164 \\
 & NAMI-HTE & 0.5 & $\gamma=0$ & 0.170 & 0.170 & 0.176 \\
 & NAMI-HTE & 0.5 & $\gamma=0.25$ & 0.166 & 0.166 & 0.172 \\
 & NAMI-HTE & 0.5 & $\gamma=0.50$ & 0.163 & 0.162 & 0.170 \\
\arrayrulecolor{gray!35}\midrule\arrayrulecolor{black}
binary & MI & 0 & $\gamma=0$ & 0.224 & 0.223 & 0.225 \\
 & MI & 0 & $\gamma=0.25$ & 0.224 & 0.223 & 0.225 \\
 & MI & 0 & $\gamma=0.50$ & 0.224 & 0.223 & 0.224 \\
 & MI & 0.5 & $\gamma=0$ & 0.227 & 0.227 & 0.227 \\
 & MI & 0.5 & $\gamma=0.25$ & 0.227 & 0.227 & 0.226 \\
 & MI & 0.5 & $\gamma=0.50$ & 0.227 & 0.227 & 0.228 \\
 & NAMI-HTE & 0 & $\gamma=0$ & 0.181 & 0.182 & 0.182 \\
 & NAMI-HTE & 0 & $\gamma=0.25$ & 0.178 & 0.179 & 0.180 \\
 & NAMI-HTE & 0 & $\gamma=0.50$ & 0.176 & 0.176 & 0.177 \\
 & NAMI-HTE & 0.5 & $\gamma=0$ & 0.185 & 0.185 & 0.188 \\
 & NAMI-HTE & 0.5 & $\gamma=0.25$ & 0.182 & 0.182 & 0.182 \\
 & NAMI-HTE & 0.5 & $\gamma=0.50$ & 0.180 & 0.180 & 0.180 \\
\arrayrulecolor{gray!35}\midrule\arrayrulecolor{black}
survival & MI & 0 & $\gamma=0$ & 0.230 & 0.229 & 0.232 \\
 & MI & 0 & $\gamma=0.25$ & 0.230 & 0.229 & 0.234 \\
 & MI & 0 & $\gamma=0.50$ & 0.230 & 0.229 & 0.236 \\
 & MI & 0.5 & $\gamma=0$ & 0.236 & 0.235 & 0.240 \\
 & MI & 0.5 & $\gamma=0.25$ & 0.236 & 0.235 & 0.242 \\
 & MI & 0.5 & $\gamma=0.50$ & 0.236 & 0.235 & 0.241 \\
 & NAMI-HTE & 0 & $\gamma=0$ & 0.180 & 0.179 & 0.187 \\
 & NAMI-HTE & 0 & $\gamma=0.25$ & 0.175 & 0.175 & 0.182 \\
 & NAMI-HTE & 0 & $\gamma=0.50$ & 0.172 & 0.171 & 0.178 \\
 & NAMI-HTE & 0.5 & $\gamma=0$ & 0.186 & 0.186 & 0.194 \\
 & NAMI-HTE & 0.5 & $\gamma=0.25$ & 0.182 & 0.181 & 0.191 \\
 & NAMI-HTE & 0.5 & $\gamma=0.50$ & 0.178 & 0.178 & 0.186 \\
\end{longtable}
 }
}

%% --------------------------------------------------------------
\clearpage
\subsubsection*{Continuous outcome}

\begin{knitrout}
\definecolor{shadecolor}{rgb}{0.969, 0.969, 0.969}\color{fgcolor}\begin{figure}[H]

{\centering \includegraphics[width=0.95\linewidth]{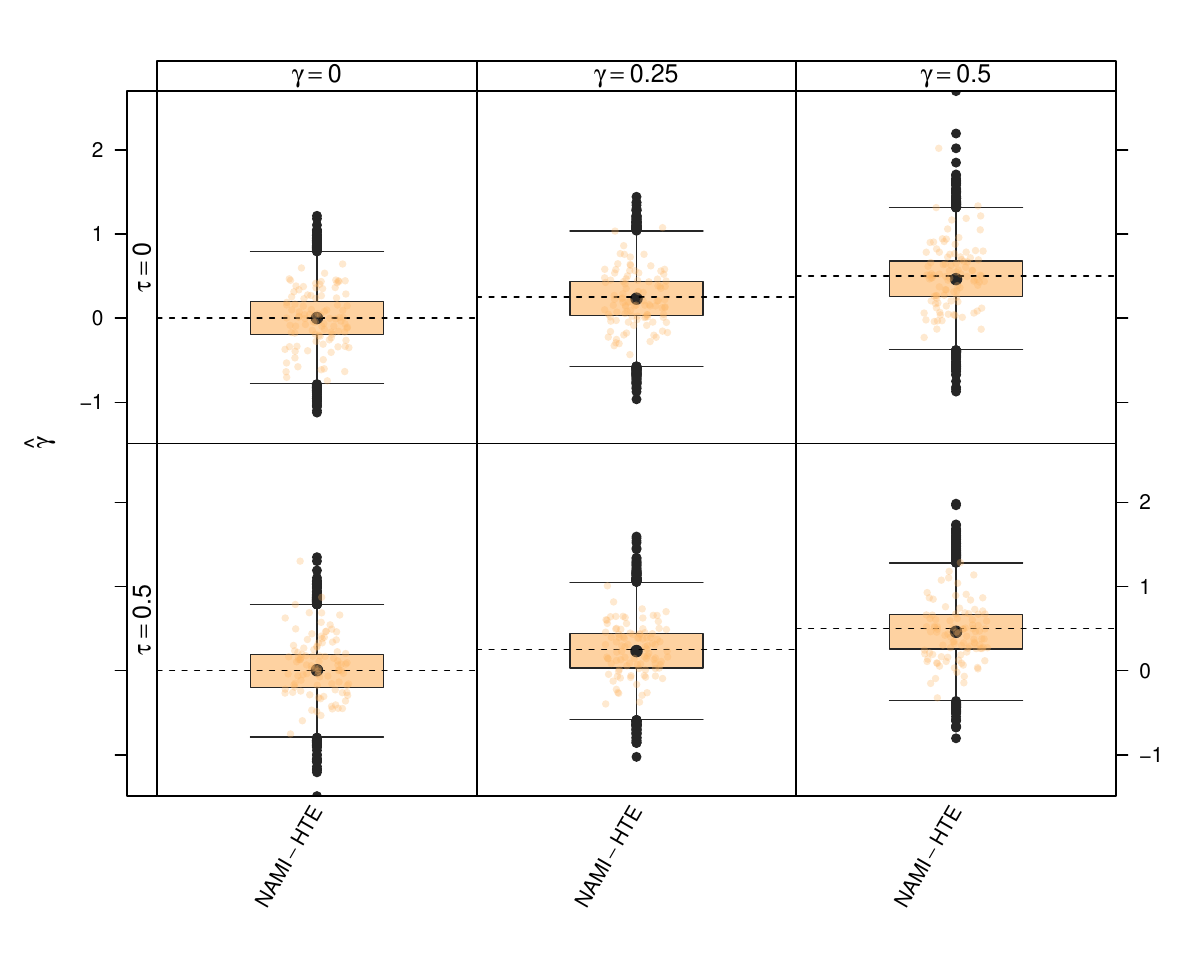} 

}

\caption[Estimated predictive effect \(\hat{\gamma}\) of \(X_1\) in simulations with a continuous normally distributed outcome]{Estimated predictive effect \(\hat{\gamma}\) of \(X_1\) in simulations with a continuous normally distributed outcome. Rows correspond to the true treatment effect \(\tau\) and columns to the true predictive effect \(\gamma\) of \(X_1\). Results are shown for the nonparanormal adjusted marginal inference model with heterogeneous treatment effects (NAMI-HTE). The dashed horizontal line indicates the true value of \(\gamma\).}\label{fig:fig-gammahat_continuous}
\end{figure}

\end{knitrout}

%% --------------------------------------------------------------
\clearpage
\subsubsection*{Binary outcome}

\vspace*{-5cm}
\begin{knitrout}
\definecolor{shadecolor}{rgb}{0.969, 0.969, 0.969}\color{fgcolor}\begin{figure}[H]

{\centering \includegraphics[width=0.95\linewidth]{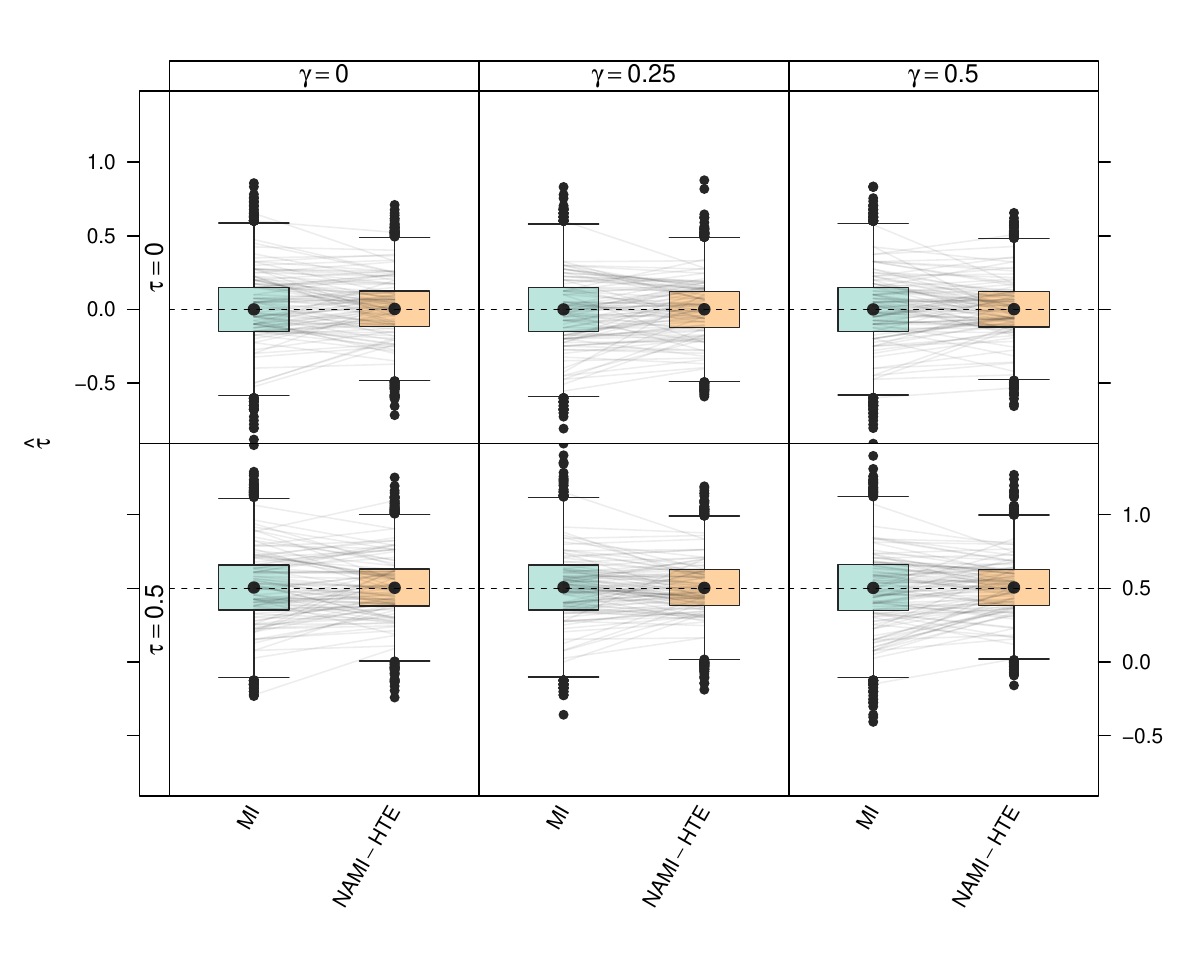} 

}

\caption[Estimated marginal treatment effect \(\hat{\tau}\) (log-odds ratio) in simulations with a binary outcome]{Estimated marginal treatment effect \(\hat{\tau}\) (log-odds ratio) in simulations with a binary outcome. Rows correspond to the true treatment effect \(\tau\) and columns to the true predictive effect \(\gamma\) of \(X_1\). Results are shown for unadjusted marginal inference (MI) and the nonparanormal adjusted marginal inference model with heterogeneous treatment effects (NAMI-HTE). The dashed horizontal line indicates the true value of \(\tau\).}\label{fig:fig-tauhat_binary}
\end{figure}

\end{knitrout}

\begin{knitrout}
\definecolor{shadecolor}{rgb}{0.969, 0.969, 0.969}\color{fgcolor}\begin{figure}[H]

{\centering \includegraphics[width=0.95\linewidth]{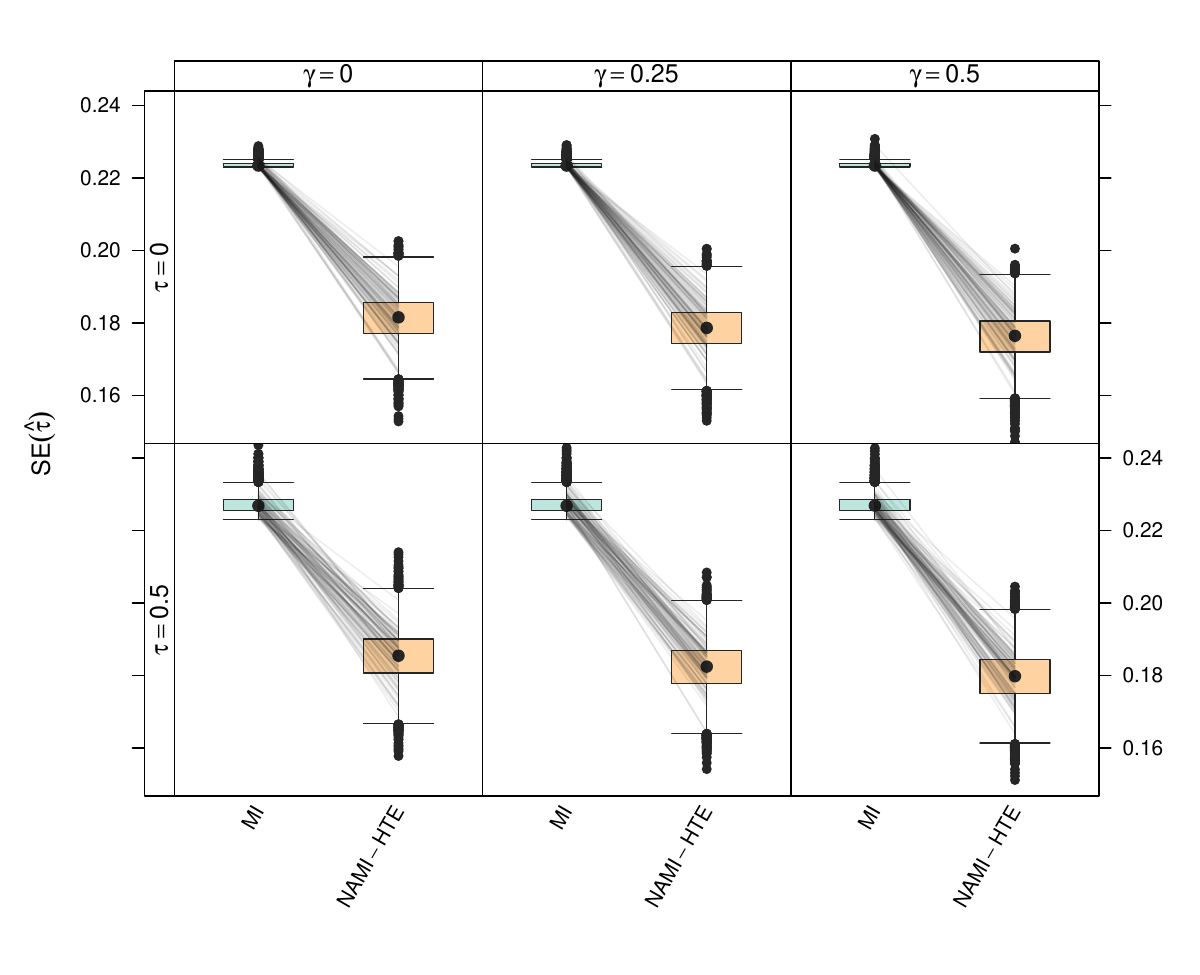} 

}

\caption[Estimated standard error \(\mathrm{SE}(\hat{\tau})\) of the marginal treatment effect \(\tau\) (log odds ratio) in simulations with a binary outcome]{Estimated standard error \(\mathrm{SE}(\hat{\tau})\) of the marginal treatment effect \(\tau\) (log odds ratio) in simulations with a binary outcome. Rows correspond to the true treatment effect \(\tau\) and columns to the true predictive effect \(\gamma\) of \(X_1\). Results are shown for unadjusted marginal inference (MI) and the nonparanormal adjusted marginal inference model with heterogeneous treatment effects (NAMI-HTE).}\label{fig:fig-SEtau_binary}
\end{figure}

\end{knitrout}

\begin{knitrout}
\definecolor{shadecolor}{rgb}{0.969, 0.969, 0.969}\color{fgcolor}\begin{figure}[H]

{\centering \includegraphics[width=0.95\linewidth]{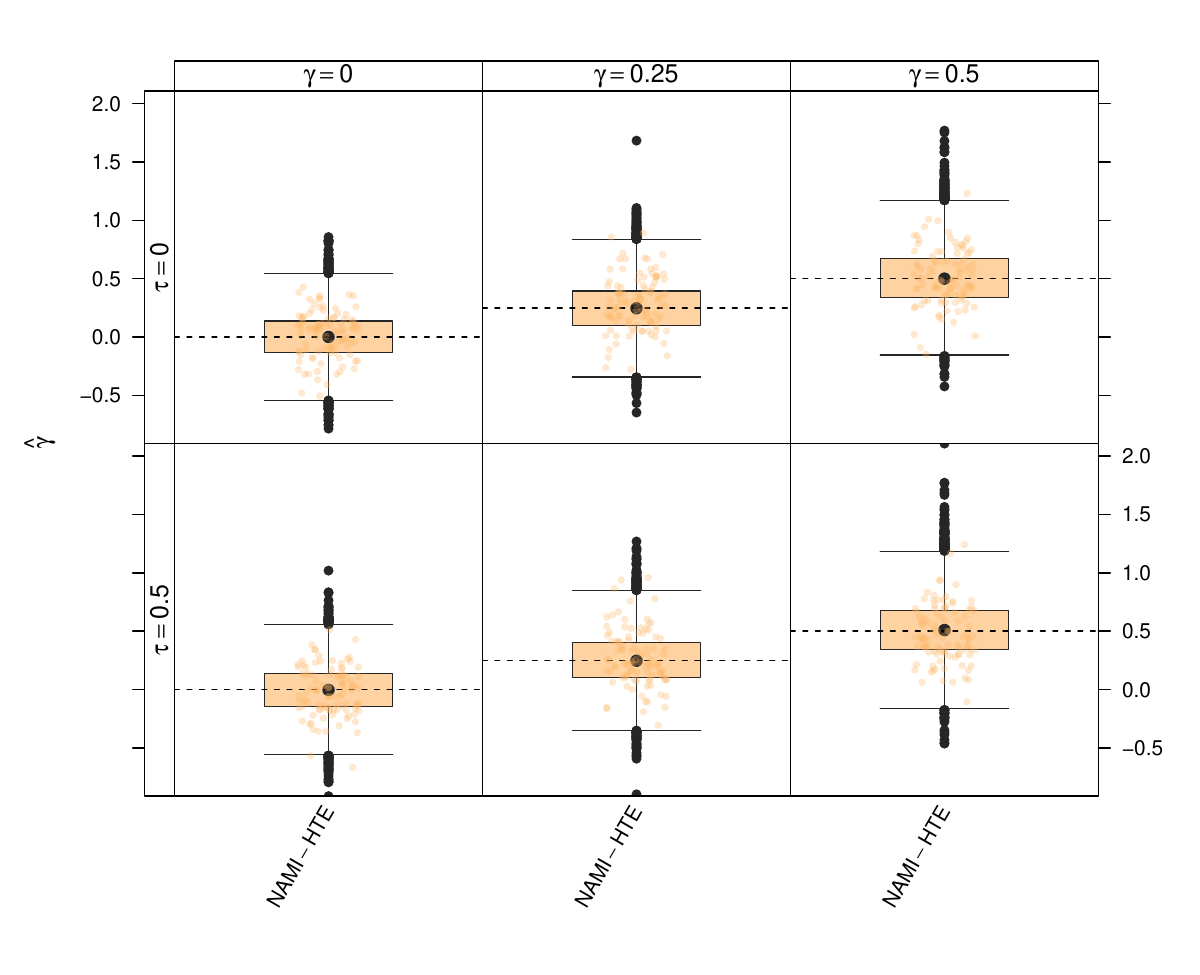} 

}

\caption[Estimated predictive effect \(\hat{\gamma}\) of \(X_1\) in simulations with a binary outcome]{Estimated predictive effect \(\hat{\gamma}\) of \(X_1\) in simulations with a binary outcome. Rows correspond to the true treatment effect \(\tau\) and columns to the true predictive effect \(\gamma\) of \(X_1\). Results are shown for the nonparanormal adjusted marginal inference model with heterogeneous treatment effects (NAMI-HTE). The dashed horizontal line indicates the true value of \(\gamma\).}\label{fig:fig-gammahat_binary}
\end{figure}

\end{knitrout}

%% --------------------------------------------------------------
\clearpage
\subsubsection*{Survival outcome}

\vspace*{-5cm}
\begin{knitrout}
\definecolor{shadecolor}{rgb}{0.969, 0.969, 0.969}\color{fgcolor}\begin{figure}[H]

{\centering \includegraphics[width=0.95\linewidth]{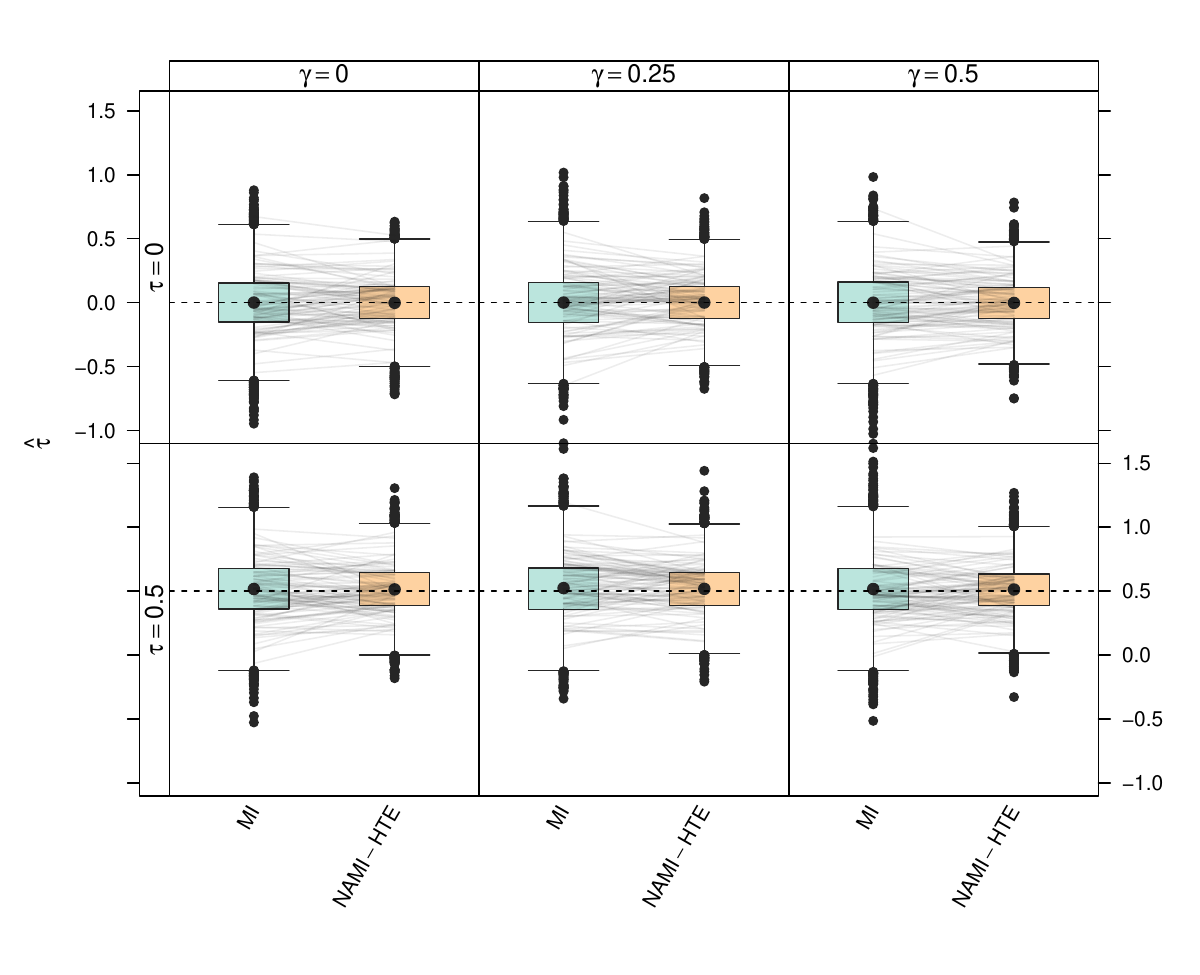} 

}

\caption[Estimated marginal treatment effect \(\hat{\tau}\) (log-hazard ratio) in simulations with a survival outcome]{Estimated marginal treatment effect \(\hat{\tau}\) (log-hazard ratio) in simulations with a survival outcome. Rows correspond to the true treatment effect \(\tau\) and columns to the true predictive effect \(\gamma\) of \(X_1\). Results are shown for unadjusted marginal inference (MI) and the nonparanormal adjusted marginal inference model with heterogeneous treatment effects (NAMI-HTE). The dashed horizontal line indicates the true value of \(\tau\).}\label{fig:fig-tauhat_survival}
\end{figure}

\end{knitrout}

\begin{knitrout}
\definecolor{shadecolor}{rgb}{0.969, 0.969, 0.969}\color{fgcolor}\begin{figure}[H]

{\centering \includegraphics[width=0.95\linewidth]{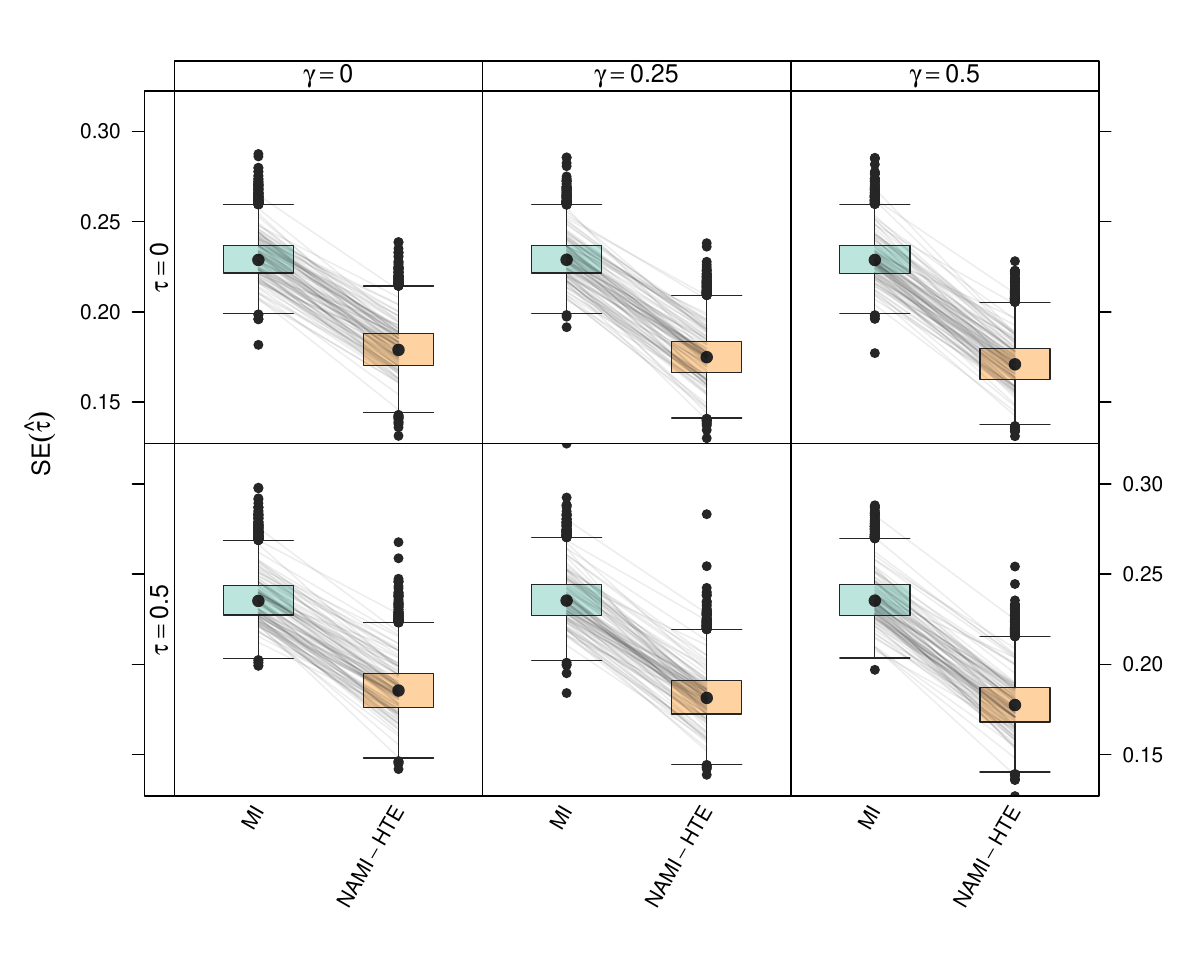} 

}

\caption[Estimated standard error \(\mathrm{SE}(\hat{\tau})\) of the marginal treatment effect \(\tau\) (log-hazard ratio) in simulations with a survival outcome]{Estimated standard error \(\mathrm{SE}(\hat{\tau})\) of the marginal treatment effect \(\tau\) (log-hazard ratio) in simulations with a survival outcome. Rows correspond to the true treatment effect \(\tau\) and columns to the true predictive effect \(\gamma\) of \(X_1\). Results are shown for unadjusted marginal inference (MI) and the nonparanormal adjusted marginal inference model with heterogeneous treatment effects (NAMI-HTE).}\label{fig:fig-SEtau_survival}
\end{figure}

\end{knitrout}

\begin{knitrout}
\definecolor{shadecolor}{rgb}{0.969, 0.969, 0.969}\color{fgcolor}\begin{figure}[H]

{\centering \includegraphics[width=0.95\linewidth]{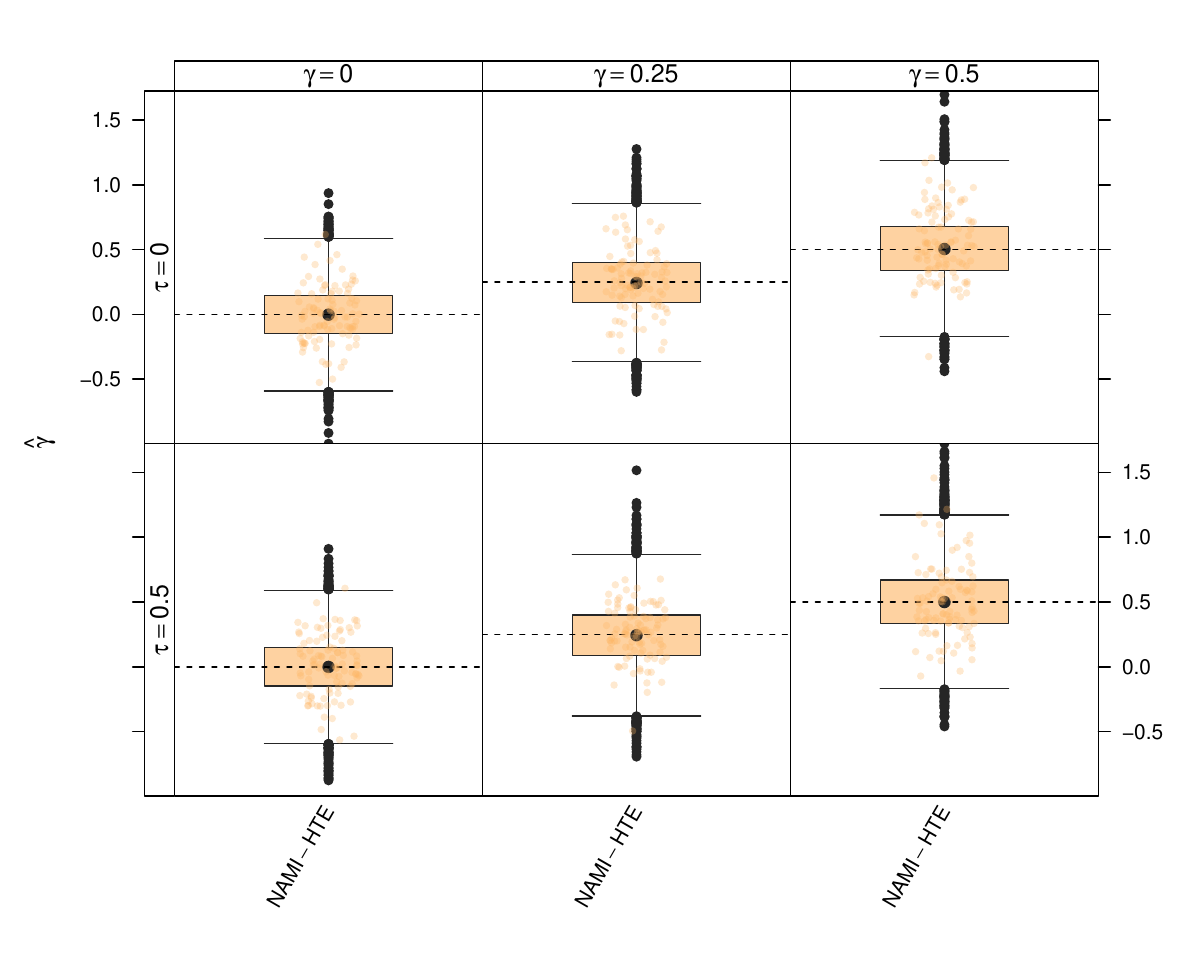} 

}

\caption[Estimated predictive effect \(\hat{\gamma}\) of \(X_1\) in simulations with a survival outcome]{Estimated predictive effect \(\hat{\gamma}\) of \(X_1\) in simulations with a survival outcome. Rows correspond to the true treatment effect \(\tau\) and columns to the true predictive effect \(\gamma\) of \(X_1\). Results are shown for the nonparanormal adjusted marginal inference model with heterogeneous treatment effects (NAMI-HTE). The dashed horizontal line indicates the true value of \(\gamma\).}\label{fig:fig-gammahat_survival}
\end{figure}

\end{knitrout}

%% --------------------------------------------------------------
\clearpage
\subsection*{Size and test statistic distribution for larger N}
\label{subsec:A_sim_qqplot_rate}
%% --------------------------------------------------------------

For the following results, we used the same simulation setup as for the normally distributed outcome, but set \(\tau = 0\) and increased the per-arm sample size to \(N = 500\).

\begin{table}[H]
\centering
\caption{Empirical size for testing $H_0: \tau = 0$ using a Wald test ($\alpha=0.05$) under true effect $\tau=0$ and per-arm sample size $N=500$. Results are shown across configurations of the true predictive effect $\gamma$ of $X_1$, only for continuous outcomes and the NAMI-HTE (nonparanormal adjusted marginal inference with heterogeneous treatment effects) model.}
\label{tbl-tau_sizeN}
\small
\renewcommand{\arraystretch}{1.15}
\setlength{\tabcolsep}{4.5pt}
\begin{tabular}{llccc}
\toprule
Outcome & Model & \multicolumn{3}{c}{Size} \\
\cmidrule(lr){3-5}
 &  & $\gamma=0$ & $\gamma=0.25$ & $\gamma=0.50$ \\
\midrule
continuous & NAMI-HTE & 0.051 & 0.049 & 0.049 \\
\bottomrule
\end{tabular}
\end{table}

\begin{knitrout}
\definecolor{shadecolor}{rgb}{0.969, 0.969, 0.969}\color{fgcolor}\begin{figure}[H]

\includegraphics[width=1\linewidth]{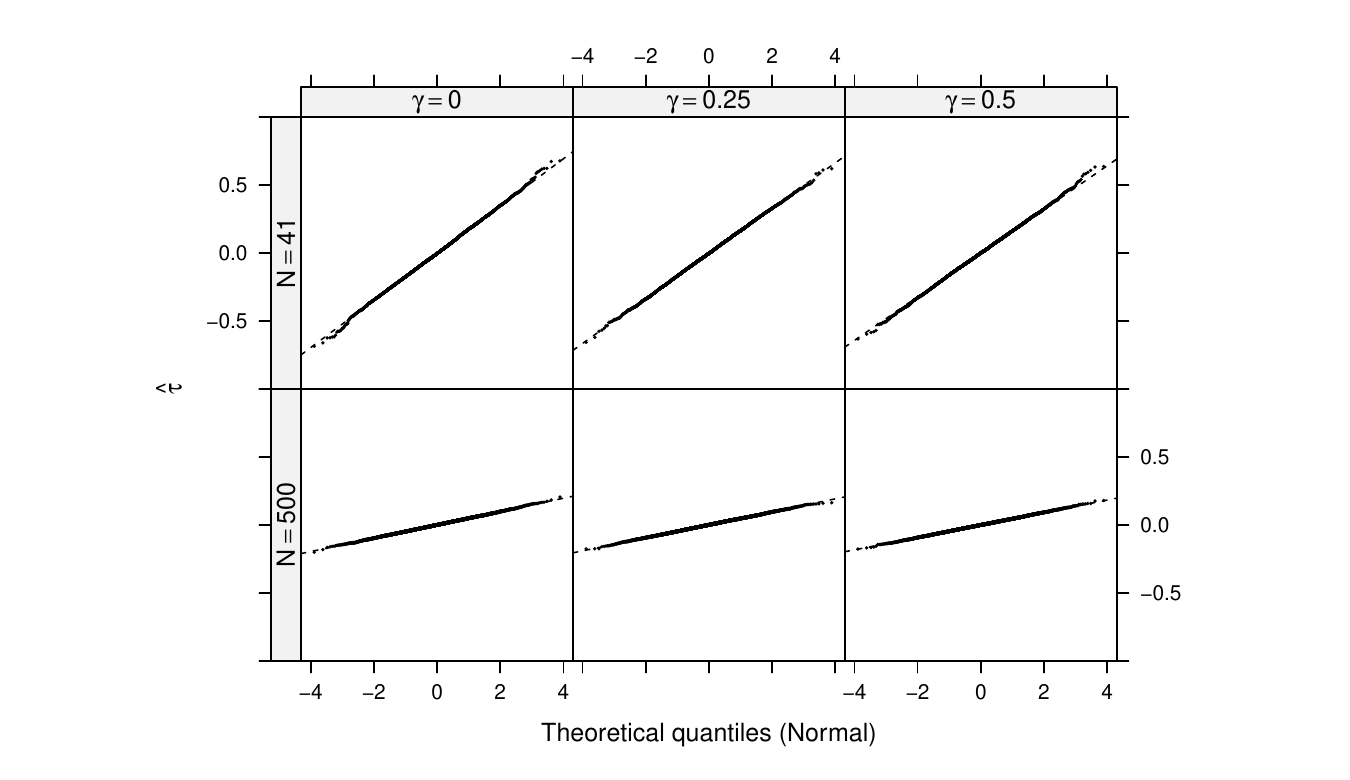} \hfill{}

\caption[QQ-plot comparing quantiles of the estimated unstandardized marginal treatment effect \(\hat{\tau}\) (Cohen's \(d\)) in simulations with a continuous normally distributed outcome under true treatment effect \(\tau = 0\)]{QQ-plot comparing quantiles of the estimated unstandardized marginal treatment effect \(\hat{\tau}\) (Cohen's \(d\)) in simulations with a continuous normally distributed outcome under true treatment effect \(\tau = 0\). Rows correspond to the sample size per arm \(N\) and columns to the true predictive effect \(\gamma\) of \(X_1\).}\label{fig:fig-tstatQQ_continuous2}
\end{figure}

\end{knitrout}

%%%%%%%%%%%%%%%%%%%%%%%%%%%%%%%%%%%%%%%%%%%%%%%%
\clearpage
\section{Consistency between theory and software}
\label{sec:A_sim_consistency}
%%%%%%%%%%%%%%%%%%%%%%%%%%%%%%%%%%%%%%%%%%%%%%%%

In the case of a single normally distributed covariate and a normally distributed outcome, closed-form expressions for the standard errors of the marginal treatment effect \(\tau\), the prognostic effect \(\lambda\), and the predictive effect \(\gamma\) can be derived (see diagonal elements of Equation~\eqref{eq:var_matrix} of Section~\ref{sec:A_seproofs}). The goal of this second simulation study is to validate the implementation of NAMI-HTE in the \texttt{tram} package by comparing these theoretical standard errors with those estimated from the software.

%% --------------------------------------------------------------
\subsection*{Data-generating process and models}
%% --------------------------------------------------------------

Data was generated according to the conditional distribution

\[
\Prob(Y \le y \mid W = w, X = x)
=
\Phi\!\left[
\omega_{21}^{(w)} h_1(x)
+ \omega_{22}^{(w)}\bigl(\eparm_1 + \eparm_2 y - \tau w \bigr)
\right],
\]

with both the covariate \(X\) and the outcome \(Y\) being standard normally distributed. As before, we set \(\eparm_1 = 0\) and \(\eparm_2 = 1\), and the treatment indicator \(W\) followed a Bernoulli distribution with \(W \sim B(1,0.5)\). Each simulation scenario was replicated 10,000 times and only the NAMI-HTE model was fitted. Both the outcome and covariate distributions were fitted using linear transformation models with \texttt{tram::Lm} as in Equation~\eqref{eq:normal-tm} and used as input to the joint model fitted with \texttt{tram::Mmlt}.

The true marginal treatment effect was set to \(\tau \in \{0, 0.5, 1\}\), the prognostic effect to \(\lambda \in \{0, 0.25, 0.5\}\), and the predictive effect to \(\gamma \in \{0, 0.25, 0.5\}\). A sample size of \(N=500\) per arm was used to assess the accuracy of the point estimates and standard errors for \(\tau\), \(\lambda\) and \(\gamma\).

%% --------------------------------------------------------------
\clearpage
\subsection*{Simulation results}
%% --------------------------------------------------------------

\vspace*{-5cm}
\begin{knitrout}
\definecolor{shadecolor}{rgb}{0.969, 0.969, 0.969}\color{fgcolor}\begin{figure}[H]

{\centering \includegraphics[width=0.95\linewidth]{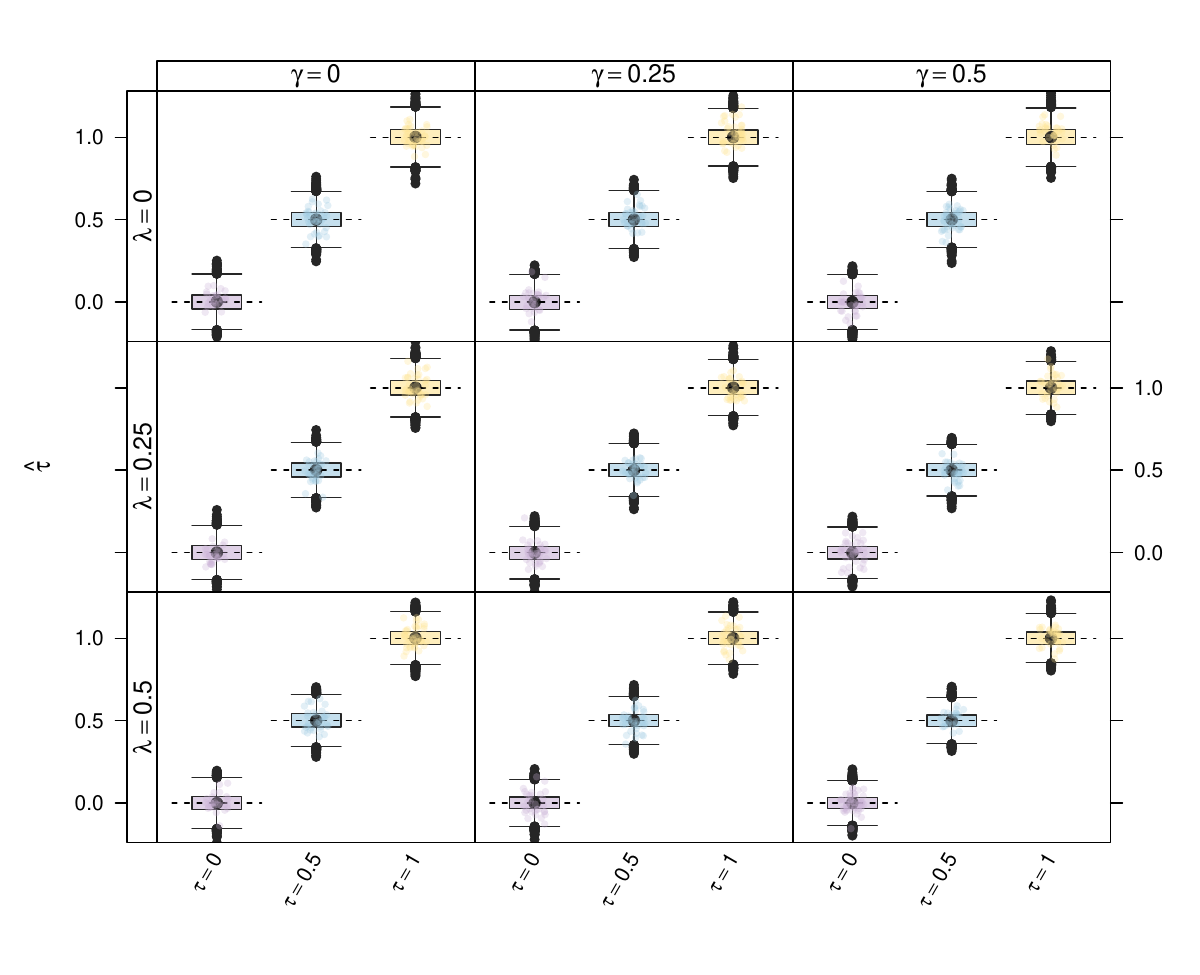} 

}

\caption[Estimated marginal treatment effect \(\hat{\tau}\) (Cohen's \(d\)) in simulations with a normally distributed outcome and a single normally distributed covariate]{Estimated marginal treatment effect \(\hat{\tau}\) (Cohen's \(d\)) in simulations with a normally distributed outcome and a single normally distributed covariate. Rows correspond to the true prognostic effect \(\lambda\) of \(X\), columns to the true predictive effect \(\gamma\) of \(X\), and the horizontal axis to the true treatment effect \(\tau\). Results are shown for the nonparanormal adjusted marginal inference model with heterogeneous treatment effects (NAMI-HTE). The dashed horizontal line indicates the true value of \(\tau\).}\label{fig:fig-tauhat_continuous_1P}
\end{figure}

\end{knitrout}

\begin{knitrout}
\definecolor{shadecolor}{rgb}{0.969, 0.969, 0.969}\color{fgcolor}\begin{figure}[H]

{\centering \includegraphics[width=0.95\linewidth]{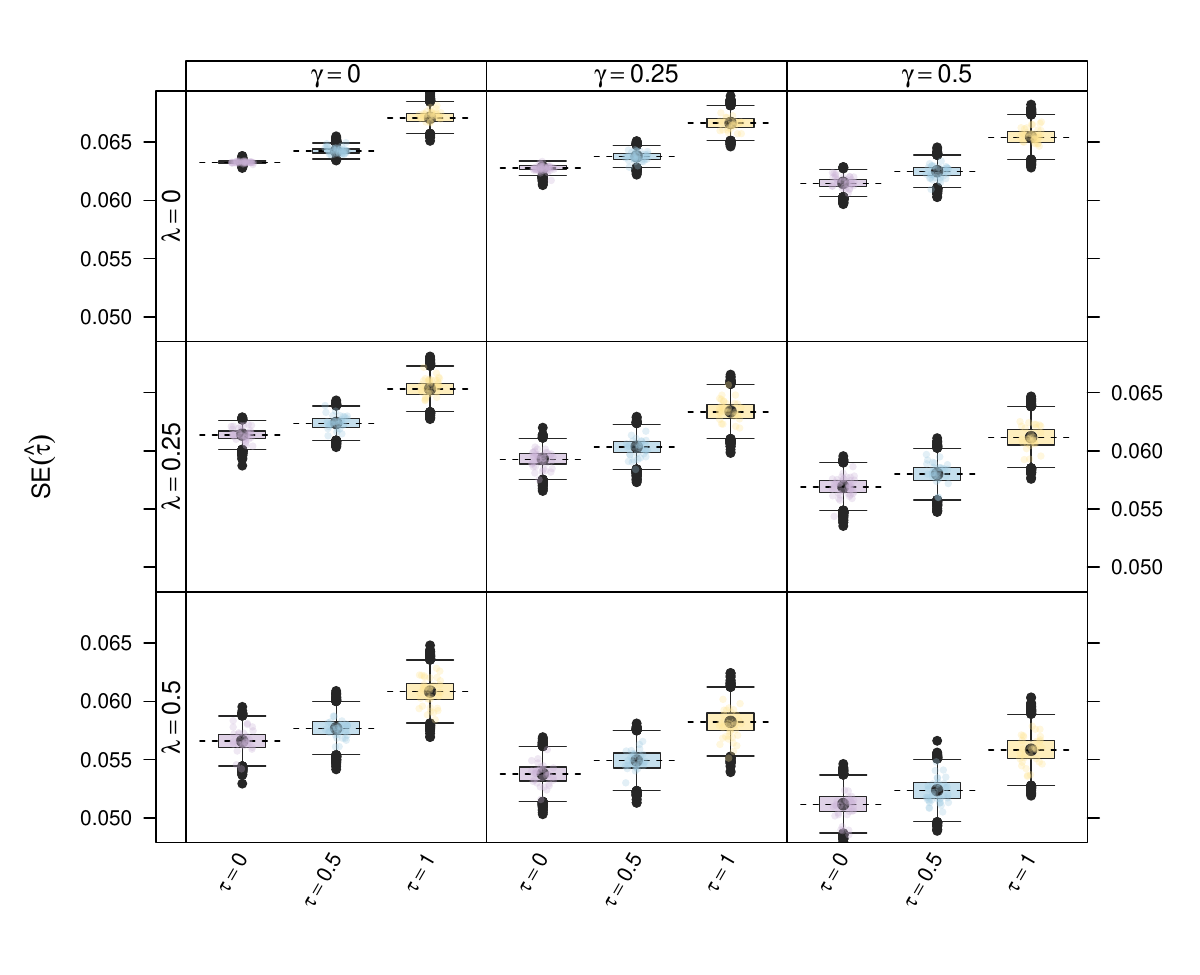} 

}

\caption{Estimated standard error \(\mathrm{SE}(\hat{\tau})\) of the marginal treatment effect \(\tau\) (Cohen's \(d\)) in simulations with a normally distributed outcome and a single normally distributed covariate. Rows correspond to the true prognostic effect \(\lambda\) of \(X\), columns to the true predictive effect \(\gamma\) of \(X\), and the horizontal axis to the true treatment effect \(\tau\). Results are shown for the nonparanormal adjusted marginal inference model with heterogeneous treatment effects (NAMI-HTE). The dashed horizontal line indicates the true theoretical value of \(\mathrm{SE}(\tau)\).}\label{fig:fig-SEtau_continuous_1P}
\end{figure}

\end{knitrout}

\begin{knitrout}
\definecolor{shadecolor}{rgb}{0.969, 0.969, 0.969}\color{fgcolor}\begin{figure}[H]

{\centering \includegraphics[width=0.95\linewidth]{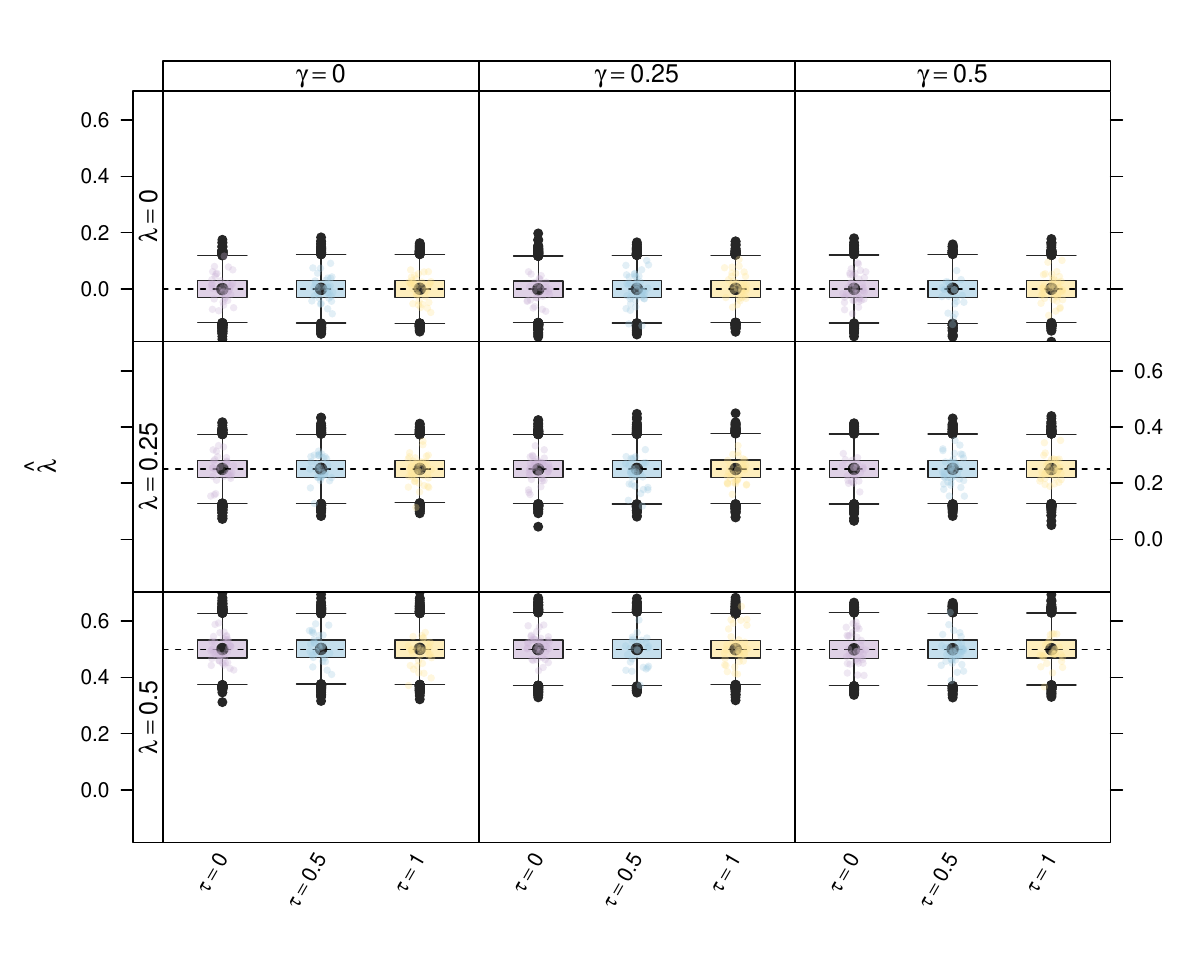} 

}

\caption[Estimated prognostic effect \(\hat{\lambda}\) of \(X\) in simulations with a normally distributed outcome and a single normally distributed covariate]{Estimated prognostic effect \(\hat{\lambda}\) of \(X\) in simulations with a normally distributed outcome and a single normally distributed covariate. Rows correspond to the true prognostic effect \(\lambda\) of \(X\), columns to the true predictive effect \(\gamma\) of \(X\), and the horizontal axis to the true treatment effect \(\tau\). Results are shown for the nonparanormal adjusted marginal inference model with heterogeneous treatment effects (NAMI-HTE). The dashed horizontal line indicates the true value of \(\lambda\).}\label{fig:fig-lambdahat_continuous_1P}
\end{figure}

\end{knitrout}

\begin{knitrout}
\definecolor{shadecolor}{rgb}{0.969, 0.969, 0.969}\color{fgcolor}\begin{figure}[H]

{\centering \includegraphics[width=0.95\linewidth]{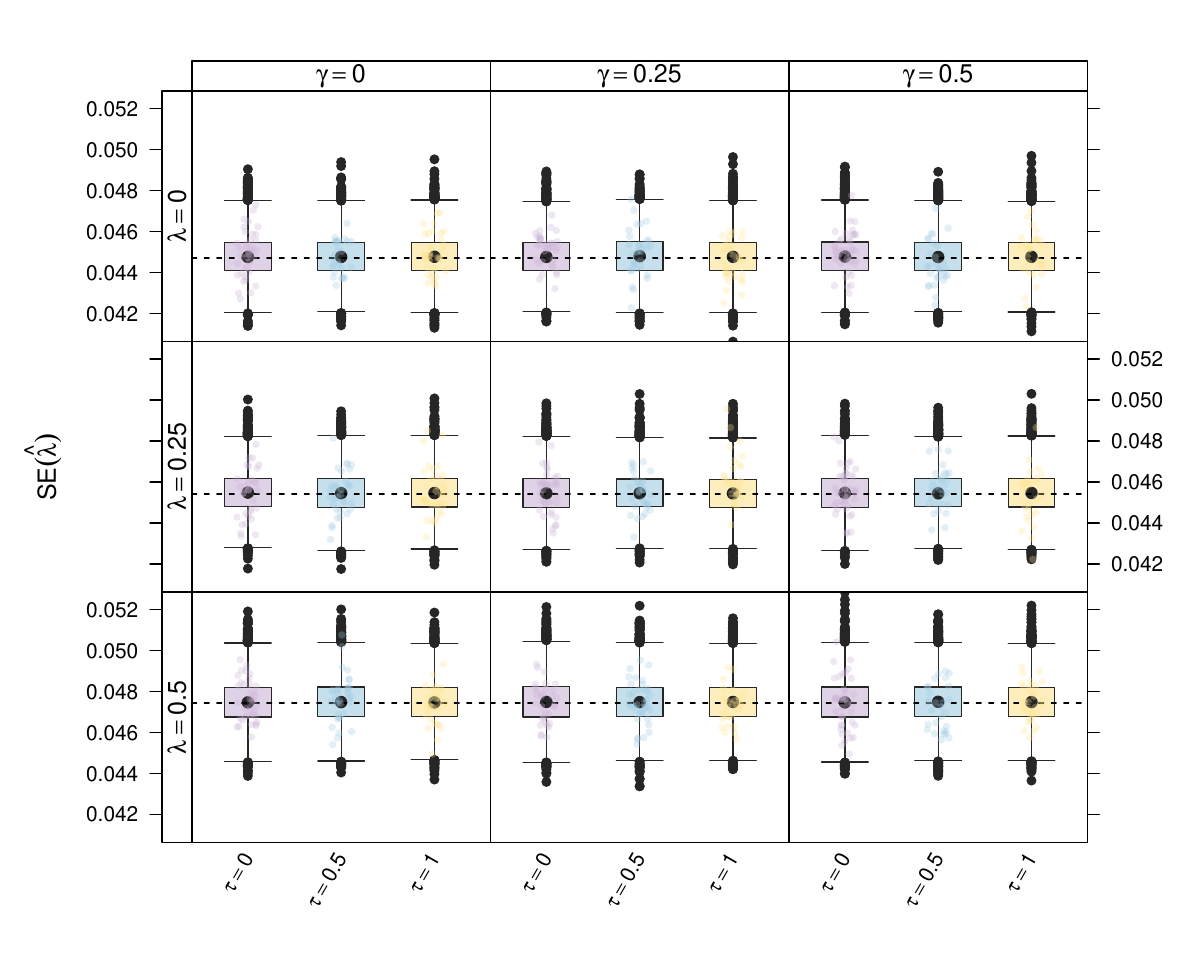} 

}

\caption{Estimated standard error \(\mathrm{SE}(\hat{\lambda})\) of the prognostic effect \(\lambda\) of \(X\) in simulations with a normally distributed outcome and a single normally distributed covariate. Rows correspond to the true prognostic effect \(\lambda\) of \(X\), columns to the true predictive effect \(\gamma\) of \(X\), and the horizontal axis to the true treatment effect \(\tau\). Results are shown for the nonparanormal adjusted marginal inference model with heterogeneous treatment effects (NAMI-HTE). The dashed horizontal line indicates the true theoretical value of \(\mathrm{SE}(\lambda)\).}\label{fig:fig-SElambda_continuous_1P}
\end{figure}

\end{knitrout}

\begin{knitrout}
\definecolor{shadecolor}{rgb}{0.969, 0.969, 0.969}\color{fgcolor}\begin{figure}[H]

{\centering \includegraphics[width=0.95\linewidth]{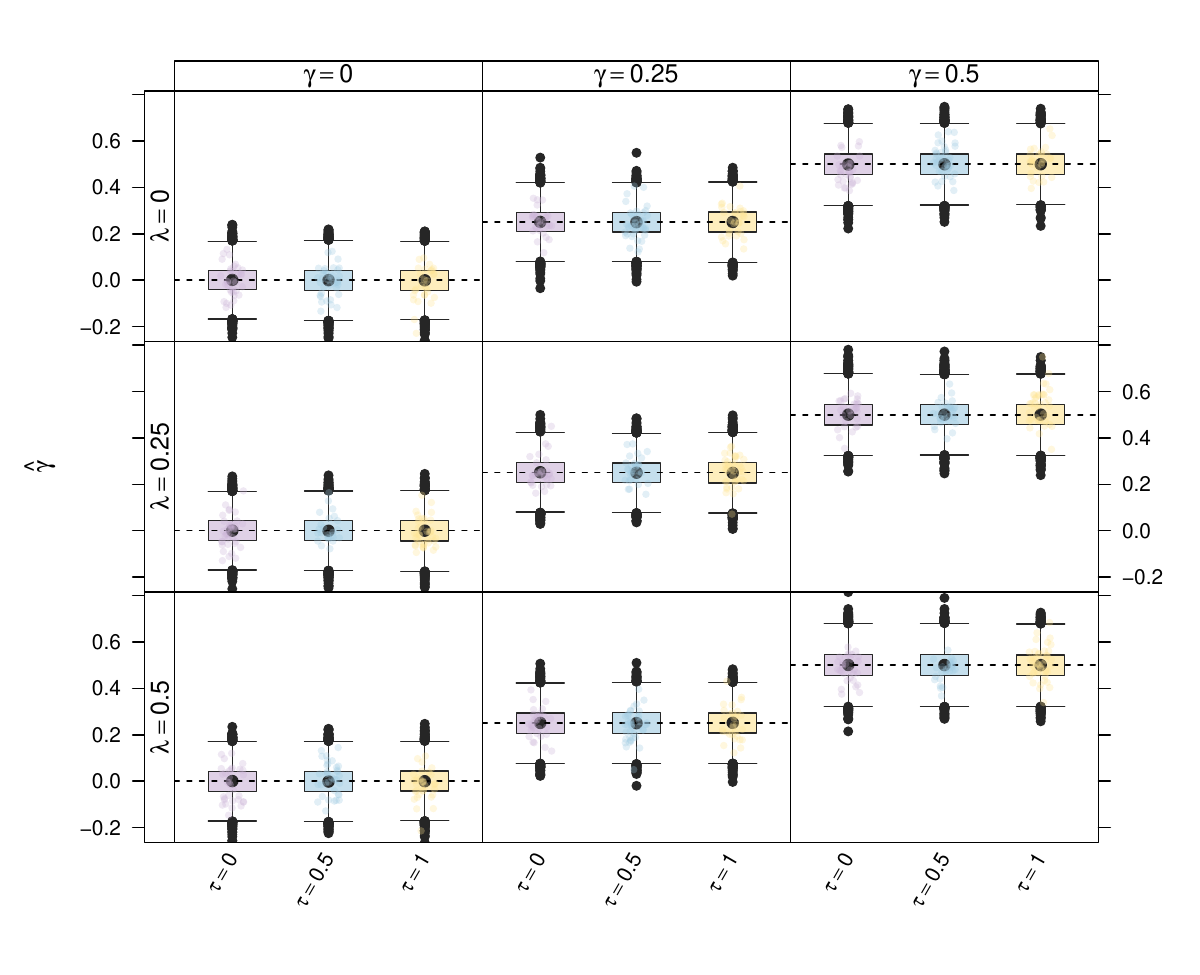} 

}

\caption[Estimated predictive effect \(\hat{\gamma}\) of \(X\) in simulations with a normally distributed outcome and a single normally distributed covariate]{Estimated predictive effect \(\hat{\gamma}\) of \(X\) in simulations with a normally distributed outcome and a single normally distributed covariate. Rows correspond to the true prognostic effect \(\lambda\) of \(X\), columns to the true predictive effect \(\gamma\) of \(X\), and the horizontal axis to the true treatment effect \(\tau\). Results are shown for the nonparanormal adjusted marginal inference model with heterogeneous treatment effects (NAMI-HTE). The dashed horizontal line indicates the true value of \(\gamma\).}\label{fig:fig-gammahat_continuous_1P}
\end{figure}

\end{knitrout}

\begin{knitrout}
\definecolor{shadecolor}{rgb}{0.969, 0.969, 0.969}\color{fgcolor}\begin{figure}[H]

{\centering \includegraphics[width=0.95\linewidth]{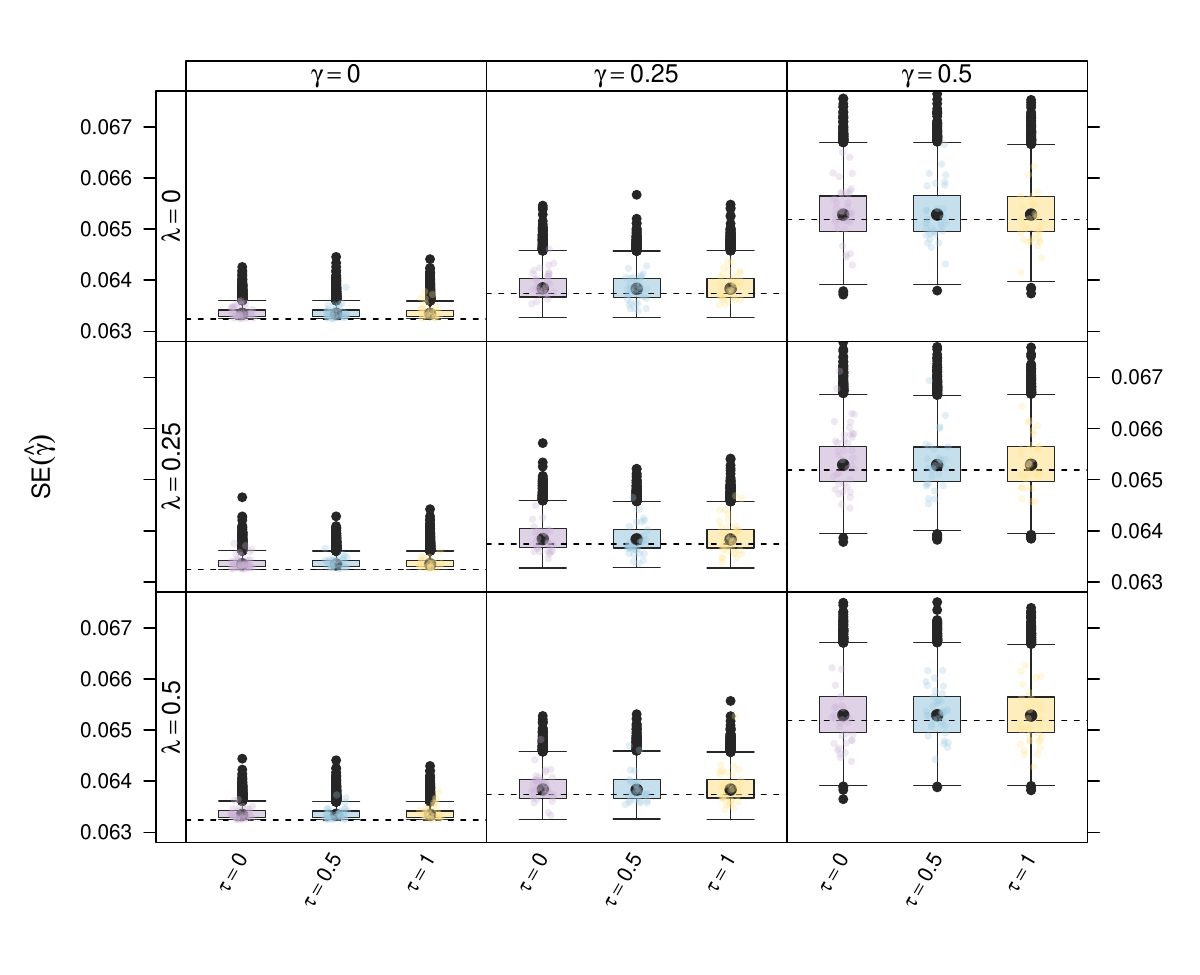} 

}

\caption{Estimated standard error \(\mathrm{SE}(\hat{\gamma})\) of the predictive effect \(\gamma\) of \(X\) in simulations with a normally distributed outcome and a single normally distributed covariate. Rows correspond to the true prognostic effect \(\lambda\) of \(X\), columns to the true predictive effect \(\gamma\) of \(X\), and the horizontal axis to the true treatment effect \(\tau\). Results are shown for the nonparanormal adjusted marginal inference model with heterogeneous treatment effects (NAMI-HTE). The dashed horizontal line indicates the true theoretical value of \(\mathrm{SE}(\gamma)\).}\label{fig:fig-SEgamma_continuous_1P}
\end{figure}

\end{knitrout}

Overall, the estimated marginal treatment effect \(\hat{\tau}\), its standard error \(\mathrm{SE}(\hat{\tau})\), the estimated prognostic effect \(\hat{\lambda}\), its standard error \(\mathrm{SE}(\hat{\lambda})\), and the estimated predictive effect \(\hat{\gamma}\) were centered around their true values and approximately symmetrically distributed. The estimated standard error \(\mathrm{SE}(\hat{\gamma})\) of the predictive effect \(\gamma\) showed slight overestimation for small values of \(\gamma\), although the deviation was minor.

%% --------------------------------------------------------------
%%\subsection*{Power curves of tau and gamma}
%% --------------------------------------------------------------

%% FOR NOW: DON'T SHOW POWER CURVES OF TAU AND GAMMA
%In this second setup, the true marginal treatment effect set to \(\tau = taus2[[1]]\), the prognostic effect to \(\lambda \in \{lambdas2[[1]][1], lambdas2[[1]][2], lambdas2[[1]][3]\}\), and the predictive effect to \(\gamma \in \{gammas2[[1]][1], gammas2[[1]][2], gammas2[[1]][3], gammas2[[1]][4], gammas2[[1]][5], gammas2[[1]][6]\}\). A sample size of \(N=Ns[[1]]\) per arm was used to assess the power of detecting \(\tau\) and \(\gamma\). 

%%%%%%%%%%%%%%%%%%%%%%%%%%%%%%%%%%%%%%%%%%%%%%%%
\section{Acupuncture application}
%%%%%%%%%%%%%%%%%%%%%%%%%%%%%%%%%%%%%%%%%%%%%%%%

%% --------------------------------------------------------------
\subsection*{Model specifications}
\label{subsec:A_app_models}
%% --------------------------------------------------------------

All models in Section~\ref{sec:application} were fitted using maximum likelihood under the nonparanormal framework, implemented via the \texttt{Mmlt} function of the \texttt{tram} package \citep{pkg:tram}. Inference for the prognostic (\(\lambda\)) and predictive (\(\gamma\)) copula parameters was based on Wald tests with multiplicity adjustment using \texttt{glht} from the \texttt{multcomp} package \citep{pkg:multcomp}.

For models \texttt{m1}--\texttt{m4}, which assume a continuous, normally distributed outcome, the marginal model of \(Y\mid W\) was fitted using \texttt{tram::Lm}, corresponding to a linear transformation model as in~\eqref{eq:normal-tm}. Model \texttt{m5} relaxes the normality assumption by using \texttt{tram::BoxCox} for the marginal model of \(Y\mid W\), corresponding to \(F_w(y) \;=\; \Phi\!\bigl(h(y) - \tau w\bigr)\) where \(h(y)\) is a flexible Bernstein polynomial of order six. 

Model \texttt{m6} treats the outcome as ordinal; the weekly average headache score (both at baseline and follow-up) was first divided by 7 to obtain a daily average and then categorized into six ordered levels: \([0,1)\), \([1,2)\), \([2,3)\), \([3,4)\), \([4,5)\), and \([5,\infty)\), thereby aligning the outcome more closely with the original 6-point Likert scale reported by patients. The marginal model of \(Y\mid W\) was then fitted using \texttt{tram::Polr} with a probit link, corresponding to \(F_w(y_k) \;=\; \Phi\!\bigl(\vartheta_k - \tau w\bigr)\) where the transformation function \(h(y_k) = \vartheta_k\) is a step function with jumps at the cutpoints \(\vartheta_k,\; k = 1, ..., 5\) that represent the first five ordered categories of the outcome. This model specification is equivalent to an ordinal probit regression model.

Marginal models for the baseline covariates were also fitted using transformation models. Continuous covariates (pk1 for \texttt{m1}--\texttt{m5}, age, chronicity) were modeled with \texttt{tram::BoxCox}, corresponding to \(F(x) \;=\; \Phi\!\bigl(h(x)\bigr)\) where \(h(x)\) is a flexible Bernstein polynomial of order six. Discrete covariates (pk1 for \texttt{m6}, migraine, sex) were modeled with \texttt{tram::Polr} using a probit link, corresponding to \(F(x_k) \;=\; \Phi\!\bigl(\vartheta_k\bigr)\) where \(h(y_k) = \vartheta_k\) is a step function.

Models \texttt{m1}--\texttt{m3} were fitted on complete cases, including only participants with observed 12-month headache scores. Models \texttt{m4}--\texttt{m6} were fitted on the full dataset, therefore incorporating all randomized participants, using the \texttt{na.action = na.pass} argument in the marginal models of \(Y\mid W\) to handle missing outcome values.

%% --------------------------------------------------------------
\subsection*{Estimates of prognostic and predictive effects}
\label{subsec:A_app_results}
%% --------------------------------------------------------------

The treatment-dependent correlation matrix is parameterized through the unconstrained matrix \(\mLambda(w)\), with \(w=0\) denoting control and \(w=1\) denoting acupuncture. Table~\ref{tab:lambda_gamma_results} presents the estimated prognostic and predictive parameters from the last row of \(\mLambda(w)\), including standard errors and multiplicity-adjusted \(p\)-values. Here, \(\mLambda(0)\) contains the prognostic parameters \(\lambda_j\) for each covariate \(X_j\), while the difference between \(\mLambda(1)\) and \(\mLambda(0)\) gives the predictive parameters \(\gamma_j\). These are the parameters used for hypothesis testing, with \(H_0: \lambda_j=0\) testing whether \(X_j\) is prognostic and \(H_0: \gamma_j=0\) testing whether it is predictive. The covariate rankings reported in Table~\ref{tab:rankings} are instead based on the corresponding entries of the inverse Cholesky factor

\[
\mOmega(w)=
  \mLambda(w)\left(\operatorname{diag}\left\{\mLambda(w)^{-1}\mLambda(w)^{-\top}\right\}\right)^{1/2},
\]

using the absolute values for prognostic importance and the absolute differences between treatment groups for predictive importance.

{\small
\renewcommand{\arraystretch}{1.15}
\setlength{\tabcolsep}{4.5pt}
\begin{longtable}{ll|ccl|ccl}
\caption{\normalsize Estimated prognostic effects $\hat{\lambda}$ and predictive effects $\hat{\gamma}$ across models, with standard errors and multiplicity-adjusted $p$-values. Entries marked with ``--'' indicate that the corresponding effect is not included or not available for that model. Covariates are abbreviated as follows: pk1 = baseline headache score, chr = headache chronicity, and mig = headache diagnosis (migraine or tension-type).}
\label{tab:lambda_gamma_results}\\
\toprule
Covariate & Model & $\hat{\lambda}$ & SE($\hat{\lambda}$) & $p$-value & $\hat{\gamma}$ & SE($\hat{\gamma}$) & $p$-value \\
\midrule
\endfirsthead
\multicolumn{8}{l}{\small\itshape Table~\thetable\ (continued)} \\
\toprule
Covariate & Model & $\hat{\lambda}$ & SE($\hat{\lambda}$) & $p$-value & $\hat{\gamma}$ & SE($\hat{\gamma}$) & $p$-value \\
\midrule
\endhead
\midrule
\multicolumn{8}{r}{\small\itshape Continued on next page} \\
\endfoot
\bottomrule
\endlastfoot
\arrayrulecolor{gray!35}
pk1 & \texttt{m2} & -0.91 & 0.09 & < 0.0001 & -- & -- & -- \\
\hline
 & \texttt{m3} & -1.10 & 0.12 & < 0.0001 & 0.33 & 0.15 & 0.11 \\
\hline
 & \texttt{m4} & -1.07 & 0.12 & < 0.0001 & 0.30 & 0.15 & 0.17 \\
\hline
 & \texttt{m5} & -0.99 & 0.11 & < 0.0001 & 0.39 & 0.14 & 0.028 \\
\hline
 & \texttt{m6} & -1.05 & 0.15 & < 0.0001 & 0.39 & 0.19 & 0.16 \\
\arrayrulecolor{black}
\midrule
\arrayrulecolor{gray!35}
age & \texttt{m2} & -0.09 & 0.07 & 0.64 & -- & -- & -- \\
\hline
 & \texttt{m3} & -0.12 & 0.10 & 0.69 & 0.03 & 0.16 & 1.00 \\
\hline
 & \texttt{m4} & -0.11 & 0.10 & 0.76 & -0.01 & 0.16 & 1.00 \\
\hline
 & \texttt{m5} & -0.05 & 0.11 & 0.99 & -0.01 & 0.15 & 1.00 \\
\hline
 & \texttt{m6} & -0.07 & 0.12 & 0.98 & -0.03 & 0.17 & 1.00 \\
\arrayrulecolor{black}
\midrule
\arrayrulecolor{gray!35}
sex & \texttt{m2} & 0.05 & 0.10 & 0.98 & -- & -- & -- \\
\hline
 & \texttt{m3} & 0.19 & 0.14 & 0.54 & -0.29 & 0.21 & 0.53 \\
\hline
 & \texttt{m4} & 0.19 & 0.12 & 0.44 & -0.25 & 0.19 & 0.58 \\
\hline
 & \texttt{m5} & 0.11 & 0.13 & 0.88 & -0.20 & 0.18 & 0.77 \\
\hline
 & \texttt{m6} & 0.22 & 0.16 & 0.51 & -0.36 & 0.21 & 0.34 \\
\arrayrulecolor{black}
\midrule
\arrayrulecolor{gray!35}
mig & \texttt{m2} & 0.16 & 0.13 & 0.70 & -- & -- & -- \\
\hline
 & \texttt{m3} & 0.04 & 0.21 & 1.00 & 0.26 & 0.28 & 0.84 \\
\hline
 & \texttt{m4} & 0.05 & 0.21 & 1.00 & 0.25 & 0.28 & 0.86 \\
\hline
 & \texttt{m5} & 0.12 & 0.22 & 0.98 & 0.16 & 0.29 & 0.98 \\
\hline
 & \texttt{m6} & 0.07 & 0.24 & 1.00 & 0.31 & 0.33 & 0.85 \\
\arrayrulecolor{black}
\midrule
\arrayrulecolor{gray!35}
chr & \texttt{m2} & -0.09 & 0.07 & 0.66 & -- & -- & -- \\
\hline
 & \texttt{m3} & -0.04 & 0.11 & 1.00 & -0.09 & 0.14 & 0.97 \\
\hline
 & \texttt{m4} & -0.04 & 0.10 & 0.99 & -0.07 & 0.14 & 0.99 \\
\hline
 & \texttt{m5} & -0.11 & 0.11 & 0.79 & -0.04 & 0.14 & 1.00 \\
\hline
 & \texttt{m6} & -0.10 & 0.12 & 0.90 & -0.07 & 0.15 & 0.99 \\
\arrayrulecolor{black}
\end{longtable}
}

%% --------------------------------------------------------------
\subsection*{Joint and conditional densities for model \texttt{m5}}
\label{subsec:A_app_cf}
%% --------------------------------------------------------------

Figure~\ref{fig:fig-app_condensity} in the main text shows the predicted conditional density of the follow-up headache score \(Y_{12}\) given the baseline headache score \(Y_1\) and treatment group \(W\). This corresponds to the model-based estimate of \(f_{Y_{12} \mid Y_1, W}(y_{12} \mid y_1, w)\), derived from the joint model \texttt{m5} by first marginalizing over the other baseline covariates and then conditioning on \(Y_1\). In contrast, Figure~\ref{fig:fig-app_jointdensity} displays the corresponding joint density \(f_{Y_1, Y_{12} \mid W}(y_1, y_{12} \mid w)\), obtained from the same model after marginalizing over the other baseline covariates. A positive association between baseline and follow-up headache scores is visible in both treatment arms, indicating that patients with higher baseline scores tend to have higher follow-up scores. The association is weaker in the acupuncture group. From the shape of the conditional density, this reduction appears to be driven primarily by patients with higher baseline scores having lower follow-up scores (better improvement) under acupuncture treatment.

\begin{knitrout}
\definecolor{shadecolor}{rgb}{0.969, 0.969, 0.969}\color{fgcolor}\begin{figure}[H]

{\centering \includegraphics[width=0.9\linewidth]{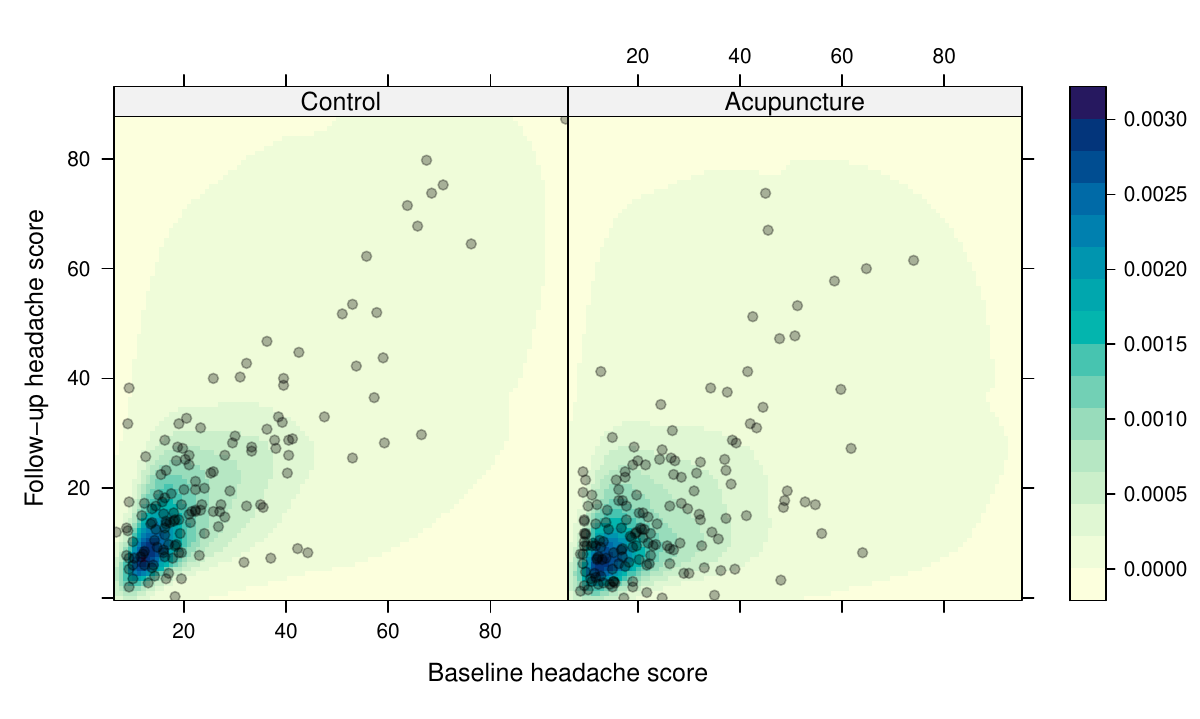} 

}

\caption[Estimated joint density \(f_{Y_1, Y_{12} \mid W}(y_1, y_{12} \mid w)\) of baseline headache score \(Y_1\) and follow-up headache score \(Y_{12}\) given treatment group \(W\), derived from model \texttt{m5} after marginalization over the other baseline covariates]{Estimated joint density \(f_{Y_1, Y_{12} \mid W}(y_1, y_{12} \mid w)\) of baseline headache score \(Y_1\) and follow-up headache score \(Y_{12}\) given treatment group \(W\), derived from model \texttt{m5} after marginalization over the other baseline covariates. Color shading represents the model-based density, with darker regions indicating higher density. Dots show the observed data.}\label{fig:fig-app_jointdensity}
\end{figure}

\end{knitrout}

\end{appendix}

\end{document}